\documentclass[longauth]{aa} 
\usepackage{natbib}
\usepackage{graphicx}
\usepackage{comment}
\usepackage{array}
\usepackage{hyperref}
\usepackage{threeparttable}
\usepackage{txfonts}
\usepackage{soul}
\usepackage[normalem]{ulem}

\usepackage[dvipsnames]{xcolor}
\usepackage{hyperref}
\usepackage{textcomp}
\def\clj{CL~J1226.9+3332}

\def\xmm{XMM-{\it Newton}}

\begin{document} 

   \title{Multi-probe analysis of the galaxy cluster CL~J1226.9+3332}

   \subtitle{Hydrostatic mass and hydrostatic-to-lensing bias}

   \author{M.~Mu\~noz-Echeverr\'ia
         \inst{\ref{LPSC}}\fnmsep\thanks{miren.munoz@lpsc.in2p3.fr}
          \and
          J. F. Mac\'ias-P\'erez \inst{\ref{LPSC}}
          \and
          G.~W.~Pratt \inst{\ref{CEA}}
          \and
          R.~Adam \inst{\ref{LLR}}
          \and
          P.~Ade \inst{\ref{Cardiff}}
          \and
          H.~Ajeddig \inst{\ref{CEA}}
          \and
          P.~Andr\'e \inst{\ref{CEA}}
          \and
          M.~Arnaud \inst{\ref{CEA}}
          \and
          E.~Artis \inst{\ref{LPSC}}
          \and
          H.~Aussel \inst{\ref{CEA}}
          \and
          I.~Bartalucci\inst{\ref{Milano}}
          \and
          A.~Beelen \inst{\ref{LAM}}
          \and
          A.~Beno\^it \inst{\ref{Neel}}
          \and
          S.~Berta \inst{\ref{IRAMF}}
          \and
          L.~Bing \inst{\ref{LAM}}
          \and
          O.~Bourrion \inst{\ref{LPSC}}
          \and
          M.~Calvo \inst{\ref{Neel}}
          \and
          A.~Catalano \inst{\ref{LPSC}}
          \and
          M.~De~Petris \inst{\ref{Roma}}
          \and
          F.-X.~D\'esert \inst{\ref{IPAG}}
          \and
          S.~Doyle \inst{\ref{Cardiff}}
          \and
          E.~F.~C.~Driessen \inst{\ref{IRAMF}}
          \and
          A.~Ferragamo \inst{\ref{Roma}}
          \and
          A.~Gomez \inst{\ref{CAB}}
          \and
          J.~Goupy \inst{\ref{Neel}}
          \and
          C.~Hanser \inst{\ref{LPSC}}
          \and
          F.~K\'eruzor\'e \inst{\ref{Argonne}}
          \and
          C.~Kramer \inst{\ref{IRAMF},\ref{IRAME}}
          \and
          B.~Ladjelate \inst{\ref{IRAME}}
          \and
          G.~Lagache \inst{\ref{LAM}}
          \and
          S.~Leclercq \inst{\ref{IRAMF}}
          \and
          J.-F.~Lestrade \inst{\ref{LERMA}}
          \and
          A.~Maury \inst{\ref{CEA}}
          \and
          P.~Mauskopf\inst{\ref{Cardiff},\ref{Arizona}}
          \and
          F.~Mayet \inst{\ref{LPSC}}
          \and
          J.-B.~Melin\inst{\ref{IRFU-CEA}}
          \and
          A.~Monfardini \inst{\ref{Neel}}
          \and
          A.~Paliwal \inst{\ref{Roma}}
          \and
          L.~Perotto \inst{\ref{LPSC}}
          \and
          G.~Pisano \inst{\ref{Roma}}
          \and
          E.~Pointecouteau\inst{\ref{Toulouse}}
          \and
          N.~Ponthieu \inst{\ref{IPAG}}
          \and
          V.~Rev\'eret \inst{\ref{CEA}}
          \and
          A.~J.~Rigby \inst{\ref{Cardiff}}
          \and
          A.~Ritacco \inst{\ref{ENS}, \ref{INAF}}
          \and
          C.~Romero \inst{\ref{Pennsylvanie}}
          \and
          H.~Roussel \inst{\ref{IAP}}
          \and
          F.~Ruppin \inst{\ref{Lyon}}
          \and
          K.~Schuster \inst{\ref{IRAMF}}
          \and
          S.~Shu \inst{\ref{Caltech}}
          \and
          A.~Sievers \inst{\ref{IRAME}}
          \and
          C.~Tucker \inst{\ref{Cardiff}}
         \and
         G.~Yepes \inst{\ref{Madrid}}
        }
          
   \institute{
     Univ. Grenoble Alpes, CNRS, LPSC-IN2P3, 53, avenue
des Martyrs, 38000 Grenoble, France
     \label{LPSC}
     \and
    Universit\'e Paris-Saclay, Universit\'e Paris Cit\'e, CEA, CNRS, AIM, 91191, Gif-sur-Yvette, France
     \label{CEA}
     \and
     LLR (Laboratoire Leprince-Ringuet), CNRS, École Polytechnique, Institut Polytechnique de Paris,
Palaiseau, France
     \label{LLR}
     \and
     School of Physics and Astronomy, Cardiff University, Queen’s
Buildings, The Parade, Cardiff, CF24 3AA, UK
     \label{Cardiff}
     \and
     INAF, IASF-Milano, Via A. Corti 12, 20133 Milano, Italy
     \label{Milano}
     \and
     Aix Marseille Univ, CNRS, CNES, LAM (Laboratoire d'Astrophysique
     de Marseille), Marseille, France
     \label{LAM}
     \and
     Institut N\'eel, CNRS, Universit\'e Grenoble Alpes, France
     \label{Neel}
     \and
     Institut de RadioAstronomie Millim\'etrique (IRAM), Grenoble, France
     \label{IRAMF}
     \and
     Dipartimento di Fisica, Sapienza Universit\`a di Roma, Piazzale
Aldo Moro 5, I-00185 Roma, Italy
     \label{Roma}
     \and
     Univ. Grenoble Alpes, CNRS, IPAG, 38000 Grenoble, France
     \label{IPAG}
     \and
     Centro de Astrobiolog\'ia (CSIC-INTA), Torrej\'on de Ardoz, 28850
Madrid, Spain
\label{CAB}
     \and
      High Energy Physics Division, Argonne National Laboratory, 9700 South Cass Avenue, Lemont, IL 60439, USA
     \label{Argonne}
     \and
     Instituto de Radioastronom\'ia Milim\'etrica (IRAM), Granada, Spain
     \label{IRAME}
     \and
     LERMA, Observatoire de Paris, PSL Research University, CNRS,
Sorbonne Universit\'e, UPMC, 75014 Paris, France
     \label{LERMA}
     \and
     School of Earth and Space Exploration and Department of Physics,
Arizona State University, Tempe, AZ 85287, USA
     \label{Arizona}
     \and
     IRFU, CEA, Universit\'e Paris-Saclay, 91191 Gif-sur-Yvette, France
     \label{IRFU-CEA}
     \and
     Univ. de Toulouse, UPS-OMP, CNRS, IRAP, 31028 Toulouse, France
     \label{Toulouse}
     \and
     Laboratoire de Physique de l’\'Ecole Normale Sup\'erieure, ENS, PSL
Research University, CNRS, Sorbonne Universit\'e, Universit\'e de Paris,
75005 Paris, France
      \label{ENS}
      \and
      INAF-Osservatorio Astronomico di Cagliari, Via della Scienza 5,
09047 Selargius, IT
     \label{INAF}
     \and
     Department of Physics and Astronomy, University of Pennsylvania,
209 South 33rd Street, Philadelphia, PA, 19104, USA
     \label{Pennsylvanie}
     \and
     Institut d'Astrophysique de Paris, CNRS (UMR7095), 98 bis boulevard
Arago, 75014 Paris, France
     \label{IAP}
     \and
     Univ. Lyon, Univ. Claude Bernard Lyon 1, CNRS/IN2P3, IP2I Lyon, F‐69622, Villeurbanne, France
     \label{Lyon}
     \and
     Caltech, Pasadena, CA 91125, USA
     \label{Caltech}
     \and
     Departamento de F\'isica Te\'orica and CIAFF, Facultad de Ciencias,
Modulo 8, Universidad Aut\'anoma de Madrid, 28049 Madrid, Spain
     \label{Madrid}
   }

   \date{Received ...; accepted ...}

 
  \abstract{The precise estimation of the mass of galaxy clusters is a major issue for cosmology. Large galaxy cluster surveys rely on scaling laws that relate cluster observables to their masses. From the high resolution observations of $\sim 45$ galaxy clusters with NIKA2 and XMM-\textit{Newton} instruments, the NIKA2 SZ Large Program should provide an accurate scaling relation between the thermal Sunyaev-Zel'dovich effect and the hydrostatic mass. In this paper, we present an exhaustive analysis of the hydrostatic mass of the well known galaxy cluster CL~J1226.9+3332, the highest-redshift cluster in the NIKA2 SZ Large Program at $z=0.89$. We combine the NIKA2 observations with thermal Sunyaev-Zel'dovich data from NIKA, Bolocam and MUSTANG instruments and XMM-\textit{Newton} X-ray observations and test the impact of the systematic effects on the mass reconstruction. We conclude that slight differences in the shape of the mass profile can be crucial when defining the integrated mass at $R_{500}$, which demonstrates the importance of the modeling in the mass determination. We prove the robustness of our hydrostatic mass estimates by showing the agreement with all the results found in the literature. Another key information for cosmology is the bias of the masses estimated assuming hydrostatic equilibrium hypothesis. Based on the lensing convergence maps from the Cluster Lensing And Supernova survey with Hubble (CLASH) data, we obtain the lensing mass estimate for CL~J1226.9+3332. From this we are able to measure the hydrostatic-to-lensing mass bias for this cluster, that spans from  $1 - b_{\mathrm{HSE/lens}} \sim 0.7$ to $1$, presenting the impact of data-sets and mass reconstruction models on the bias.}

   \maketitle

%
\section{Introduction}
\label{sec:intro}

Galaxy clusters are formed by gravitational collapse at the last step of the hierarchical structure formation process 
\citep{KravtsovBorgani2012}. Thus, they are tracers of the large-scale structure formation physics. Their abundance in mass and redshift is sensitive to the initial conditions in the primordial Universe and its expansion history and matter content \citep{huterer2015}. Therefore, galaxy clusters are probes of the underlying cosmology \citep{allen2011}.  

Large catalogs of clusters \citep[e.g.][]{planck2016,rykoff2016, adami2018,bleem2020} have enabled, in the past few years, to constrain cosmological parameters. Nevertheless, these results show some tension with respect to the cosmology obtained from the Cosmic microwave background (CMB) power spectrum analysis \citep{planck2014b, planck2016a,salvati2018}, in line with a more general problem in that early- and late-Universe probes are giving different results \citep{Verde2019}. Cluster-based cosmological analyses rely on their distribution in mass and one source of the discrepancy may be the inaccuracy on those mass estimates \citep[e.g.][]{pratt2019,salvati2020}.

About $85 \%$ of the total mass of clusters of galaxies is composed of dark matter. The remaining $15 \%$ corresponds to the hot intra-cluster medium (ICM) and the galaxies in the cluster. For this reason, most of the mass content of the clusters is not directly observable and it needs to be estimated either from the gravitational potential reconstruction or via scaling relations that link cluster observables to their masses \citep{pratt2019}.
The gravitational potential of clusters can be inferred from their gravitational lensing effect on background sources \citep{bartelmann2010}, from the dynamics of member galaxies \citep{biviano2003,aguado2021} or from the combination of the thermodynamical properties of the gas in the ICM under hydrostatic equilibrium (HSE) hypothesis. 
Scaling relations that allow to recover the mass have been measured and calibrated for various cluster observables at different wavelengths: in the optical and infrared domains the galaxy member richness \citep{andreon2009}, in X-ray the emitted luminosity of the hot gaseous ICM \citep{pratt2009} and in millimeter wavelengths the amplitude of the Sunyaev-Zel'dovich (SZ) effect \citep{sunyaev,planck2011} proportional to the thermal energy in the ICM. 

The thermal SZ (tSZ) effect \citep{sunyaev}, corresponding to the spectral distortion of the CMB photons due to the inverse Compton scattering on the hot thermal electrons of the ICM, is an excellent way of detecting clusters as its amplitude is not affected by cosmological dimming.  
Large surveys, in particular the \textit{Planck} satellite observations \citep{planck2016} and the ground-based Atacama Cosmology Telescope \citep[ACT,][]{hilton2018,Hilton2020} and South Pole Telescopes \citep[SPT,][]{bleem2020}, have obtained large SZ-detected galaxy cluster catalogs. However, they need to rely on the aforementioned SZ-mass scaling relation in order to carry out cosmological analyses. So far, most of the SZ-mass scaling relations have been mainly determined from low-redshift ($z < 0.5$) cluster samples with masses obtained from X-ray observations \citep{arnaud10, planck2011}. Other scaling relations, with optical data for example \citep{saro2017}, are also used.

In any case, it is essential to study the redshift evolution of these scaling relations, as they would impact the cosmological results \citep{salvati2020}. When building scaling relations, another key aspect is the impact of the assumptions on which the mass computations rely, such as the HSE hypothesis. High spatial resolution cluster observations can assess these assumptions by studying the impact of the dynamical states of clusters on the mass-observable scaling relations.

The NIKA2 SZ Large program \citep{mayet2020, perotto2022}, described in Sect.~\ref{sec:nika2}, seeks to address the aforementioned issues. The work presented in this paper constitutes the third analysis of a cluster in the NIKA2 SZ Large Program. The first analysis on PSZ2~G144.83+25.11 comprised a science verification study, as well as the proof of the impact of substructures in the reconstruction of the physical cluster properties \citep{ruppin1}. The second, the worst-case scenario for the NIKA2 SZ Large Program, analysed the ACT-CL~J0215.4+0030 galaxy cluster, proving the quality of NIKA2 camera \citep{adam1,NIKA2-electronics,calvo, perotto} in the most challenging case of a high-redshift and low mass cluster \citep{keruzore}. In this work we present a study on the HSE mass of the CL~J1226.9+3332 galaxy cluster, the impact of different systematic effects on the recovered mass, as well as a thorough comparison to previous works. Moreover, we re-estimate its lensing mass and compare it to the mass obtained under the hydrostatic assumption, therefore allowing us to compute the hydrostatic-to-lensing mass bias.


This paper is organized as follows. In Sect.~\ref{clj1227} we present a summary of the results in the literature for CL~J1226.9+3332. 
In Sect.~\ref{sec:icmobservations} we describe the observations of the ICM with NIKA2 and XMM-\textit{Newton} instruments.
The reconstruction of the thermodynamical properties of the ICM is presented in  Sect.~\ref{sec:icmprofiles}. We detail the pressure profile reconstruction from NIKA2 maps, accounting for systematic effects due to the data reduction process and point source contamination. We compare these profiles to previous results as well as to the X-ray pressure profile. In Sect.~\ref{sec:hsemassgeneral} we describe the HSE mass estimates from the combination of SZ and X-ray data and test the robustness of the results against data processing and modeling effects. 
In Sect.~\ref{othermasses} we present the lensing masses that we estimate for CL~J1226.9+3332.
All the results are put together in Sect.~\ref{sec:discussions} for discussion, where we compute the hydrostatic-to-lensing mass bias for this cluster. The conclusions are given in Sect.~\ref{conclusion}.

Throughout this paper we assume a flat $\Lambda$CDM cosmology with $H_{0} = 70 \hspace{2pt} \mathrm{km/s/Mpc}$ and $\Omega_{m,0} = 0.3$. With this assumption, 1 arcmin corresponds to a distance of 466~kpc at the cluster redshift, $z=0.89$.


\section{The CL~J1226.9+3332 galaxy cluster}
\label{clj1227}
This paper focuses on the CL~J1226.9+3332 galaxy cluster, also known as PSZ2-G160.83+81.66.  Discovered by the Wide Angle ROSAT Pointed Survey \citep[WARPS,][]{ebeling}, it has already been studied at different wavelengths: in X-ray \citep{maughan2, bonamente}, visible \citep{jeetyson} and millimeter \citep{bonamente, mroczkowski2009, korngut, adam2} wavelengths. Located at redshift 0.89 \citep{planck2016,aguado2021}, it is the highest-redshift cluster of the NIKA2 SZ Large Program sample, with the X-ray peak at (R.A., Dec.)$_\mathrm{J2000}$ = (12h26m58.37s, +33d32m47.4s)  according to \citet{cavagnolo}. Less than 2 arcseconds away from this peak, its brightest cluster galaxy (BCG) is located at (R.A., Dec.)$_\mathrm{J2000}$ = (12h26m58.25s, +33d32m48.57s) according to \citet{holden}.

\subsection{Previous observations}
Since the first SZ observations with BIMA \citep{joy}, the projected morphology of CL~J1226.9+3332 appeared to have a quite circular symmetry. Nevertheless, the combination of XMM-\textit{Newton} and \textit{Chandra} X-ray data \citep{maughan1} showed a region, at $\sim$~40’’ to the south-west of the X-ray peak, with much higher temperature than the average in the ICM. This substructure was also confirmed by posterior SZ analyses with MUSTANG \citep{korngut} and NIKA \citep{monfardini,adam2, adam2018_2}. \citet{romero}, hereafter R18, performed a study combining SZ data from NIKA, MUSTANG and Bolocam instruments. Their different capabilities enabled to probe different angular scales in the reconstruction of ICM properties and agreed with a non-relaxed cluster core description for CL~J1226.9+3332. In this work we will make use of the pressure profiles obtained from NIKA, MUSTANG and Bolocam data summarized in Table 2 in \citet{romero}. 

Lensing data from the Cluster Lensing And Supernova survey with Hubble \citep[CLASH,][]{zitrin1}, as well as the galaxy distribution in the cluster \citep{jeetyson}, agree on the existence of a main clump centered on the BCG and a secondary clump on its south-west. 

However, this second region does not appear as a structure in X-ray surface brightness \citep{maughan1}. One hypothesis presented in \citet{jeetyson} suggests that the mass of the south-western galaxy group is not big enough to be observed as an X-ray overdensity. Motivated by the slight elongation of the X-ray peak toward the southwest, \citet{jeetyson} also hypothesize that the two-halo system is being observed after the less massive cluster has passed through the central one. A previous study \citep{maughan2} showed also a region of cooler emission in the west side of the BCG, that is, in the north of the mentioned hot region. This was seen using \textit{Chandra} data and it was explained as a possible infall of some cooler body. Additionally, from the diffuse radio emission analysis with LOFAR data, \citet{digennaro} showed that CL~J1226.9+3332 hosts the most distant radio halo discovered to date: a radio emission with a size of 0.7~Mpc that follows the thermal gas distribution. In brief, CL~J1226.9+3332 shows evidence of disturbance in the core, but a relaxed morphology at large scales.

\subsection{The mass of CL~J1226.9+3332}
\label{sec:allmasses}
Regarding the mass of CL~J1226.9+3332, which constitutes the main topic in this study, we present here the results obtained in previous works (summarized in Table~\ref{tab-masses}). These masses have not been homogenised, nor scaled to the same cosmology and are the values extracted directly from different analyses. We define $M_{\Delta}$ as the mass of the cluster inside a sphere of radius $R_{\Delta}$. $R_{\Delta}$ is the radius at which the mean mass density of the cluster is $\Delta$ times the critical density of the Universe at its redshift, $\rho_{c} = 3 H (z)^{2}/(8 \pi G)$ where $H (z)$ is the Hubble function. 

The first SZ mass analysis of this cluster was done in \citet{joy} and they estimated $M (r < 340 \; h_{100}^{-1} \; \mathrm{kpc}) = (3.9 \pm 0.5) \times 10^{14}\hspace{2pt} \mathrm{M}_{\odot}$. Using \textit{Chandra} X-ray data from \citet{cagnoni} and assuming an isothermal $\beta$-model, \citet{jeetyson} obtained the hydrostatic projected mass $M (r < 1 \hspace{2pt}\mathrm{Mpc}) = 1.4^{+0.6}_{-0.4} \times 10^{15} \hspace{2pt}\mathrm{M}_{\odot}$. Also assuming an isothermal $\beta$-model and hydrostatic equilibrium, \citet{maughan2} obtained with XMM-\textit{Newton} data $M_{1000}$ 
$ = 6.1^{+0.9}_{-0.8} \times 10^{14} \hspace{2pt}\mathrm{M}_{\odot}$ and 
$M_{200}$ 
$ = (1.4 \pm 0.5) \times 10^{15} \hspace{2pt}\mathrm{M}_{\odot}$. The subsequent analysis of three dimensional hydrodynamical properties with \textit{Chandra} and XMM-\textit{Newton} by \citet{maughan1}, again under the assumptions of spherical symmetry and hydrostatic equilibrium, concluded that $M_{500}$ 
$ = 5.2^{+1.0}_{-0.8} \times 10^{14} \hspace{2pt}\mathrm{M}_{\odot}$. According to the X-ray analysis in \citet{mantz2010}, $M_{500}$ 
$ = (7.8 \pm 1.1) \times 10^{14} \hspace{2pt}\mathrm{M}_{\odot}$. 
\begin{figure}
    \centering
    \includegraphics[trim={0cm 0cm 0.0cm 0cm},scale=0.43]{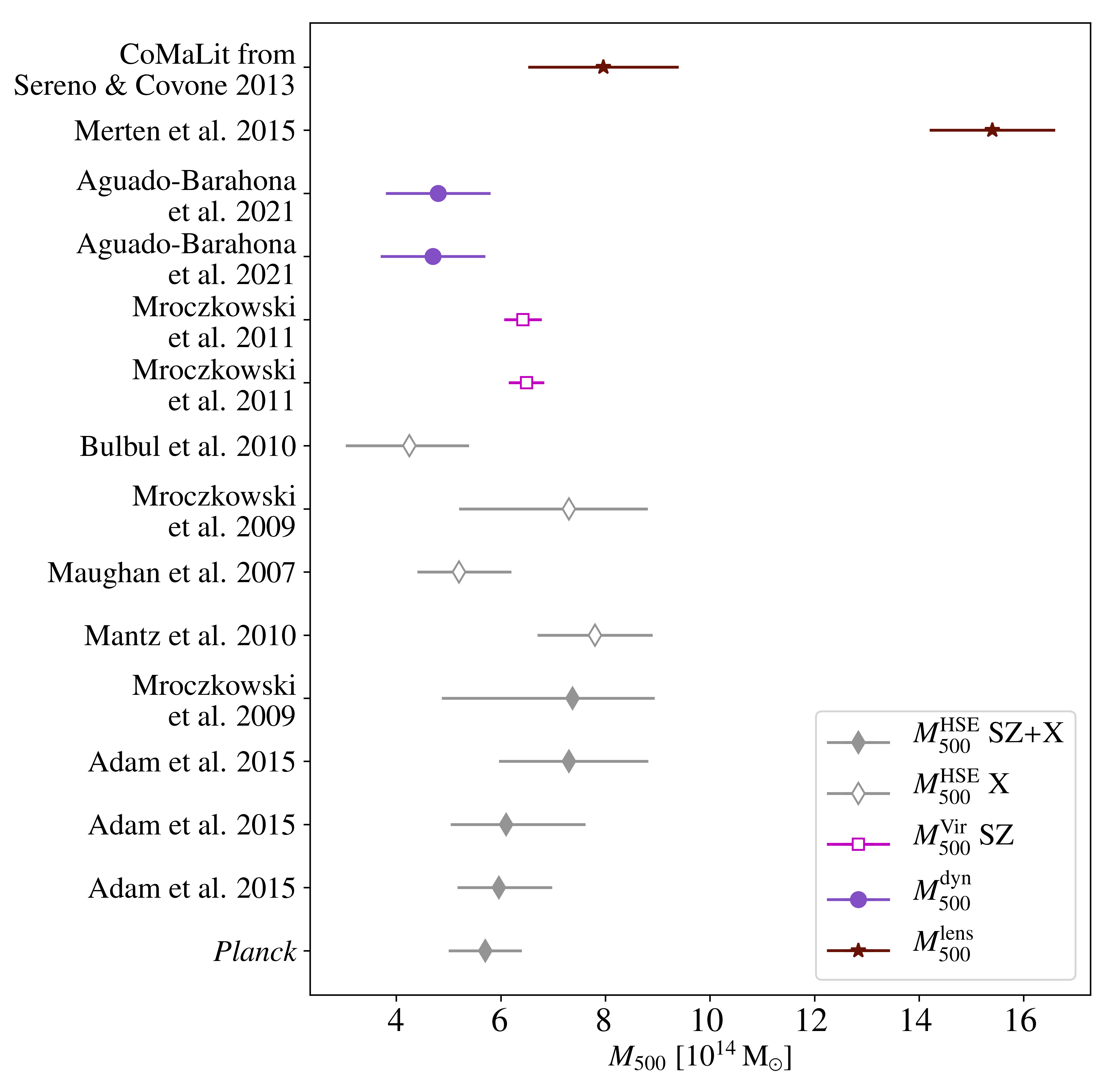}
    \caption{$M_{500}$ estimates for CL~J1226.9+3332 in literature. Full grey diamonds represent HSE masses from the combination of SZ and X-ray data and empty ones correspond to X-ray-only results. Pink squares show the SZ-only mass assuming Virial relation, purple circles are dynamical mass estimates and brown stars correspond to lensing $M_{500}$.} 
    \label{fig:masses_literature}
\end{figure} 
From the combination of Sunyaev-Zel’dovich Array \citep[SZA,][]{muchovej} interferometric data and the \textit{Chandra} X-ray observations, under hydrostatic equilibrium hypothesis, \citet{mroczkowski2009} obtained $M_{500}$ 
$ = 7.37^{+2.50}_{-1.57} \times 10^{14} \hspace{2pt}\mathrm{M}_{\odot}$ and $M_{2500}$ 
$ = 2.67^{+0.29}_{-0.27} \times 10^{14} \hspace{2pt}\mathrm{M}_{\odot}$.  This was compared to the results using only the X-ray data and assuming an isothermal $\beta$-model: $M_{500}$ 
$ = 7.30^{+2.10}_{-1.51} \times 10^{14} \hspace{2pt}\mathrm{M}_{\odot}$, $M_{2500}$ 
$ = 2.98^{+0.90}_{-0.63} \times 10^{14} \hspace{2pt}\mathrm{M}_{\odot}$. Using a new approach that instead relies on the Virial relation and the Navarro-Frenk-White \citep[NFW,][]{navarro} density profile, \citet{mroczkowski2011} and \citet{mroczkowski2012} estimated the mass for CL~J1226.9+3332 using only SZ data from SZA:  $M_{500}$ 
$ = 6.49^{+0.34}_{-0.34} \times 10^{14} \hspace{2pt}\mathrm{M}_{\odot}$ and $M_{2500}$ 
$ = 2.35^{+0.15}_{-0.16} \times 10^{14} \hspace{2pt}\mathrm{M}_{\odot}$ assuming a pressure described by a generalized Navarro-Frenk-White \citep[gNFW,][]{nagai} profile with $(a,b,c) = (0.9,5.0,0.4) $ parameters and $M_{500}$ 
$ = 6.42^{+0.36}_{-0.36} \times 10^{14} \hspace{2pt}\mathrm{M}_{\odot}$ and $M_{2500}$ 
$ = 2.53^{+0.14}_{-0.15} \times 10^{14} \hspace{2pt}\mathrm{M}_{\odot}$ with $(a, b, c) = (1.0510, 5.4905, 0.3081)$ as in \citet{arnaud10}. Some years before, \citet{muchovej} fitted the temperature decrement due to the cluster's SZ effect to the SZA data and assuming hydrostatic equlibrium and isothermality estimated: $M_{200}$ 
$= 7.19^{+1.33}_{-0.92} \times 10^{14} \hspace{2pt}\mathrm{M}_{\odot}$ and $M_{2500}$ 
$= 1.68^{+0.37}_{-0.26} \times 10^{14} \hspace{2pt}\mathrm{M}_{\odot}$.
 
Another approach was considered in \citet{bulbul} to compute the hydrostatic mass, with the polytropic equation of state and using only \textit{Chandra} X-ray observations, $M_{500}$ 
$ = 4.25^{+1.22}_{-1.14} \times 10^{14} \hspace{2pt}\mathrm{M}_{\odot}$ and $M_{2500}$ 
$ = 2.16^{+0.69}_{-0.63} \times 10^{14} \hspace{2pt}\mathrm{M}_{\odot}$. According to \citet{planck2016} results, the hydrostatic mass of the cluster is $M_{500} = 5.70^{+0.63}_{-0.69} \times 10^{14} \hspace{2pt}\mathrm{M}_{\odot}$. This mass was obtained using the SZ-mass scaling relation  given in the Eq.~7 of \citet{planck2014a}. 
 
In addition, combining SZ data from NIKA and \textit{Planck} with the X-ray electron density from the \textit{Chandra} ACCEPT data \citep{cavagnolo}, \citet{adam2} obtained three hydrostatic mass estimates for different parameters in their gNFW pressure profile modeling: $M_{500}$ 
$ = 5.96^{+1.02}_{-0.79} \times 10^{14} \hspace{2pt}\mathrm{M}_{\odot}$ using $(a, b,c) = (1.33, 4.13, 0.014)$, 
$M_{500}$ 
$ = 6.10^{+1.52}_{-1.06} \times 10^{14} \hspace{2pt}\mathrm{M}_{\odot}$ with $(b,c) = (4.13, 0.014)$ and $M_{500}$ 
$ = 7.30^{+1.52}_{-1.34} \times 10^{14} \hspace{2pt}\mathrm{M}_{\odot}$ with $(a, b,c) = (0.9, 5.0, 0.4)$.

The weak-lensing analysis in \citet{jeetyson} realized by fitting a NFW density profile found that $M_{200}$ 
$ = (1.38 \pm 0.20) \times 10^{15} \hspace{2pt}\mathrm{M}_{\odot}$. Similarly, they computed the weak-lensing mass estimate at $R_{500}$ from \citet{maughan1}: $M (r < (0.88 \pm 0.05) \hspace{2pt}\mathrm{Mpc}) = (7.34 \pm 0.71) \times 10^{14} \hspace{2pt}\mathrm{M}_{\odot}$, therefore finding a $30 \%$ higher mass than the X-ray estimate in \citet{maughan1}. This discrepancy was explained in \citet{jeetyson} as a sign of an on-going merger in the cluster that would create an underestimation of the hydrostatic mass with X-rays without altering the lensing estimate. \citet{jeetyson} also estimated the projected mass in each of the two big substructures within $r <$~20'': for the most massive and central clump they found $M (r < 20") = (1.3 \pm 0.1) \times 10^{14} \hspace{2pt}\mathrm{M}_{\odot}$ and for the structure at $\sim  40"$ to the southwest of the BCG, $M (r < 20") = (8.5 \pm 0.6) \times 10^{13} \hspace{2pt}\mathrm{M}_{\odot}$. \citet{merten2015} performed a lensing analysis and obtained $M_{200} =  (2.23 \pm 0.14) \times 10^{15} \hspace{2pt}\mathrm{M}_{\odot}$, $M_{500} =  (1.54 \pm 0.12) \times 10^{15} \hspace{2pt}\mathrm{M}_{\odot}$ and $M_{2500} =  (0.61 \pm 0.10) \times 10^{15} \hspace{2pt}\mathrm{M}_{\odot}$ by fitting a NFW density profile to the CLASH data. In addition, based on the weak and strong lensing analysis from \citet{serenocovone2013}, \citet{serenocomalit} followed the same procedure as for all clusters in the CoMaLit\footnote{\label{comalitref}Comparing masses in literature. Cluster lensing mass catalog available at \url{http://pico.oabo.inaf.it/\~sereno/CoMaLit/}} sample and obtained $M_{500} =  (7.96 \pm 1.44) \times 10^{14} \hspace{2pt}\mathrm{M}_{\odot}$.

Moreover, a recent study based on the velocity dispersion of galaxy members in \citet{aguado2021} obtained two dynamical mass estimates for CL~J1226.9+3332: $M_{500} = (4.7 \pm 1.0) \times 10^{14} \hspace{2pt}\mathrm{M}_{\odot}$ and $M_{500} = (4.8 \pm 1.0) \times 10^{14} \hspace{2pt}\mathrm{M}_{\odot}$, from the velocities of 52 and 49 member galaxies, respectively.

 We display in Fig.~\ref{fig:masses_literature} the different $M_{500}$ estimates found in the literature. Grey diamonds with error bars correspond to the HSE mass estimates. We distinguish the HSE masses obtained from the combination of SZ and X-ray data (full diamonds) and the X-ray-only results (empty diamonds). The mass given by \citet{planck2016} is considered here a SZ+X result, but it is important to keep in mind that this mass was obtained applying a scaling relation (derived from X-ray data) to the \textit{Planck} SZ signal. The empty pink squares show the $M_{500}$ assuming the Virial relation and using only SZ data \citep{mroczkowski2011,mroczkowski2012}. The purple circles are the dynamical masses from \citet{aguado2021} and the brown stars the lensing estimates from \citet{merten2015} and \citet{serenocomalit}. We decide not to present in the same figure the projected masses, as it would be misleading to compare them to the masses integrated in a sphere. The figure shows that both HSE and lensing masses among them vary more than $40 \%$ from one analysis to another. 

 All these mass estimates for CL~J1226.9+3332 are hindered by systematic effects, which are difficult to deal with. Comparing properly masses obtained from different observables, methods or modeling approaches is crucial, but very challenging. 
Moreover, as the shape of the HSE mass profile varies depending on the data and the analysis procedure that is considered, the value of $R_{500}$ is not the same for all estimates presented in Fig.~\ref{fig:masses_literature}. Comparisons are thus delicate due to the correlation between the mass and the radius at which it is estimated. As mentioned, an accurate knowledge of the mass of galaxy clusters will be essential for cosmological purposes \citep{pratt2019}. This motivates the study in this paper, which follows previous work in \citet{ferragamo}.

\section{ICM observations}
\label{sec:icmobservations}
\subsection{NIKA2}
\label{sec:nika2general}
    \subsubsection{NIKA2 and the SZ Large Program}
    \label{sec:nika2}
    The NIKA2 camera \citep{adam1,NIKA2-electronics,calvo} is a millimeter camera operating at the IRAM 30-meter telescope on Pico Veleta (Sierra Nevada, Spain). It observes simultaneously in two bands centered at 150 and 260~GHz with three arrays of kinetic inductance detectors \citep[KIDs,][]{day2003,shu2018}, two operating at 260~GHz and one at 150~GHz. These frequency bands are adapted to detect the tSZ effect characteristic distortion, which shows a decrement of the CMB brightness at frequencies lower than $\sim$~217 GHz and an increment at higher frequencies. As described in the instrument performance analysis review \citep{perotto}, NIKA2 maps the sky with a 6.5' field of view and high angular resolution, 17.6” and 11.1” full width at half maximum at 150 and 260~GHz, respectively. These capabilities, combined with its high sensitivity (9 mJy·s$^{1/2}$ at 150 GHz and 30 mJy·s$^{1/2}$ at 260 GHz), enable to study the ICM of galaxy clusters in detail, as it has already been proved in previous works \citep{ruppin1, keruzore}. Substructures in clusters as well as contaminant sources can also be identified.
    
    Part of the NIKA2 guaranteed time was allocated to the NIKA2 SZ Large Program. This program is a high angular resolution follow-up of $\sim$ 45 galaxy clusters detected with \textit{Planck} and ACT \citep{planck2016,hasselfield2013}. These clusters have been chosen at intermediate to high redshift ranges (0.5 $\leq$ $z$ $\leq$ 0.9) and covering a wide range of masses (3 $\leq M_{500}/10^{14} \hspace{1pt}\mathrm{M}_{\odot} \leq$ 10) estimated according to their tSZ signal and mass-observable scaling relations. The improvement in resolution with respect to the previous instruments, going below the arcminute scale, allows us to resolve distant clusters for which the apparent angular sizes are small. The SZ Large Program also benefits from the X-ray observations obtained with the XMM-\textit{Newton} satellite. From the combination of tSZ and X-ray data, the NIKA2 SZ Large Program will be able to re-estimate precise HSE masses of the galaxy clusters in the sample and improve the current SZ-mass scaling relations. A more detailed explanation of the NIKA2 SZ Large Program is given in \citet{mayet2020, perotto2022}.

    \subsubsection{NIKA2 observations and maps of \clj\ }
    \label{sec:cljnika2}
    CL~J1226.9+3332 was observed for 3.6 hours during the 15th NIKA2 science-purpose observation campaign (13-20 February 2018) as part of the NIKA2 Guaranteed Time (under project number 199-16) at the IRAM 30-m telescope. The data consists of 36 raster scans of $8 \times 4$ arcminutes in series of four scans with angles of 0, 45, 90, and 135 degrees with respect to the right-ascension axis. The scans were centered at the XMM-\textit{Newton} X-ray peak, (R.A., Dec.)$_\mathrm{J2000}$ = (12h26m58.08s, +33d32m46.6s). The mean elevation of the scans is of 58.51$^{\circ}$. Data at 150 and 260~GHz were acquired simultaneously. 

     The raw data were calibrated and reduced following the \emph{baseline} method developed for the performance assessment of NIKA2 \citep{perotto} and extended to diffuse emission as in \citet{keruzore} and \citet{rigby}. The data for the two frequency bands are processed independently. Common modes of most-correlated detectors are estimated for data outside a circular mask covering the cluster. 
     These modes are subtracted from the raw data and a new mask is defined at the positions where the significance of the 
    signal is above a threshold, S/N~$>$~3 in this case. At the next iteration the common modes are estimated outside the new mask and we repeat the procedure until convergence. We chose as initial mask a 2 arcminute diameter disk centered at the center of the scans.
    \begin{figure*}
        \begin{minipage}[b]{0.48\textwidth}
        \includegraphics[trim={0.5cm 0.1cm 0.5cm 0cm},clip,scale=0.20]{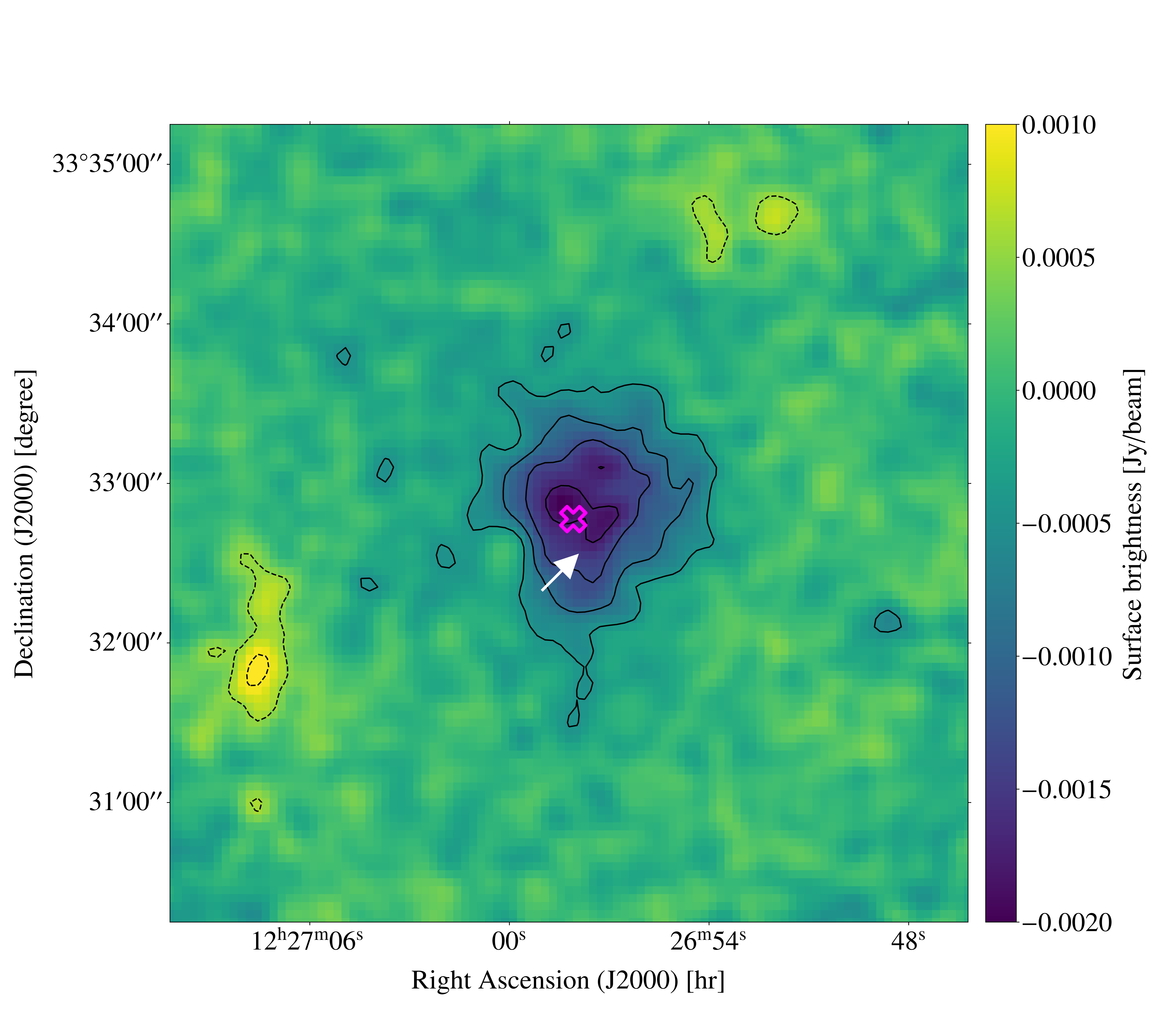}
        \end{minipage}
        \hfill
        \begin{minipage}[b]{0.48\textwidth}
        \includegraphics[trim={0.5cm 0.1cm 0.5cm 0cm},clip,scale=0.20]{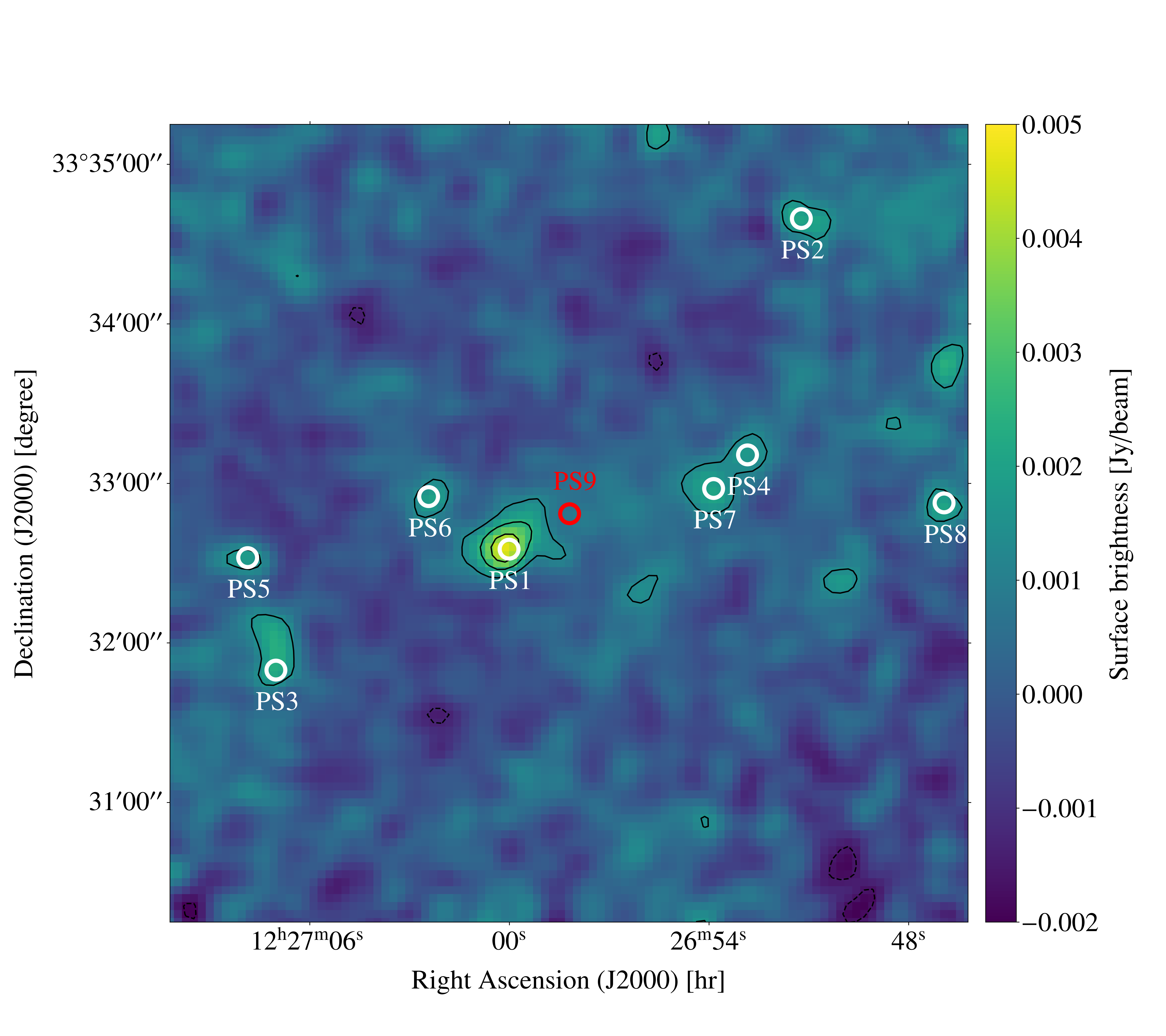}
        \end{minipage}
    \caption{NIKA2 maps of CL~J1226.9+3332 at 150~GHz (left) and 260~GHz (right) in Jy/beam units. Contours show S/N levels multiples of $\pm 3\sigma$. Both maps have been smoothed with a 10” FWHM Gaussian kernel. The position of the X-ray center is shown as a magenta cross in the 150~GHz map and the elongation of the tSZ signal towards the south west is indicated by the white arrow. White and red circles in the 260~GHz map show the submillimetric and radio point sources, respectively. }  
    \label{fig:maps}
    \end{figure*}
 
    
    We present in Fig. \ref{fig:maps} the resulting NIKA2 surface brightness maps at 150 and 260~GHz for CL~J1226.9+3332. Black contours indicate significance levels starting from $3\sigma$ with a $3\sigma$ spacing. The map at 150~GHz (left panel) shows the cluster as a negative decrement with respect to the background. This is the characteristic signature of the tSZ effect at frequencies lower than 217~GHz. In this map we also identify positive sources that can compensate the negative tSZ signal of the cluster, as it is the case for the central southeastern source. Moreover, in the 150~GHz map we observe an elongation of the tSZ peak towards the south west. Similar structures were found by \citet{maughan1, korngut,adam2, zitrin1} and \citet{jeetyson} as mentioned in Sect. \ref{clj1227}. 
    The 260~GHz map, in the right panel, is dominated by the signal of the point sources and can be used to identify them. Nonetheless, tSZ signal of the cluster is not detected at this frequency. This is expected, since the integrated tSZ signal is about three times weaker in the NIKA2 band centered at 260~GHz than in the 150~GHz one.
    
    \subsubsection{Estimation of noise residuals in the NIKA2 maps}
    \label{noise}
\begin{figure}
        \centering
        \includegraphics[trim={4cm 0cm 5cm 0cm}, scale=0.3]{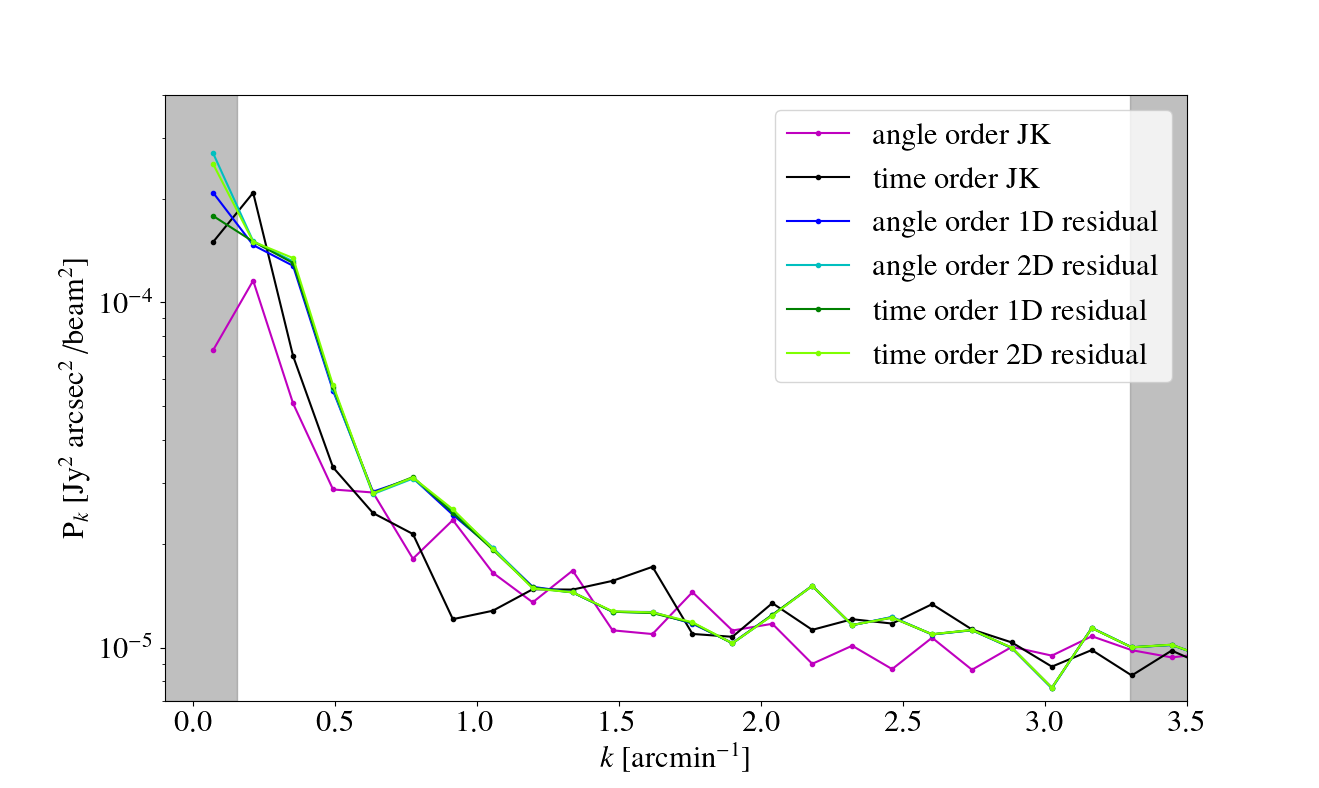}
        \caption{Power spectra of noise map estimates for the NIKA2 150 GHz data: in pink and black the spectra for the JK maps estimated with angle ordered and time ordered scans, respectively. In blue and green we show the power spectra of the different residual maps  for the best-fit models shown in Fig.~\protect\ref{fig:datamodel}. The grey regions represent the NIKA2 instrumetal limits given by the field of view (for small angular frequencies) and the beam FWHM (for large angular frequencies). }
        \label{fig:pkidl}
    \end{figure}

    The residual noise in the final 150~GHz NIKA2 map of \clj\ needs to be quantified for the reconstruction of the pressure profile of the cluster. It is usually estimated on null maps, also known as jackknives (JK), by computing half-differences of two statistically equivalent sets of scans in order to eliminate the astrophysical signal and recover the residuals. 
    We have developed two different noise estimates to evaluate possible systematic bias and uncertainties. The so called ``angle order'' (AO) noise map is computed from the half-differences of scans observed with the same angle with respect to the right-ascension axis. This ensures that signal residuals from differential filtering along the scan direction are minimized in the null maps. Alternatively, the so-called ``time order'' (TO) noise map is calculated from the half-differences of consecutive scans. This minimizes the time dependent effects that may be induced by atmospheric residual fluctuations. We present in Fig. \ref{fig:pkidl} in pink and black the power spectra of the AO and TO null maps at 150~GHz for CL~J1226.9+3332, respectively. The AO null map has a flatter power spectrum for small wave numbers, meaning that it contains less large-scale correlated noise than the TO null map. This would suggest that the TO null map might be affected by signal or atmospheric residuals differently filtered for each scanning angle. 
    The power spectra shown in Fig.~\ref{fig:pkidl} have been used to compute 1000 Monte Carlo noise realizations to estimate the pixel-pixel noise covariance matrices used in Sect.~\ref{sec:pressure} \citep[following the method developed in][]{adam2016}.
    
    \subsubsection{Pipeline Transfer function estimation}
    \label{Sect:transferfunction}
\begin{figure*}
        \begin{minipage}[b]{0.55\textwidth}
        \includegraphics[scale=0.33]{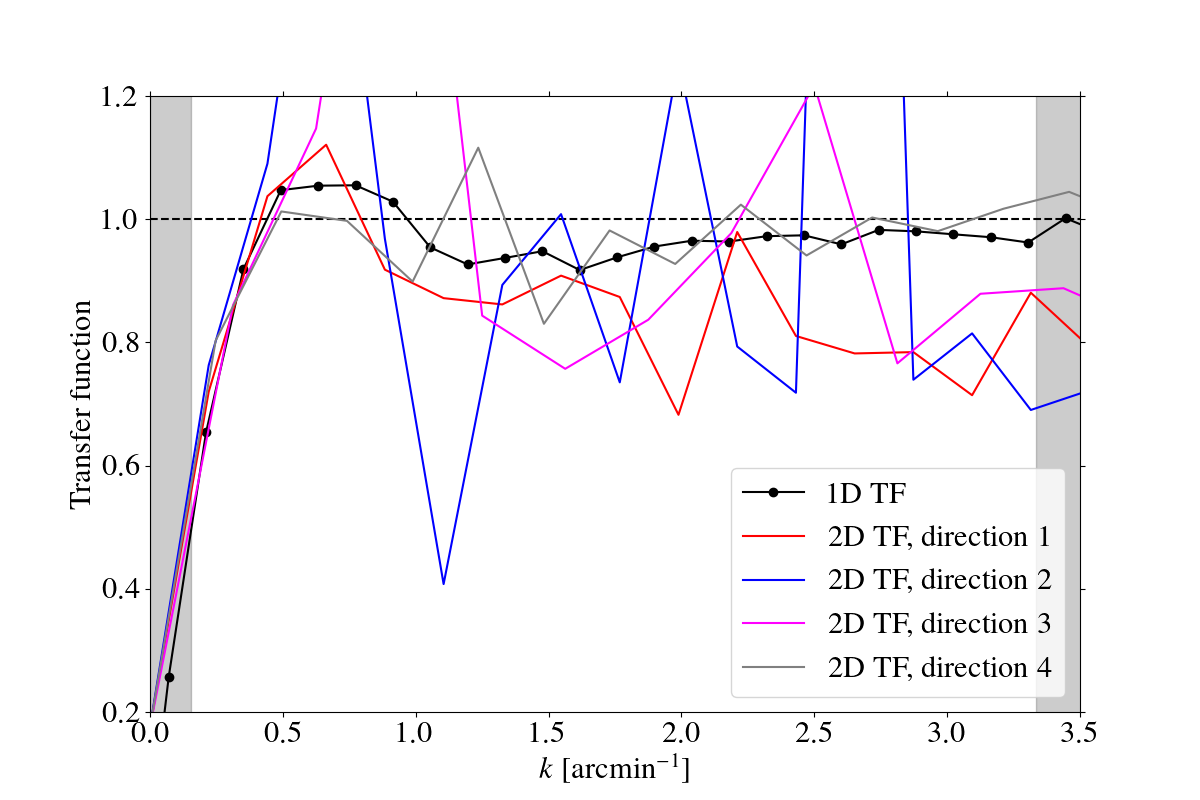}
        \end{minipage}
        \hfill
        \begin{minipage}[b]{0.42\textwidth}
        \includegraphics[trim={0cm 0.5cm 0cm 0cm},clip,scale=0.35]{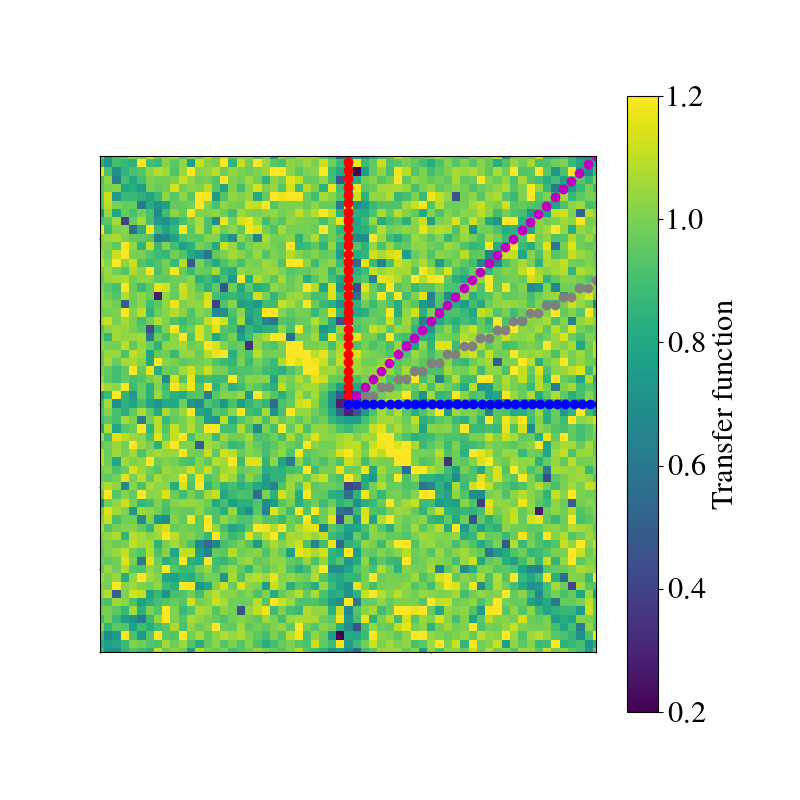}
        \end{minipage}
    \caption{1D (left) and 2D (right) transfer functions describing the filtering induced by data processing for the 150 GHz map in Fig.~\protect\ref{fig:maps}. Color lines in the left panel represent the values of the 2D transfer function for the directions shown, with the same colors, in the right panel. Grey shaded areas correspond to the NIKA2 field of view (for small angular frequencies) and beam FWHM (for large angular frequencies) instrumental limits.}  
    \label{fig:tf}
    \end{figure*}

    The filtering induced by the observations and the data reduction process on the cluster signal needs to be evaluated to be accounted for when estimating the pressure profile of the cluster. The transfer function is calculated by repeating the data processing steps discussed above on a simulation of the cluster signal. It is computed in Fourier space  as the ratio of the power spectrum of the simulation filtered by the processing and that of the input simulation. To simulate the cluster tSZ signal we used the universal pressure profile described in \citet{arnaud10} with the integrated tSZ signal\footnote{\label{y500} $Y_{500}$ is the integrated tSZ signal within a sphere of radius $R_{500}$ with $\mathcal{D}_{\mathrm{A}}$ the angular diameter distance at the cluster redshift. It is defined as: $Y_{500} = 4\pi \frac{\sigma_{\mathrm{T}}}{m_{e}c^2 \mathcal{D}_{\mathrm{A}}}\int_{0}^{R_{500}} r^2 P_{e}(r)\hspace{2pt} \mathrm{d}r$} and redshift of CL~J1226.9+3332 given in the \citet{planck2016} catalog, $Y_{500}^{Planck} \sim 3.82 \hspace{2pt}\times 10^{-4} \hspace{2pt}\mathrm{arcmin}^{2}$ and $z=0.89$. 
    We also add to the simulated cluster a Gaussian signal with flat spectrum (i.e. a random white noise) to explore angular scales at which the cluster signal is negligible. Up to now the NIKA2 SZ Large Program and NIKA analyses \citep{adam2,ruppin1,keruzore} considered one-dimensional transfer functions, hereafter 1D TF. In these cases circular symmetry is assumed and the 1D TF is obtained by averaging the power spectra ratio in Fourier-domain annuli at a fixed angular scale. Nevertheless, it is expected that the filtering is not isotropic in the map as it might depend on the scanning direction. This motivates the use of the two-dimensional transfer function, hereafter 2D TF. In the right panel of Fig.~\ref{fig:tf} we present the 2D TF describing the filtering in the NIKA2 150 GHz map of Fig.~\ref{fig:maps}. The black line in the left panel of the figure shows the 1D TF, whereas the color lines correspond to the one-dimensional cuts of the 2D TF for the different directions represented in the right plot.
    Except for the scanning directions, the 2D TF  is compatible with the 1D one and greater than 0.8 at large angular scales, meaning that the signal is well preserved. 
    On the contrary, filtering is significant for angular frequencies below $\sim 0.5 \hspace{2pt} \mathrm{arcmin}^{-1}$. At 0.4~arcmin$^{-1}$ $\lesssim$ $k$ $\lesssim$ 0.8~arcmin$^{-1}$ the transfer function is larger than unity, meaning that the signal has been slightly enhanced by the data analysis process at these scales. 
    In order to evaluate the impact of considering the anisotropy of the filtering, the analyses presented in the following sections will be carried out
    with both the 1D and 2D transfer functions (see Sect.~\ref{sec:pressure}).
\subsubsection{Point source contamination}
\label{Sect:ps}
    Point sources in the 150 GHz map contaminate the tSZ signal and also need to be considered. We start by identifying submillimetric sources by blindly searching for point sources in the NIKA2 260 GHz map of \clj. By cross-checking the detections with a S/N greater than 3 with \textit{Herschel} SPIRE\footnote{\label{spirenote}European Space Agency, Herschel SPIRE Point Source Catalogue, Version 1.0, 2007. \url{https://doi.org/10.5270/esa-6gfkpzh}} and PACS\footnote{\label{pacsnote}European Space Agency, Herschel PACS Point Source Catalogue, Version 1.0, 2007. \url{https://doi.org/10.5270/esa-rw7rbo7}} catalogs, seven submillimetric sources were identified in the region covered by the NIKA2 maps. The position and fluxes from the above-mentioned catalogs for each submillimetric point source (PS1 to PS5, PS7 and PS8) are summarized in Table~\ref{tab-1} and the corresponding \textit{Herschel} names are given in Table~\ref{tab-2}. 
    
    For each of these sources, a modified blackbody spectrum model is adjusted to the fluxes in Table~\ref{tab-1} together with the measurement of the flux at 260~GHz from the NIKA2 map. The spectra are then extrapolated to 150 GHz to obtain an estimate of the flux of each source at 150 GHz. In Table~\ref{tab-2} we summarize the fluxes at 260 GHz obtained from the NIKA2 map and the extrapolated values at 150 GHz. The contribution of the noise and the filtering in the 260~GHz map are considered when measuring the fluxes. The estimates obtained with the AO and TO noise approaches give compatible results for all point sources. 
    A more detailed explanation of the method used to deal with submillimetric point sources is given in \citet{keruzore}. The obtained probability distributions of the fluxes at 150 GHz are used as priors for the joint fit of the cluster pressure profile and the point sources fluxes described in Sect. \ref{sec:pressure}.
    \begin{table*}
    \centering
    \caption{Submillimetric point sources coordinates and fluxes identified within a radius of 2' around the center of CL~J1226.9+3332. Fluxes at 600, 860 and 1200 GHz are obtained from SPIRE catalog$^{\protect\ref{spirenote}}$. Fluxes at 1870 and 3000 GHz are given in PACS catalog$^{\protect\ref{pacsnote}}$.}
    \label{tab-1}       
    \begin{tabular}{lllllll}
      \hline
      \hline
    Source & Coordinates J2000 &  600 GHz & 860 GHz & 1200 GHz & 1870 GHz & 3000 GHz    \\
    ~&~ & [mJy] & [mJy] &[mJy] & [mJy] & [mJy] \\\hline
    PS1 & 12h27m00.01s +33d32m35.29s  &  100.3 $\pm$ 10.0 &  121.2 $\pm$ 10.0 & 109.8 $\pm$ 7.6  & 55.7 $\pm$ 6.0 & 14.6 $\pm$ 2.1\\
    PS2 & 12h26m51.22s +33d34m39.61s & 37.8 $\pm$ 9.0 &   46.4 $\pm$  9.9  & 29.1 $\pm$ 7.5  & 24.9 $\pm$ 7.4 & 8.0 $\pm$ 1.7 \\
    PS3 & 12h27m07.02s +33d31m49.79s & 34.8 $\pm$  7.9  &  32.4   $\pm$ 8.7 & 25.6  $\pm$ 7.7 &  31.1 $\pm$ 6.6 &25.6 $\pm$ 2.5\\
    PS4 & 12h26m52.84s +33d33m10.74s & ~&   33.0 $\pm$  10.3&   41.9  $\pm$ 9.7      & 31.5 $\pm$ 7.0 & 17.7 $\pm$ 1.8\\
    PS5 & 12h27m07.87s +33d32m32.08s & ~& ~& 30.0 $\pm$ 9.4  &~ &~\\
    PS6 & 12h27m02.43s +33d32m55.06s &~ &~&~&~&~\\ 
    PS7 &12h26m53.86s +33d32m58.10s &~&~&~& 21.8 $\pm$ 1.5 &14.4 $\pm$ 3.0\\
    PS8 & 12h26m46.93s +33d32m52.66s & ~ &~&19.8 $\pm$ 5.5&~&~\\\hline
    \end{tabular}
    \vspace*{0.7cm}  
    \end{table*}

    \begin{table*}

    \centering
    \caption{Submillimetric point sources fluxes and their corresponding names in \textit{Herschel} SPIRE or PACS catalogs. Fluxes at 260 GHz are estimated from the NIKA2 maps and fluxes at 150 GHz obtained from the extrapolation of the fitted spectral energy distributions.} 
    \label{tab-2}       
    \begin{tabular}{llllll}
      \hline
      \hline
    \small Source & \small \textit{Herschel} name & \small 150 GHz &\small  150 GHz & \small 260 GHz  & \small 260 GHz   \\
    ~& ~&  \small [extrap. ``angle order''] &  \small[extrap. ``time order''] &  \small[``angle order''] & \small [``time order''] \\
    ~ &~&\small [mJy] & \small[mJy] & \small[mJy] & \small[mJy] \\\hline
    \small PS1 & \small HSPSC250A$\_$J1227.00+3332.5 & \small2.8 $\pm$ 0.3 & \small2.7 $\pm$ 2.3 &\small 8.2 $\pm$ 0.5 &\small 8.1 $\pm$ 0.5\\
    \small PS2 &  \small HSPSC250A$\_$J1226.86+3334.7& \small 1.1 $\pm$ 0.3 &\small 1.0 $\pm$ 0.3 &\small 3.7 $\pm$ 0.6 & \small3.6 $\pm$ 0.6\\
    \small PS3 & \small HSPSC250A$\_$J1227.12+3331.9 &\small 1.4 $\pm$ 0.4 &\small 1.4 $\pm$ 0.4 &\small 3.9 $\pm$ 0.6 &\small 3.8 $\pm$ 0.6\\
    \small PS4 &\small HSPSC250A$\_$J1226.85+3333.2 &\small 0.5 $\pm$ 0.2  &\small 0.4 $\pm$ 0.2 &\small 2.6 $\pm$ 0.5 &\small 2.5 $\pm$ 0.5\\
    \small PS5 & \small HSPSC250A$\_$J1227.13+3332.4 &\small 0.6 $\pm$ 0.2 &\small 0.5 $\pm$ 0.2 &\small 3.5 $\pm$ 0.6 &\small 3.4 $\pm$ 0.6\\
    \small PS6 & ~&\small  0.4 $\pm$ 2.0 &\small 0.4 $\pm$ 2.0 &\small 3.1 $\pm$ 0.5 &\small 2.9 $\pm$ 0.5\\ 
    \small PS7 &\small HPPSC160A$\_$J122654.1+333253 &\small \textit{1.2 $\pm$ 0.2}$^{(a)}$ &\small \textit{1.2 $\pm$ 0.2}$^{(a)}$ &\small 3.2 $\pm$ 0.5 & \small 3.1 $\pm$ 0.5\\
    \small PS8 &\small HSPSC250A$\_$J1226.78+3332.8 &\small 0.5 $\pm$ 0.2 & \small 0.5 $\pm$ 0.2 & \small 3.4 $\pm$ 0.6 &\small 3.3 $\pm$ 0.6\\\hline
    \end{tabular}
    \begin{tablenotes}
    \small
    \item \textbf{Notes.} $^{(a)}$ For PS7 the extrapolated 150 GHz fluxes are too high and we take these values as upper limits of flat priors for the fit in Sect.~\ref{sec:pressure}.
    \end{tablenotes}

    \vspace*{0.5cm}  
    \end{table*}

    Comparing results in Table \ref{tab-2}, we notice very large uncertainties for the extrapolated flux of PS1 at 150~GHz when estimated with the TO noise map, but this will not affect substantially the following results for the simultaneous fit of the PS1 flux (see Table~\ref{tab-3}) when estimating the cluster pressure profile. Moreover, we can compare its flux at 260~GHz to the measurement in \citet{adam2}, where the source is called PS260. Using NIKA data, they obtain $F_{\mathrm{260~GHz}} = 6.8 \pm 0.7$ (stat.) $\pm 1.0$ (cal.)~mJy which is consistent with our estimates.
    
    Regarding PS6, it doesn't have a counterpart in \textit{Herschel} SPIRE and PACS catalogs. But it appears as a weak signal in the \textit{Herschel} maps, as a 3$\sigma$ detection in the 260 GHz NIKA2 map and, moreover, it compensates the extended tSZ signal at 150 GHz (also clearly observed in \citet{adam2}). For this source the modified blackbody is used to get a prior knowlegde of the flux at 150~GHz for the assumed prior distributions of the spectral index and temperature \citep{keruzore}. Another tricky point source is PS7. The extrapolated 150 GHz values shown in Table~\ref{tab-2} are clearly overestimating the flux of the source. This is understandable since we do not have enough constraints for the low frequency slope of the spectral energy distribution. We choose to use the obtained values as upper limits of a flat prior for the flux of PS7 in the estimation of the cluster pressure profile.

    In addition to submillimetric sources, according to the VLA FIRST Survey catalog \citep{vla}, a radio source of $3.60 \pm 0.13$ mJy at 1.4~GHz is present in (R.A., Dec.)$_{\rm{J2000}}$ = (12h26m58.19s, +33d32m48.61s), hereafter PS9. This galaxy corresponds to the BCG identified in \citet{holden} and the compact radio source detected with LOFAR in \citet{digennaro}. We know beforehand that the contribution of this radio source at 150~GHz is small, but given its central position it is important to consider it. Assuming a synchrotron spectrum $F(\nu) = F_{0} (\nu / \nu_{0})^{\alpha}$ with $\alpha = -0.7 \pm 0.2$, which describes the spectral energy distribution for an average radio source \citep{condon}, we get at 150 GHz, $0.1 \pm 0.2$ mJy. The obtained probability distribution of the flux at 150 GHz is also used as a prior for the fit in Sect. \ref{sec:pressure}. The extrapolation of fluxes from radio to millimeter wavelengths can be dangerous and lead to biasing the electron pressure reconstruction. However, this is not the case for our analyses (see results in Sect.~\ref{sec:pressure} and Table~\ref{tab-4}).

\subsection{\xmm\ }    
\label{xmmobservations}

Regarding X-ray data, CL~J1226.9+3332 was observed by XMM-\textit{Newton} (Obs ID 0200340101) for a total observation time of 90/74 ks (MOS/pn), reducing to 63/47 ks after cleaning. Data were reduced following standard procedures described in \citet{bartalucci2017}. The raw data were processed using the XMM-\textit{Newton} standard pipeline Science Analysis System (SAS version 16). Only standard events from the EMOS1, 2, and EPN detectors with PATTERN $<4$ and $<13$, respectively, were kept. The data were further filtered for badtime events and solar flares. Vignetting was accounted for following the weighting scheme as described in \citet{Arnaud2001}. Point sources were detected on the basis of wavelet filtered images in low ([0.3-2] keV) and high ([2-5] keV) energy bands, then subsequently masked from the events list. This process was controlled by a visual check allowing also to further extract obvious (sub-)structures present in the field. The instrumental background was modelled through the use of stacked filter-wheel closed observations, whilst the astrophysical contamination due to the Galaxy and the cosmic X-ray background were accounted for as a constant background in the 1D surface brightness analysis. This component was modelled in the spectral analysis following the method outlined in \citet{pratt2010}, using an annulus external to the target of radii $300$''$ < \theta < 480$''.

\section{ICM thermodynamical profiles}
\label{sec:icmprofiles}
\subsection{Electron pressure reconstruction from tSZ}
\label{sec:szmodel}
The spectral distortion of the CMB caused by the thermal energy in the cluster, i.e. the tSZ effect, is characterized by its amplitude or Compton parameter, $y$ \citep{sunyaev}. This is directly proportional to the thermal pressure of the electrons in the ICM, $P_{e}$, integrated along the line of sight, 
\begin{equation}
	    y = \frac{\sigma_{T}}{m_{e}c^{2}} \int P_{e} \hspace{0.1cm} \mathrm{d}l
	    \label{eq:compton}
\end{equation}
where $\sigma_{T}, m_{e}$ and $c$ are the Thomson cross section, the electron rest mass and the speed of light, respectively. Hence, the tSZ surface brightness is proportional to the Compton parameter integrated over the tSZ spectrum convolved by the NIKA2 bandpass and therefore, proportional to the integrated thermal pressure of the ICM in the cluster.

\subsubsection{Pressure profile reconstruction with NIKA2}
    \label{sec:pressure}
    
    \subsubsection*{Reconstruction procedure}
    To reconstruct the electron pressure in the ICM of CL~J1226.9+3332 we fit a model map of the surface brightness
    of the cluster to the NIKA2 150 GHz map. 

    The model map is obtained from the pressure profile of the galaxy cluster integrated along the line of sight in Compton parameter (\textit{y}) units, following Eq.~\ref{eq:compton}. We describe the pressure of the galaxy cluster with a \textit{Radially-binned} spherical model (also known as \textit{Non-parametric model} in \citet{ruppin2, romero}):
    \begin{equation}
        P_{e} (r_{i} < r < r_{i+1} ) = P_{i}\left( \frac{r}{r_{i}}\right) ^{-\alpha_{i}}
    \end{equation}
    where $P_{i}$ and $\alpha_{i}$ are the values of the pressure and the slope at the radial bin $r_i$. The slope is directly calculated as:
    \begin{equation}
        \alpha_{i} = - \frac{\mathrm{log}\hspace{2pt}P_{i+1} - \mathrm{log}\hspace{2pt}P_i}{\mathrm{log}\hspace{2pt}r_{i+1} - \mathrm{log}\hspace{2pt}r_i}
    \end{equation}
    We initialize the pressure bin values taking random values from a normal distribution centered at the corresponding pressure from the universal profile of \citet{arnaud10} at each radial bin. The radial bins are chosen to cover mainly the range between the NIKA2 resolution and field of view capabilities. We center the pressure profile at the coordinates of the X-ray peak as determined from XMM-\textit{Newton} data analysis (Sect.~\ref{sec:cljnika2} and \ref{xmmobservations}). 
    The derived \textit{y}-map is convolved with the NIKA2 beam, which is approximated by a two-dimensional Gaussian with $\mathrm{FWHM} = 17.6''$ \citep{perotto}. In order to account for the attenuation or filtering effects due to data processing in the NIKA2 150 GHz map, the model map is also convolved with the transfer function (Sect.~\ref{Sect:transferfunction}). We repeat this procedure for the 1D and 2D transfer functions. Finally, the \textit{y}-map is converted into surface brightness units with a conversion coefficient, accounting for the tSZ spectra shape convolved by the NIKA2 bandpass, that is also left as a parameter of the fit \citep[as done in][]{keruzore}.
    
    
    Furthermore, for the comparison with the 150 GHz NIKA2 map, we added the contribution of point sources to the model map. Point sources are modeled as two-dimensional Gaussian functions and we repeat the procedure detailed in \citet{keruzore} to fit the flux of each source at 150~GHz. Priors on the flux of the sources at 150~GHz are obtained from the results of the spectral fits presented in Sect.~\ref{Sect:ps}. 
    The last component in the model map is a constant zero-level that we also adjust as a nuisance parameter.
    
    The parameters ($\vartheta$) of our fit are: the pressure bins
    describing the ICM of the cluster, the fluxes of the contaminant point sources, the conversion factor from Compton to surface brightness units and the zero-level. The likelihood that we use to compare our model $\mathcal{M}$ pixel by pixel to the data $\mathcal{D}$ is given by:
    \begin{equation}
    \begin{split}
        \mathrm{log } \hspace{2pt}\mathcal{L} (\vartheta) & = - \frac{1}{2} \sum_{i=1}^{n_{\mathrm{pixels}}} \left[ \left( \mathcal{M} (\vartheta) - \mathcal{D}\right)^{T} C_{\mathrm{pix-pix}}^{-1} \left( \mathcal{M} (\vartheta) - \mathcal{D}\right) \right] _{i}\\
       & - \frac{1}{2}\left(\frac{ Y_{500}(\vartheta)-Y_{500}^{Planck}}{\Delta Y_{500}^{Planck}}\right)^{2}
    \end{split}    
    \end{equation}
    
    Here $C_{\mathrm{pix-pix}}$ is the pixel-pixel noise covariance matrix accounting for the residual noise in the NIKA2 150~GHz map (Sect.~\ref{noise}). We repeat the fit with the covariance matrices from both ``angle order'' and ``time order'' noise estimates. 
    
    We also compute the $Y_{500}$ integrated Compton parameter and compare it in the likelihood to the integrated Compton parameter measured by \citet{planck2016}, $Y_{500}^{Planck} = (3.82 \pm 0.79)\times 10^{-4}\hspace{2pt}\mathrm{arcmin}^{2}$ within an aperture of $\theta_{500} = 1.907 $~arcmin. We do not compare the integrated Compton parameter at $5\theta_{500}$ as measured by \textit{Planck} because it would require extrapolating the pressure profile far beyond the NIKA2 data.

    For the map fit we use the \texttt{PANCO2} pipeline \citep{panco2} and follow the procedure described in \citet{adam2,ruppin1} and \citet{keruzore}. This pipeline performs a Markov Chain Monte Carlo (MCMC) fit using the \texttt{emcee} python package \citep{foreman,goodman}. The sampling is performed using 40 walkers and $10^{5}$ steps, with a burn-in of $10^{3}$ samples, and convergence is monitored following the $\hat{R}$ test of \citet{gelmanrubin} and chains autocorrelation. The \texttt{PANCO2} code has been successfully tested on simulations.
    
    \begin{figure*}
        \centering
        \includegraphics[trim={0cm 0cm 0cm 0cm}, scale=0.25]{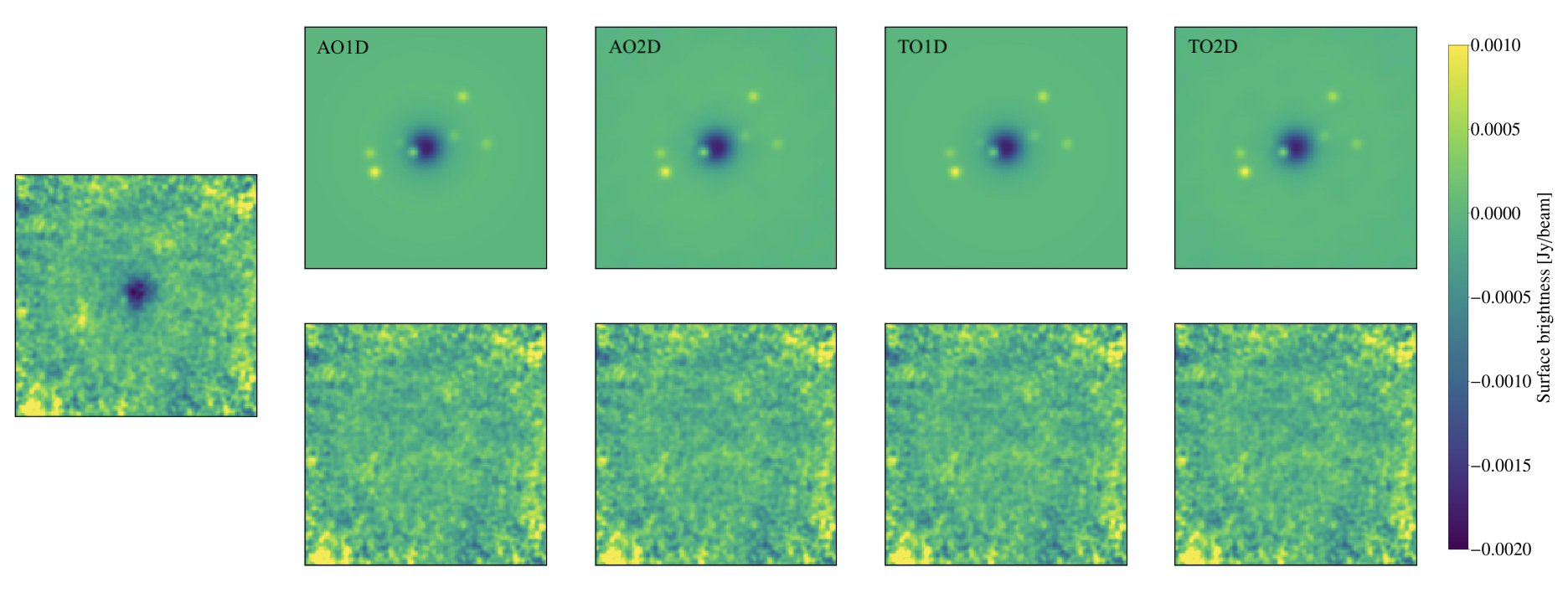}
        \caption{\textit{Left:} NIKA2 150 GHz surface brightness map of CL~J1226.9+3332. \textit{Top:} best-fitting models of the tSZ signal and point sources. \textit{Bottom:} residual maps, difference between the data map and each best fit model. From left to right results obtained with different transfer function and noise estimates: AO1D, AO2D, TO1D and TO2D. All maps have been smoothed with a 10” Gaussian kernel for display purposes and are shown in Jy/beam units. }
        \label{fig:datamodel}
    \end{figure*}

    \subsubsection*{NIKA2 pressure radial profile}
    \label{sec:NIKA2preprof}

   
    In order to estimate the robustness of the results of the above procedure, we have performed the fit to the NIKA2 data in four different cases with respect to the choice of noise residuals and transfer function estimates. Thus, we consider, AO1D (TO1D) and AO2D (TO2D) using the angle (time) order noise residual map and the 1D and 2D transfer functions, respectively.
    In Fig.~\ref{fig:datamodel} we compare the NIKA2 150 GHz map of CL~J1226.9+3332 to the obtained best fit models and their residuals for these four analyses.
    Comparing the power spectra of the residual maps to the power spectra of the noise estimate maps, we see in Fig.~\ref{fig:pkidl} that for the TO case the fit residuals and the noise estimates power spectra are consistent. However, for the AO cases there is an excess of power in the fit residuals, which could be interpreted as coming from the signal due to the differential filtering effects that are not captured in the AO noise as discussed in Sect.~\ref{noise}. 
    Regarding point sources, the reconstructed fluxes are consistent for the four analyses (see Tables~\ref{tab-3} and \ref{tab-4}). 

    We present in Fig.~\ref{fig:autressz} the {\it Radially-binned} best-fit pressure profiles obtained for the four tested cases. The blue and cyan (dark and light green) dots correspond to the AO (TO) 1D and 2D transfer function estimates, respectively. The plotted uncertainties correspond to $1sigma$ of the posterior distributions derived from the MCMC chains. Overall, the four 
    NIKA2 analyses give consistent results, specially in the radial ranges in which we expect the NIKA2 results to be reliable, i.e. between the beam and the field of view (FoV) scales, both represented with dashed vertical lines in the figure. We give the HWHM of the NIKA2 beam (17.6''/2) and half the diameter of the FoV (6.5'/2) in the physical distances corresponding to the redshift of the cluster.
    
    In terms of noise estimates we observe that the uncertainties on the pressure bin estimates are slightly larger for the time ordered case as one would expect. However, we notice no significant bias between the time and angle ordered results. 
    The effect of the transfer function is hard to evaluate: even if the 2D TF is a more precise description of the filtering in the map, when fitting a spherical cluster model the use of the 1D TF gives consistent results. In the following we use the results for the four analyses to evaluate possible systematic uncertainties induced by the NIKA2 processing.
    
    
    \subsubsection{Comparison to previous results}
    
\begin{figure*}
        \centering
        \includegraphics[trim={1cm 1cm 1cm 0.5cm},scale = 0.5]{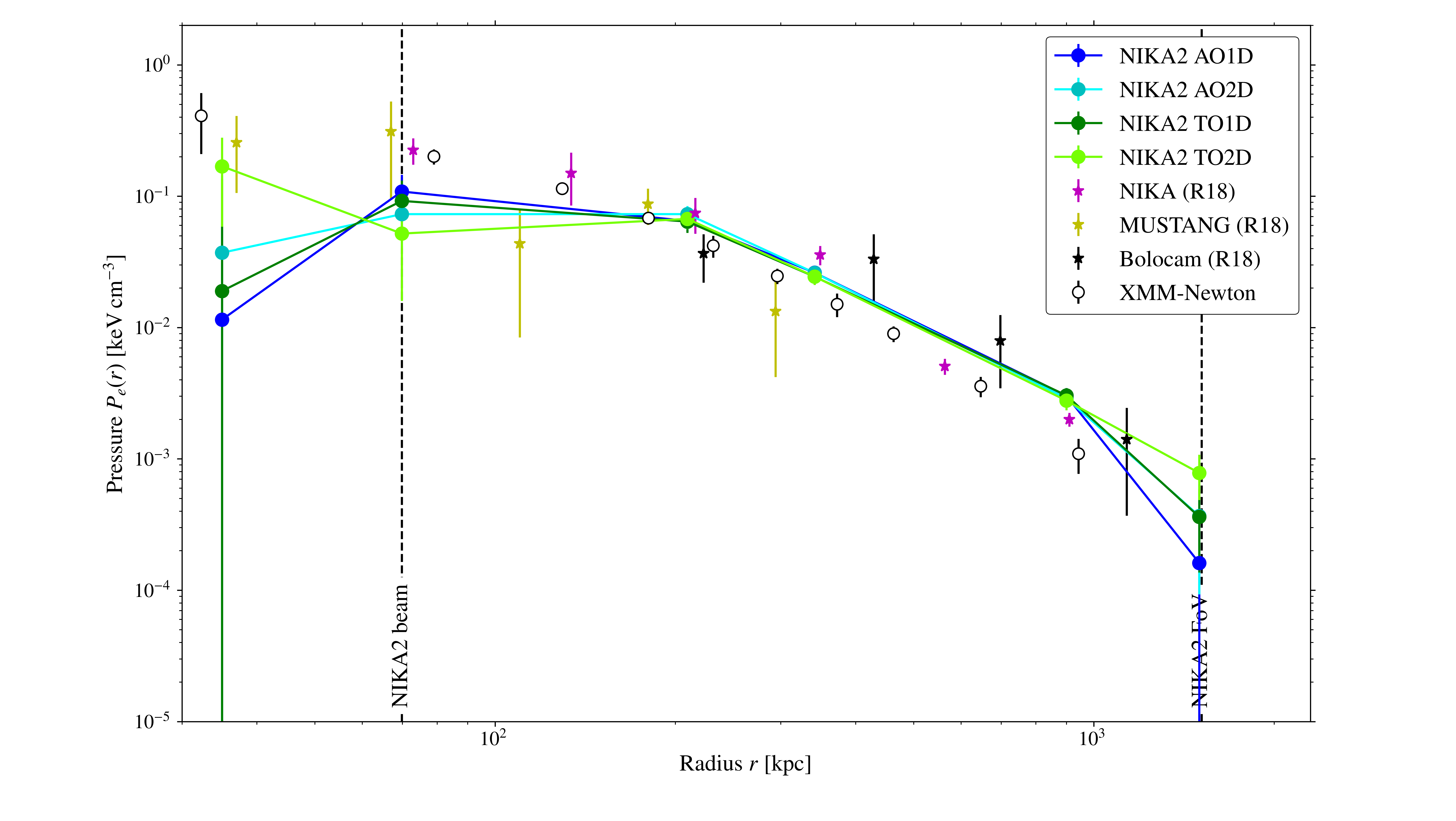}
        \caption{Pressure profile of the ICM of CL~J1226.9+3332. Blue and green markers correspond to the results obtained in this work from the NIKA2 150 GHz map. The error bar edges represent the $1\sigma$ uncertainties. Pink, yellow and black stars show the reconstructed profiles in R18 for NIKA, MUSTANG and Bolocam data, respectively. Empty markers correspond to the pressure profile obtained from the combination of XMM-\textit{Newton} electron density and temperature profiles. Vertical dashed lines indicate the instrumental limits of NIKA2 as radius of the beam and FoV.}       

        \label{fig:autressz}
    \end{figure*}

    Also in Fig.~\ref{fig:autressz}, we compare our results to the profiles obtained in R18 with tSZ data from NIKA (pink), Bolocam (black) and MUSTANG (yellow). MUSTANG's high angular resolution (9'' FWHM at 90~GHz) enables to map the core of the cluster, whereas Bolocam's large field of view (8' at 140~GHz) allows one to recover the large angular scales. NIKA and the improved NIKA2 camera are able to cover all the intermediate radii. The consistency of the different pressure bins in the radial range from the NIKA2 beam to the FoV proves the reliability of the reconstruction with NIKA2 data.
    
    Before going further, we have to consider again the effect of the filtering on the NIKA2 data. 
    The filtering due to the data processing affects mainly small angular frequencies, i.e. small $k$ numbers (Sect.~\ref{Sect:transferfunction}), which is translated into large angular scales in real space. In this case it means that the region at $\sim$~1000 kpc from the center of the cluster is strongly filtered. For this reason, we cast doubt on the results of our fits for the last NIKA2 bin in pressure. 

 \subsection{Thermodynamical profiles from X-rays}
   
\label{xray}
\begin{figure*}
    \centering
    \begin{minipage}[b]{0.48\textwidth}
    \includegraphics[trim={1cm 0.1cm 1cm 1cm},scale=0.33]{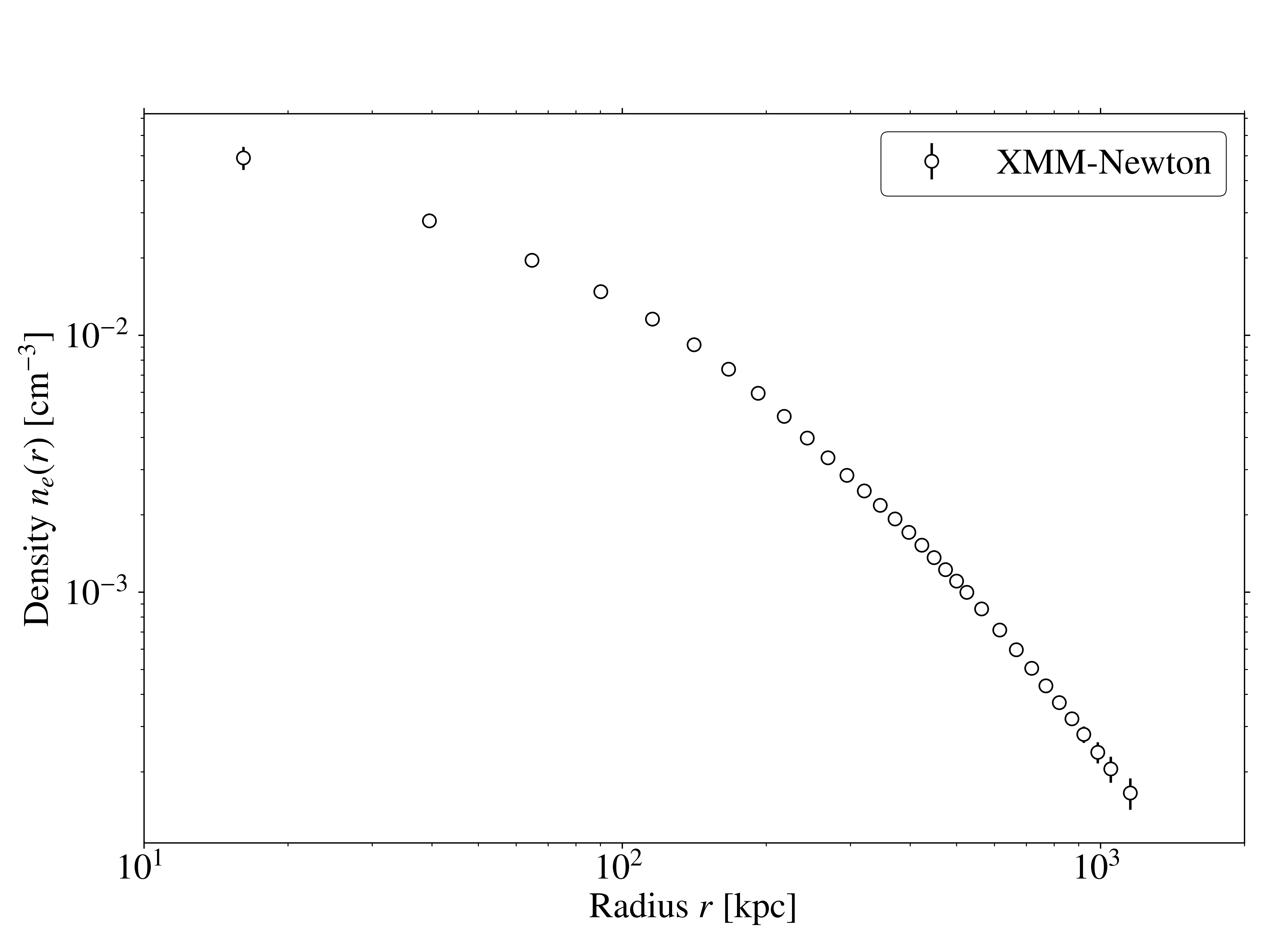}
    \end{minipage}
    \begin{minipage}[b]{0.48\textwidth}
    \includegraphics[trim={0cm 0.1cm 0cm 1cm},scale=0.33]{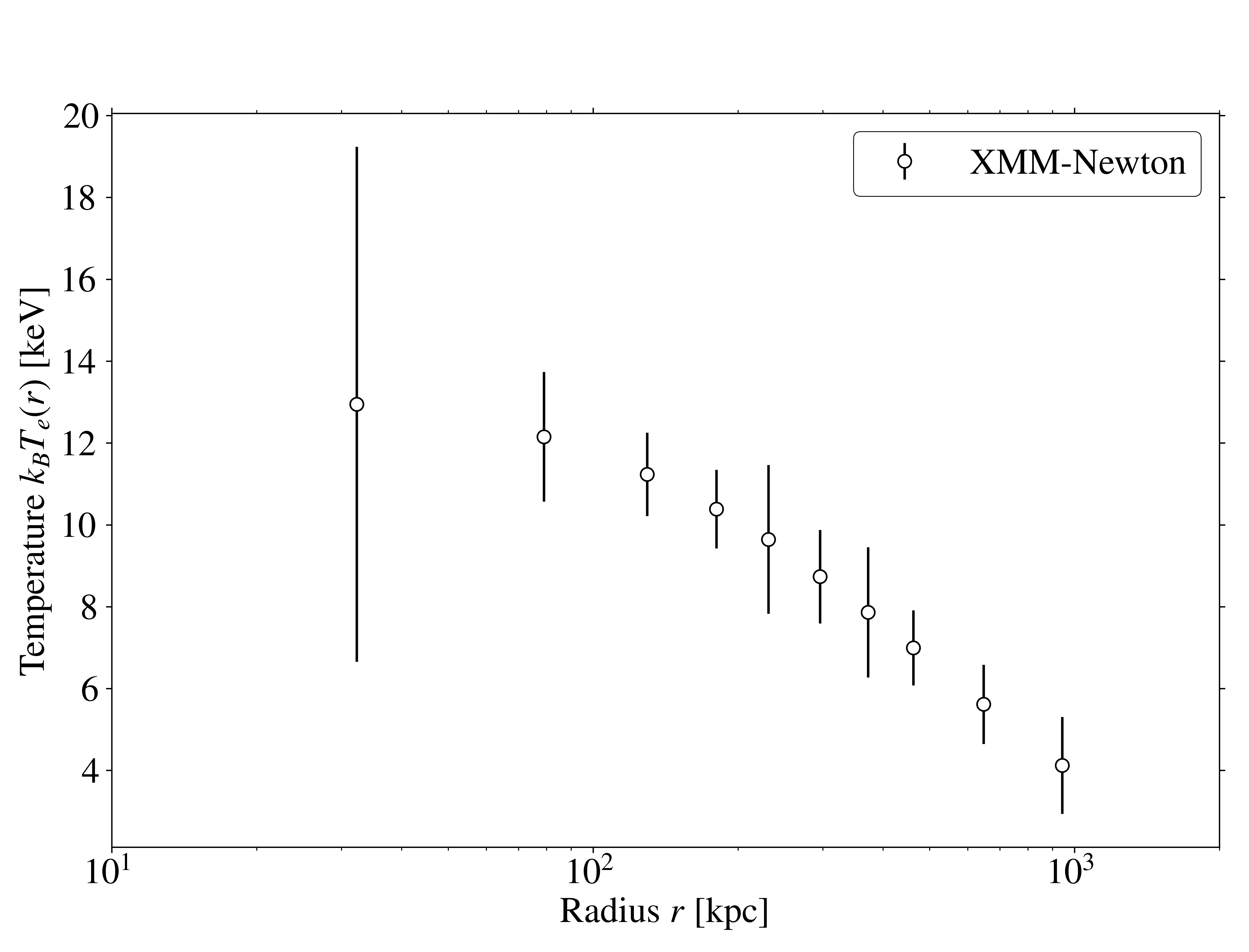}
    \end{minipage}
    \caption{Electron density (left) and temperature (right) profiles reconstructed from XMM-\textit{Newton} observations, with $1\sigma$ error bars. The profiles are centered at the X-ray peak (R.A., Dec.)$_\mathrm{J2000}$ = (12h26m58.08s, +33d32m46.6s).}
    \label{fig:ne}
\end{figure*}
The electron density and temperature profiles were extracted following the methodology described by \citet{pratt2010} and \citet{bartalucci2017}. In short, the vignetted-corrected and background-subtracted surface brightness profile obtained in concentric annuli from the X-ray peak is deconvolved from the PSF and geometrically deprojected assuming spherical symmetry using the regularisation technique described in \citet{croston}.

The temperature profile is derived in bins defined from the afore-derived binned surface brightness profile through a spectral analysis modelling the ICM emission via an absorbed MEKAL model under XSPEC\footnote{\url{https://heasarc.gsfc.nasa.gov/xanadu/xspec/}} and accounting for both the instrumental and astrophysical backgrounds. The derived 2D temperature profile is then PSF-corrected and deprojected following the ``Non-parametric-like'' method presented in \citet{bartalucci2018}.   

The gas pressure, $P$, and entropy, $K$, profiles were then derived from the deprojected density, $n_e$, and temperature, $T$, profiles assuming $P=n_e\times T$ and $K=T/n_e^{2/3}$, respectively.


In this paper we focus on the former, which is shown in Fig.~\ref{fig:autressz}. as open black circles. The electron density and temperature profiles are shown in Fig.~\ref{fig:ne}.


\section{Hydrostatic mass}
\label{sec:hsemassgeneral}


\subsection{Hydrostatic equilibrium}
Under the hydrostatic equilibrium (HSE) hypothesis 
we can compute, for a spherical cluster, the total cluster mass enclosed within the radius $r$ as:
\begin{equation}
\label{eqn:hse}
    M_{\mathrm{HSE}} (<r) = - \frac{1}{\mu m_{p} G} \frac{r^2}{n_{e}(r)} \frac{\mathrm{d}P_{e}(r)}{\mathrm{d}r}
\end{equation}
where $\mu$, $m_p$ and $G$ are the mean molecular weight of the ICM gas, the proton mass and the gravitational constant, respectively. We assume $\mu \approx 0.6$ \citep{pessah, ettori2019} for the gas. 
Combining the pressure profiles obtained from the thermal SZ or X-ray data with the electron density from the X-ray we can reconstruct the mass of the galaxy cluster as in \citet{adam2, ruppin1, keruzore}. 
\subsection{Pressure profile modeling for mass estimation}
Deriving the mass directly from the \textit{Radially-binned} profiles presented above (see Fig.~\ref{fig:autressz}) leads to non-physical
results (i.e. negative mass contributions) as no condition was imposed regarding the slope of the profile in the pressure reconstruction. This was done to prevent extra constraints on the pressure profile induced by assumptions on the model. To overcome that issue, we fit here pressure models ensuring physical mass profiles to the \textit{Radially-binned} tSZ results from Sect.~\ref{sec:pressure}.
We will consider two different approaches: 1) a direct fit of a generalized Navarro-Frenk-White (gNFW) profile to the \textit{Radially-binned} pressure, and 2) an indirect fit of a Navarro-Frenk-White (NFW) mass density model under the HSE assumption.
In both cases and aiming for a precise reconstruction of the HSE mass, which requires having accurately constrained slopes for the pressure profile, we combine the NIKA2 pressure bins with the results obtained in R18\footnote{The binned profiles in R18 and the ones in this work have been centered in positions at 3 arcsec of distance, which is the typical rms pointing error for NIKA2 \citep{perotto}, so we consider that combining them is a valid approach.}. 
    \subsubsection{
    gNFW pressure model}
    \label{Sect:gnfw}
    The first approach, which has been used in previous NIKA2 studies \citep{ruppin1, keruzore,ferragamo}, consists in fitting the widely used gNFW pressure profile model \citep{nagai} to the tSZ data:
    \begin{equation}
    \label{eqn:gnfw}
        P_{e}(r) = \frac{P_0}{\left(\frac{r}{r_p}\right)^c \left( 1 + \left(\frac{r}{r_p}\right)^{a}\right)^{(b-c)/a} }
    \end{equation}
    where $P_{0}$ is the normalization constant, $b$ and $c$ the external and internal slopes, $r_p$ the characteristic radius of slope change and $a$ the parameter describing the steepness of the slopes transition. 

    \begin{figure*}
        \begin{minipage}[b]{0.52\textwidth}
        \includegraphics[trim={1cm 0cm 0cm 0cm},scale=0.25]{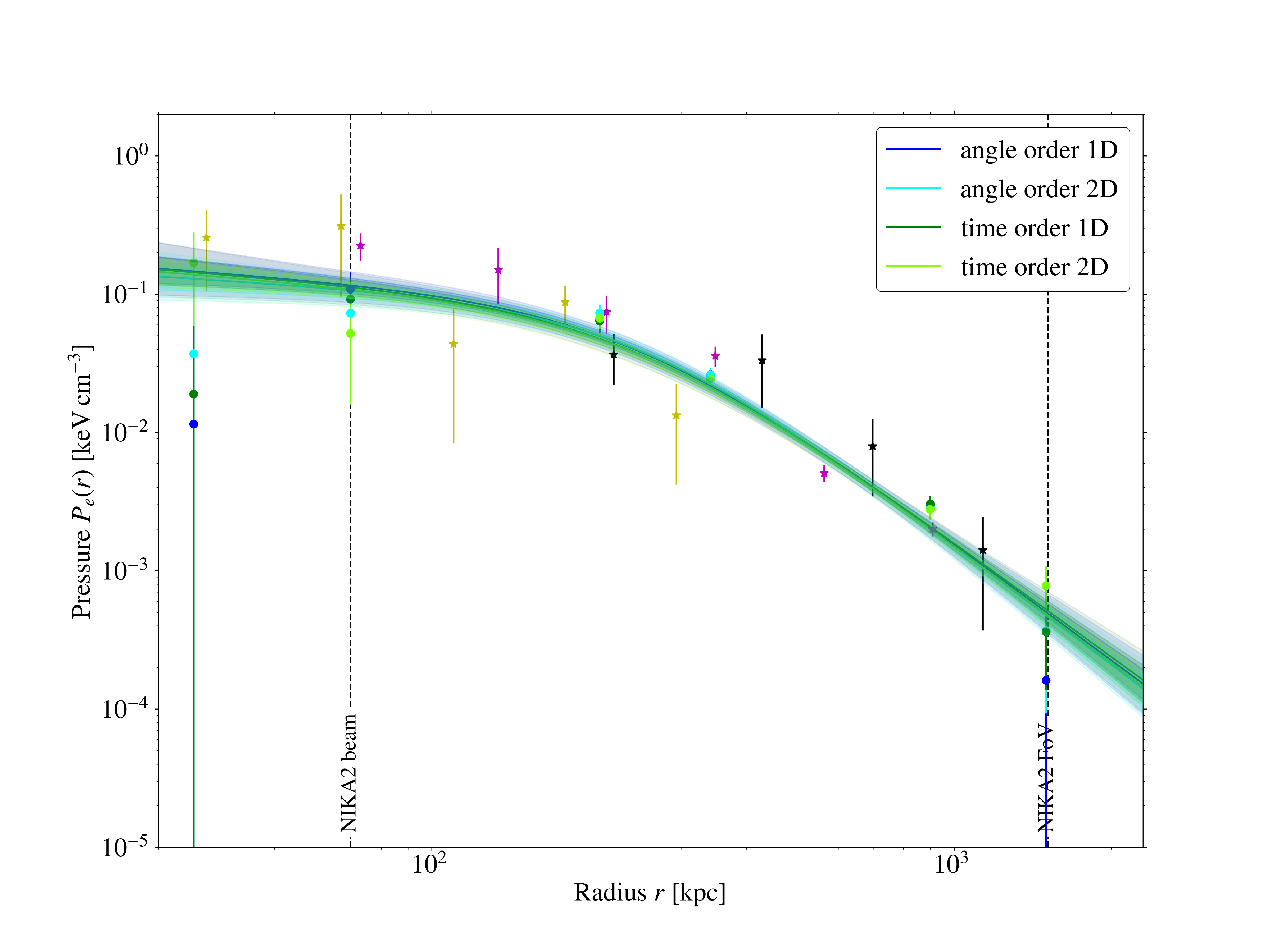}
        \end{minipage}
        \begin{minipage}[b]{0.48\textwidth}
        \includegraphics[trim={2cm 0cm 0cm 0cm},scale=0.25]{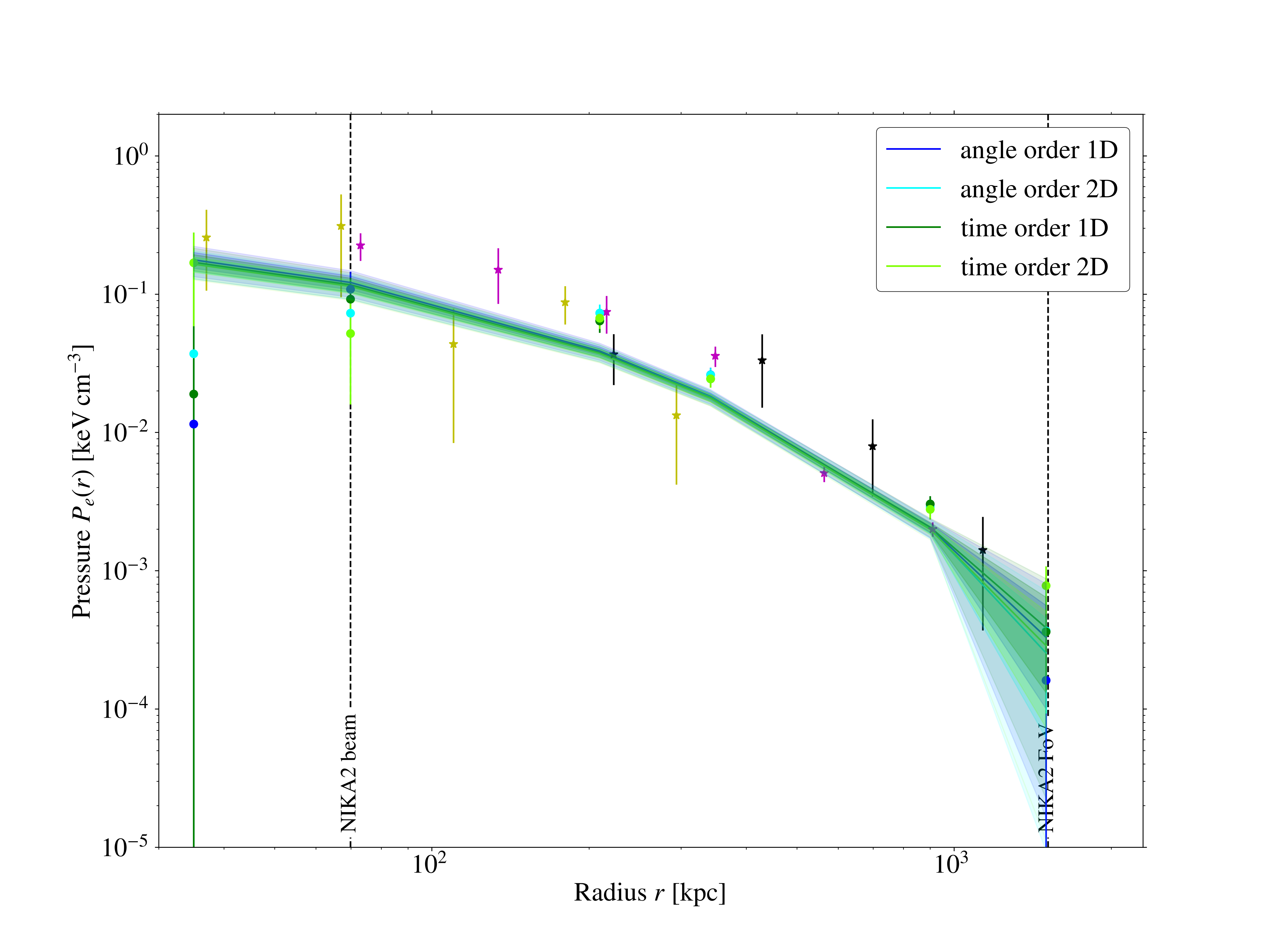}
        \end{minipage}
        \hfill
    \caption{Pressure profile and best-fit for the gNFW (left) and NFW (right) models. The data points correspond to the NIKA2 \textit{Radially-binned} results for the four data sets discussed above, and to the NIKA, MUSTANG and Bolocam bins from R18. Blue and green solid lines represent the best-fit values for the four NIKA2 pressure estimates considered. The shaded regions show the 2.5th, 16th, 84th and 97.5th percentiles.}

    \label{fig:pressprofs}
    \end{figure*}

    We perform a MCMC fit using \texttt{emcee} package.
    The likelihood function
    is given by: 
    \begin{equation}
    \label{eqn:likelihood}
        \begin{split}
        \mathrm{log } \hspace{2pt}\mathcal{L} (\vartheta) & = \\ 
         &  \hspace{10pt} - \frac{1}{2} \left( P^{gNFW}(\vartheta) - P^{N2}\right)^{T} C^{-1} \left(P^{gNFW}(\vartheta) - P^{N2}\right) \\ 
        &  \hspace{10pt} - \frac{1}{2}\sum_{k=1}^{n_{\mathrm{R18bins}}} \left(\frac{P_k^{gNFW}(\vartheta) - P_k^{\mathrm{R18}}}{\Delta P_k^{\mathrm{R18}}} \right)^{2} \\
       &  \hspace{10pt} - \frac{1}{2}\left(\frac{Y^{gNFW}_{500}(\vartheta)- Y_{500}^{Planck}}{\Delta Y_{500}^{Planck}}\right)^{2}
     \end{split} 
    \end{equation}
    where $P^{N2}$ and $C$ represent the NIKA2 \textit{Radially-binned} pressure profile bins and associated covariance matrix. $P_k^{\mathrm{R18}}$ and $\Delta P_k^{\mathrm{R18}}$ are the R18 pressure profile data and uncertainties. And $P^{gNFW}(\vartheta)$ are the gNFW pressure profile values for a set of parameters $\vartheta=[P_0,r_p, a, b, c]$.
    As we do not rely on the value of the last NIKA2 pressure bin, 
    we have chosen to modify the NIKA2 inverse covariance matrix $C^{-1}$ by setting the last diagonal term to $[C^{-1}]_{6,6}$ = 0, so that the correlation of the last bin with the others is taken into account, but not its value. 
    We also set a constraint on the integrated Compton parameter of the model $Y^{gNFW}_{500}(\vartheta)$ using again the \textit{Planck} satellite PSZ2 catalog results, $Y_{500}^{Planck}$ \citep{planck2016}.
    However, we find that the impact of this constraint is completely negligible for this cluster. 
    Furthermore, we have added an extra condition to the fit to ensure increasing HSE mass profiles with radius: $\frac{r^2}{n_{e}(r)}\frac{\mathrm{d}P_{e}(r)}{\mathrm{d}r} < 0$.
    
    The best-fit gNFW pressure profiles (solid lines) and uncertainties (shaded area) are presented in the left panel of Fig.~\ref{fig:pressprofs} for the four sets of NIKA2 data discussed in Section~\ref{sec:NIKA2preprof}. We observe that the best-fit models are a good representation of the data over the full range in radius as demonstrated by the corresponding reduced $\chi^{2}$, which are close to 1 for all the cases (see solid lines in Fig.~\ref{chi2}). 
    The posterior distributions of the $\vartheta_{gNFW}$ parameters 
    can be found in Appendix~\ref{sec:gnfwparamappend}. 
    
\begin{figure}[t]
        \centering
        \includegraphics[trim ={0cm 0cm 0.5cm 0cm},scale=0.31]{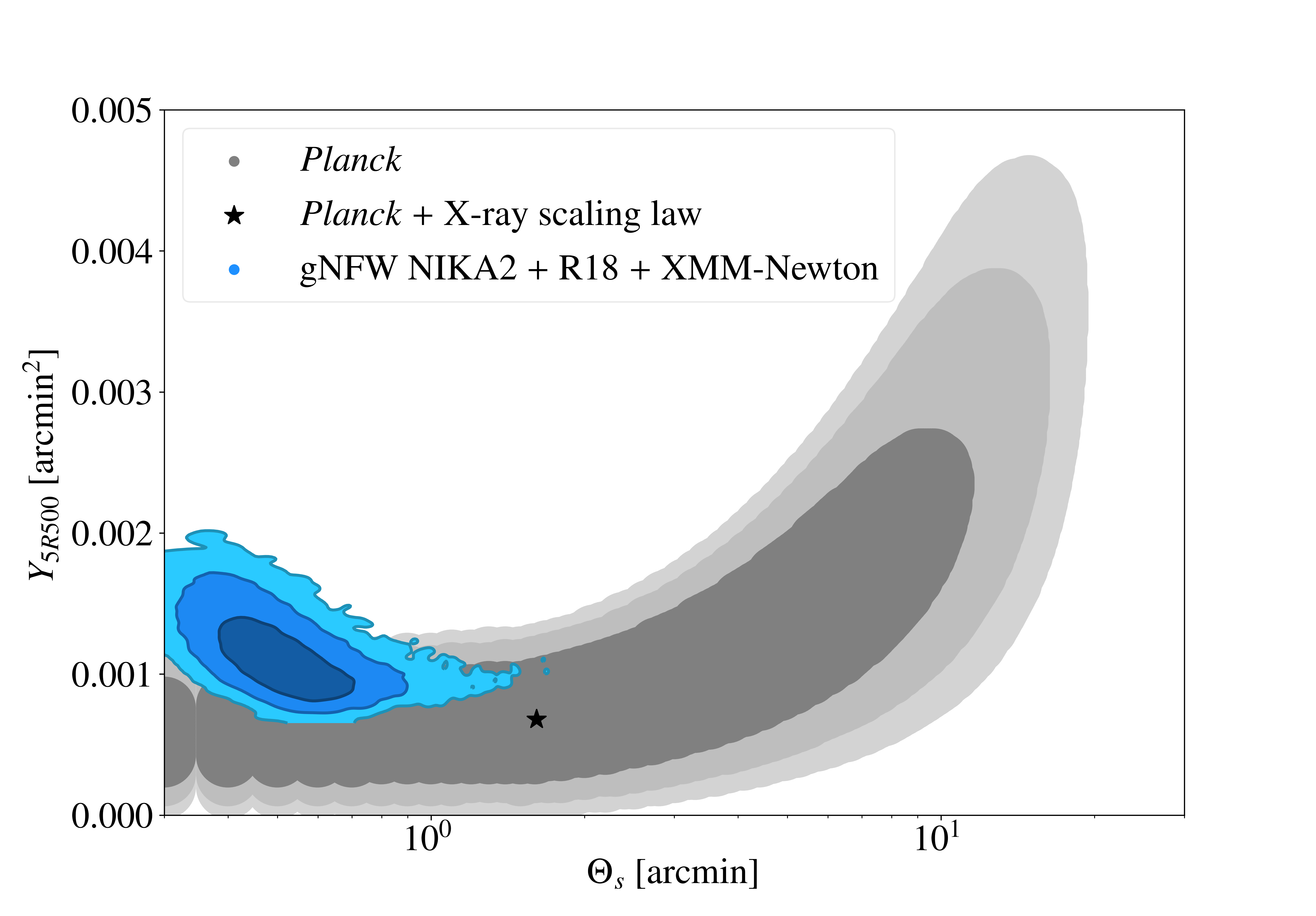}
        \caption{Distribution of $Y_{5R500}$ with respect to $\Theta_{s}$ for the gNFW pressure model fits to \textit{Planck} data (grey) in \protect\citet{planck2016} and to the NIKA2 + R18 + XMM-\textit{Newton} data (blue) in this work. Different contours show the $68\%$, $95\%$ and $99\%$ confidence intervals. The black star corresponds to the intersection between the \protect\citet{planck2016} distribution and the X-ray scaling law shown in Fig.~16 in \protect\citet{planck2016}.}
        \label{fig:planck}
    \end{figure}

    In addition, it is interesting to compare our results to those from \textit{Planck} for which a similar modeling was used. In Fig.~\ref{fig:planck} we present the 2D posterior distributions of the integrated Compton parameter at $5R_{500}$ (with $R_{500}$ calculated independently in each case) with respect to the $\Theta_{s}$ parameter of the gNFW model, at $68\%$, $95\%$ and $99\%$ C.L.. $\Theta_{s}$ and $ r_{p}$ are related via the angular diameter distance at the cluster redshift: $\mathrm{tan}(\Theta_{s}) =  r_{p} /\mathcal{D}_{\mathrm{A}}$. We compare the results obtained in \citet{planck2016} (with the MMF3 matched multi-filter, available in the \textit{Planck} Legacy Archive\footnote{\url{https://pla.esac.esa.int/\#catalogues}}) to the constraints from the gNFW profiles obtained in this work with NIKA2, R18 and XMM-\textit{Newton} data. Our contours were obtained varying all the parameters in the gNFW model fit, while for \textit{Planck} $a$, $b$ and $c$ were fixed. For simplicity, we only show the contours for the NIKA2 AO1D case. This figure illustrates the important gain in precision due to high resolution observations: resolving the galaxy cluster allows us to determine, even at such high redshift, the $\Theta_{s}$ characteristic radius. Contours are marginally in agreement.

    \subsubsection{NFW density model}
    \label{sec:nfwhsemass}
    
    We present here a different approach for the modeling of the pressure profile in the scope of estimating the HSE mass.
    Indeed, the estimation of the pressure derivative (Eq.~\ref{eqn:hse}) can be very problematic as: 1) it is very sensitive to local variations in the slope of the pressure profile and, 2) it requires, as discussed above, additional constraints to ensure recovering physical masses. To overcome these issues we model the pressure profile starting from a mass density model and assuming HSE.
    An equivalent idea is the ``backward process'' to fit X-ray temperatures described in \citet{ettori2013} and references therein. This method was used for the mass reconstruction in \citet{ettori2019} and in \citet{eckert2022}. From the HSE defined in Eq.~\ref{eqn:hse}, we can write: 
    \begin{equation}
        P(r_b) - P(r_a) =  
        \int_{r_a}^{r_b} -\mu m_{p}Gn_{e}(r) \frac{M_{\mathrm{HSE}}( < r)}{r^2} \mathrm{d}r
    \end{equation}
    Moreover, we can relate a radial mass density profile $\rho (R)$ to the mass by,
    \begin{equation}
    \label{eqn:intmass}
    M(<r) = \int_{0}^{r} 4 \pi R^2 \rho(R)\hspace{2pt} \mathrm{d}R
    \end{equation}
    that allows us to relate the pressure directly to a mass density profile. 
    We 
    use here the NFW model, 
    which is a good description of dark matter halos \citep{navarro} and has been widely used in the literature \citep[e.g.][]{ettori2019}:
    \begin{equation}
         \rho_{\mathrm{NFW}} (R) = \frac{\rho_{c} \delta_{c_{200}} (c_{200})}{R/r_{s}(1+R/r_{s})^{2}}
         \label{eq:nfw}
    \end{equation}
    where $\rho_{c}$ is the critical density of the Universe at the cluster redshift and $\delta_{c_{200}}$\footnote{We define: $\delta_{c_{200}} = \frac{200}{3}\frac{c_{200}^{3}}{\mathrm{ln}(1+c_{200}) - c_{200}/(1+c_{200})}$} is a function that depends only on $c_{200}$,
    the concentration parameter (we switch here from an overdensity of 500 to 200 in order to conform to most of previous works). Finally, $r_{s}$ represents a characteristic radius and it is also a free parameter of the model.
   Using this definition we obtain,
    \begin{equation}
    \label{eqn:nfwintegral}
    \begin{split}
        P_{zero} - P(r_a) & = -\mu m_{p}G 4 \pi \rho_{c} \delta_{c_{200}}(c_{200}) r_{s}^3 \\
        &\int_{r_a}^{r_{zero}} \frac{ n_{e}(r)}{r^2} \left[ \frac{1}{1+r/r_{s}} + \ln (1+r/r_{s}) -1  \right] \mathrm{d}r
    \end{split}    
    \end{equation}
    where $r_{zero}$ is a radius at which we are dominated by a zero level component.
    
    We perform a MCMC analysis similar to the one 
    described above for the gNFW pressure profile model. In this case the free parameters of the model 
    are $\vartheta$ = [$c_{200}$, $r_{s}$, $P_{zero}$]. 
    At each step of the MCMC 
    we compute the integral in Eq.~\ref{eqn:nfwintegral} to evaluate $P(\vartheta)$ as needed for the  
    likelihood  function in Eq.~\ref{eqn:likelihood}.
    Calculating the integral 
    can be computationally very expensive. As the result of this integral depends only on $r_s$ and $r_a$, we create a grid of the integrals for a range of $r_s$ values (from 100 to 2000 kpc) and $r_a$ the radial bins of interest. 
    We use this grid to interpolate the values of the integrals at each step. Flat priors are given for the concentration, $0 < c_{200} < 4$. We also make use of the python \texttt{NFW} package\footnote{Jörg Dietrich, 2013-2017. DOI:10.5281/zenodo.50664 }.
    
    
    The best-fit 
    pressure profiles and uncertainties are presented in the right panel of Fig.~\ref{fig:pressprofs} for the four NIKA2 \textit{Radially-binned} data sets discussed above. The posterior probability distributions of the free parameters of the models are shown in Appendix~\ref{sec:gnfwparamappend}. The posterior distributions of the $c_{200}$ and $r_s$ parameters can be compared to the results for the analysis of clusters in X-rays in \citet{pointecouteu2005, ettori2019, eckert2022}. In these studies, $c_{200}$ spans from 1 to 6 and $r_{s}$ from 200 kpc to 1200 kpc, which is compatible with our results. We note that our constraints on $c_{200}$ do not significantly differ from our prior on this parameter. We find that the NFW model is overall a good fit to the data as shown by the reduced $\chi^{2} \sim 1$ (see Fig.~\ref{chi2} for the distributions). However, we observe that the uncertainties increase significantly in the outskirt of the cluster with respect to the gNFW pressure profile model discussed in the previous section. This can be probably explained by the flexibility of the NFW-based approach, which is high enough to show that the last point in the profile is not well-constrained by the data.

    \subsection{HSE mass estimates}
    \label{sec:hsemass}
\begin{figure}
        \centering
        \includegraphics[trim={0.5cm 0cm 0cm 0cm},scale=0.3]{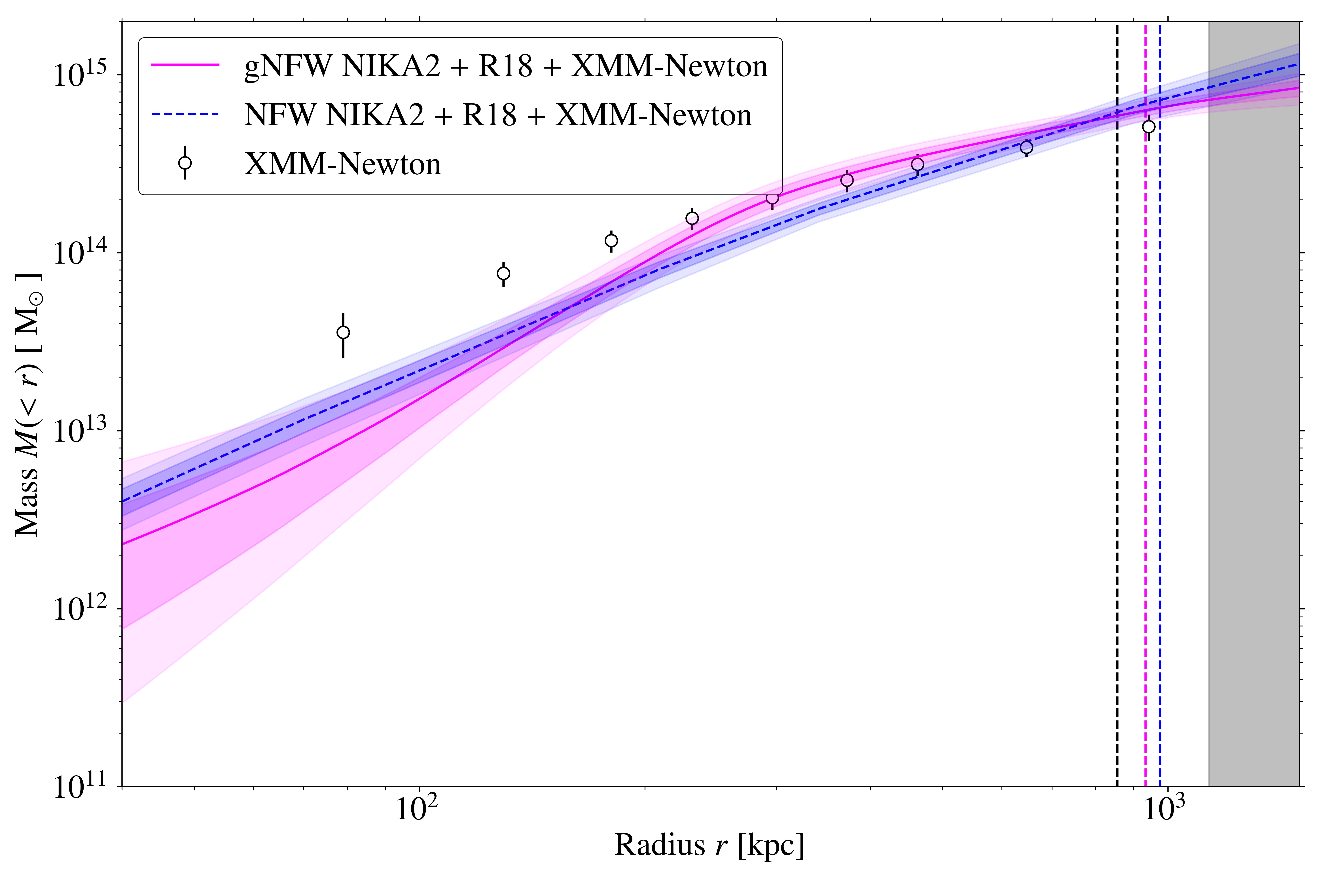}
        \caption{HSE mass profile estimates for CL~J1226.9+3332 obtained with the NIKA2 (angle order 1D) and R18 tSZ data combined with the \xmm\ electron density profile. The solid magenta and dashed blue lines correspond to gNFW and NFW methods, respectively. The shaded areas show the 2.5th, 16th, 84th and 97.5th percentiles. Empty dots correspond to the HSE mass profile obtained from XMM-\textit{Newton} data only. Vertical dashed lines show the $R_{500}^{\mathrm{HSE}}$ obtained from each mass profile. The grey region represents the radial ranges at which the profiles are extrapolated.}
        \label{fig:hsemasses}
    \end{figure}

    We present in Fig.~\ref{fig:hsemasses}
    the HSE mass profiles inferred from the gNFW best-fit pressure profile in combination with the \xmm\ electron density, and, from the NFW density best-fit model. Uncertainties (shaded areas) are obtained directly from the MCMC chains by computing the HSE mass profile for each sample from the model parameters. 
    For the sake of clarity we only present the masses obtained with NIKA2 AO1D estimates, but we change the color code for gNFW profile so that we can differentiate both results.
    
    The capability of the pressure model to describe the shape of the profile slopes is the key element for a good mass reconstruction. From the above results it seems 
    that the NFW approach does not have enough degrees of freedom to fully describe slope variations in the reconstructed pressure profile. 
    Nevertheless, the resulting HSE mass profiles for both models are compatible within $2\sigma$.
    The vertical dashed lines in the figure represent the $R_{500}^{\mathrm{HSE}}$ for each mass profile. Slight differences in the shape of the pressure profile at these radial ranges are critical for defining $R_{500}^{\mathrm{HSE}}$. 
    \begin{figure*}
        \begin{minipage}[b]{0.32\textwidth}
        \includegraphics[trim={0.5cm 0.5cm 0cm 0cm},scale=0.23]{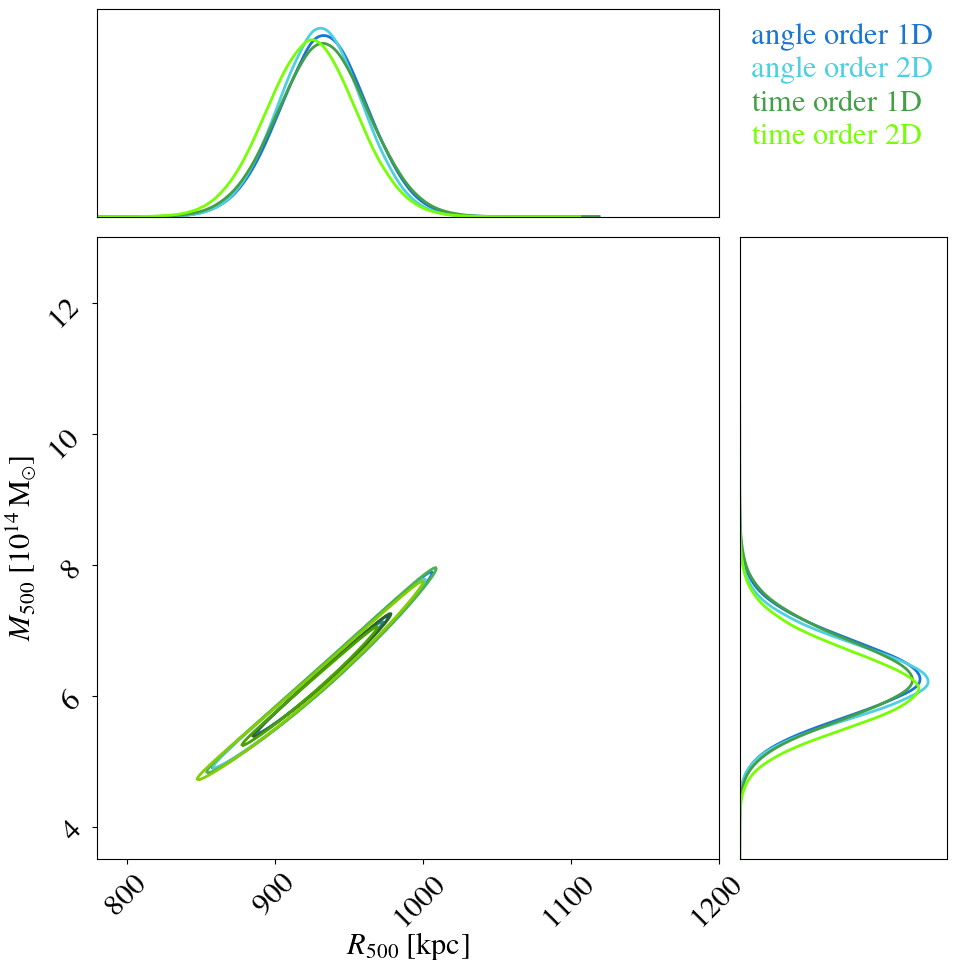}
        \end{minipage}
        \hfill
        \begin{minipage}[b]{0.32\textwidth}
        \includegraphics[trim={0.2cm 0.5cm 0cm 0cm},scale=0.23]{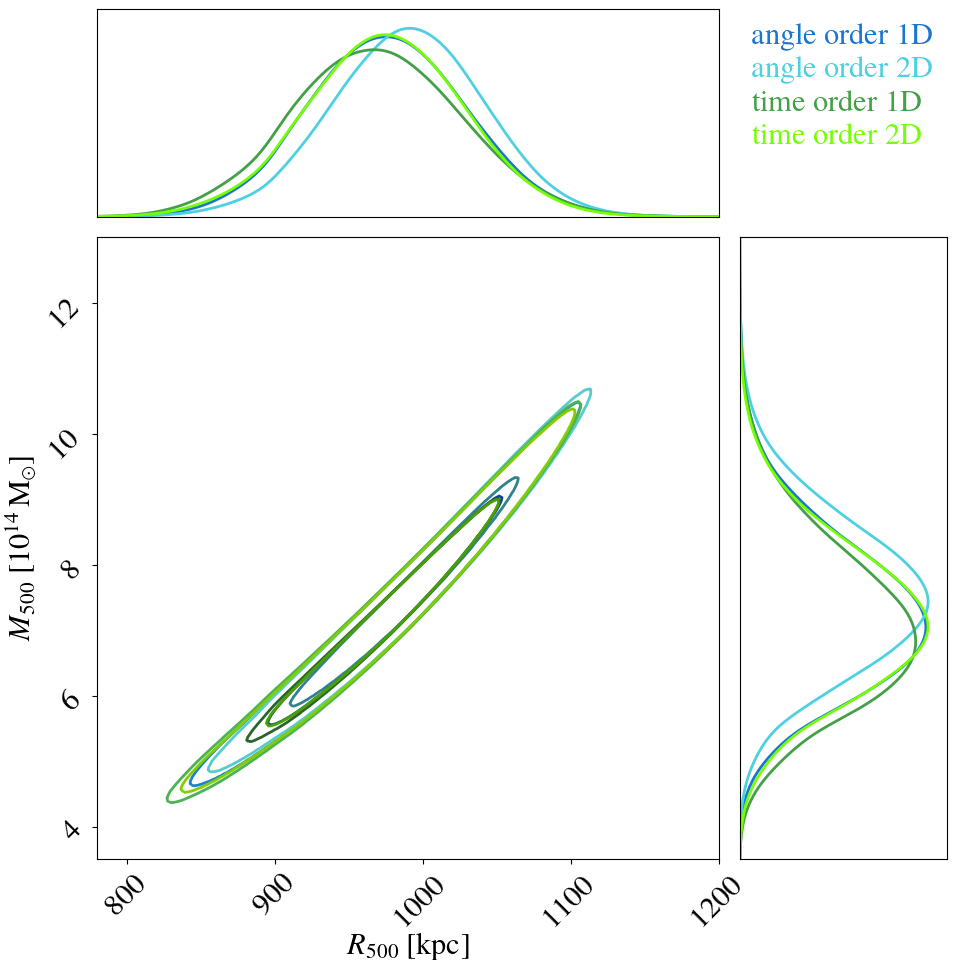}
        \end{minipage}
        \hfill
        \begin{minipage}[b]{0.32\textwidth}
        \includegraphics[trim={0.cm 0.8cm 0cm 0cm},scale=0.165]{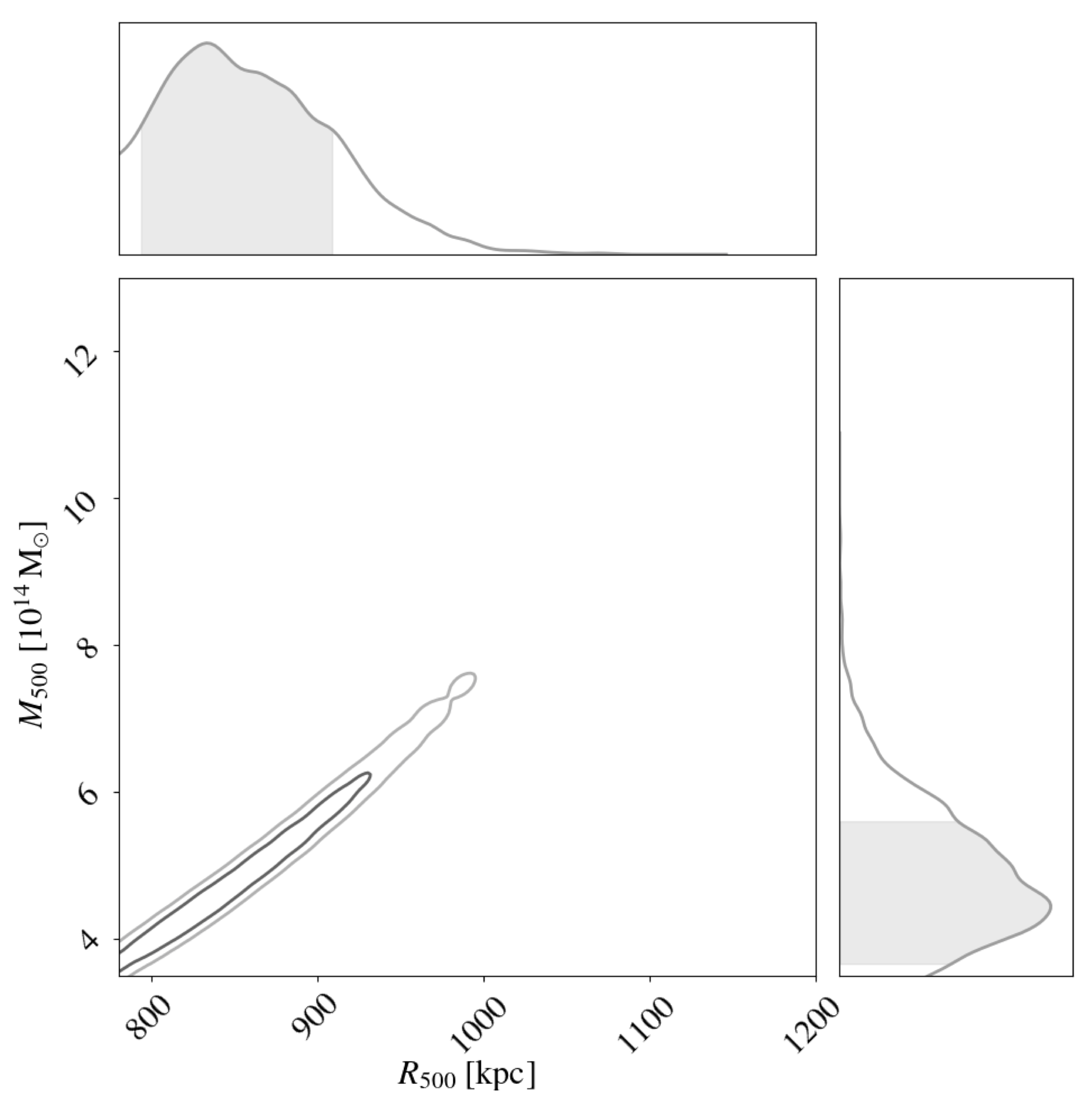}
        \end{minipage}
     \caption{1D and 2D probability distributions for $M_{500}^{\mathrm{HSE}}$ and $R_{500}^{\mathrm{HSE}}$ from the gNFW (left) and NFW (center) models in the combined \xmm\ and NIKA2 and R18 data, and, from the XMM-\textit{Newton} X-ray data only (right). The different blue and green lines correspond to results for the four NIKA2 test cases considered.}  
    \label{fig:m500-nika2charles}
    \end{figure*}       

    These mass profiles, obtained from the combination of tSZ and X-ray data, are also compared to the X-ray-only HSE mass estimate in Fig~\ref{fig:hsemasses}. Assuming spherical symmetry, the X-ray mass profile was derived, following the Monte Carlo procedure detailed in \citet{democles2010} and \citet{bartalucci2017} with the XMM-\textit{Newton} electron density and temperature profiles presented in Sect.~\ref{xray}. 
    Despite the different behaviour of the X-ray-only profile in the cluster core, 
    it is consistent  with the tSZ+X estimates at around $ R_{500}^{\mathrm{HSE}}$. 
    
    From the reconstructed HSE mass profiles we can obtain $R_{500}^{\mathrm{HSE}}-M_{500}^{\mathrm{HSE}}$ probability distributions for each of the considered cases. We present in the left and central panel of Fig.~\ref{fig:m500-nika2charles} the $R_{500}^{\mathrm{HSE}}-M_{500}^{\mathrm{HSE}}$ distributions for the gNFW and NFW models. 
    They have been obtained from the MCMC chains in the same way uncertainties in Fig.~\ref{fig:hsemasses} have been computed. The width of the ellipses is an artifact from the display procedure and each value of $M_{500}^{\mathrm{HSE}}$ is associated to a single value of $R_{500}^{\mathrm{HSE}}$.
    
    We present here results for the four NIKA2 analyses (AO1D/2D and TO1D/2D). The results are consistent, with small dependency on the chosen TF estimate. From the comparison of left and central panels in Fig.~\ref{fig:m500-nika2charles} we verify that the largest uncertainty in the HSE mass estimates comes from the modeling of the pressure profile. In spite of this effect, the reconstructed HSE mass profiles are compatible within $1\sigma$. The right panel of Fig.~\ref{fig:m500-nika2charles} shows the $R_{500}^{\mathrm{HSE}}-M_{500}^{\mathrm{HSE}}$ probability distribution obtained with XMM-\textit{Newton} data only. Even if it is compatible with the gNFW and NFW results, the X-ray-only results prefer lower HSE masses. 
    A similar effect was observed for ACT-CL~J0215.4+0030 cluster \citep{keruzore}, but not for PSZ2~G144.83+25.11 \citep{ruppin1}. We summarize in Table~\ref{tab:hsemasses} the marginalized $M_{500}^{\mathrm{HSE}}$ masses obtained in this work. We give the mean value and the 84th and 16th percentiles. For gNFW and NFW we combine the probability distributions obtained for the four NIKA2 results so that the results account for the systematic effects from NIKA2 data processing.
    
    \renewcommand{\arraystretch}{1.4}    
    \begin{table}
    \centering

    \caption{HSE masses for different estimates at $R_{500}^{\mathrm{HSE}}$. }
    \label{tab:hsemasses}       
    \begin{tabular}{ll}
      \hline
      \hline
    HSE mass estimates &  $M_{500}^{\mathrm{HSE}}$ [$10^{14} \hspace{2pt} \mathrm{M}_{\odot}$]\\ \hline 
    (tSZ+X-ray)$_{\mathrm{gNFW}}$ & $6.25^{+0.59}_{-0.60}$ \\
    (tSZ+X-ray)$_{\mathrm{NFW}}$ &  $7.30^{+1.17}_{-1.17}$\\
    X-ray &  $4.83^{+0.98}_{-0.96}$\\ \hline
    \end{tabular}
    \vspace*{0.2cm}  
    \end{table}
 In general, we conclude that minor variations in the HSE mass profiles at $\sim R_{500}$ can lead to large variations on $M_{500}$. We also stress that the degeneracy between $M_{500}$ and $R_{500}$ is intrinsic and not induced by systematic or stochastic uncertainties.
    \begin{figure*}
        \begin{minipage}[b]{0.5\textwidth}
        \includegraphics[trim={1cm 1cm 1cm 1cm},scale=0.3]{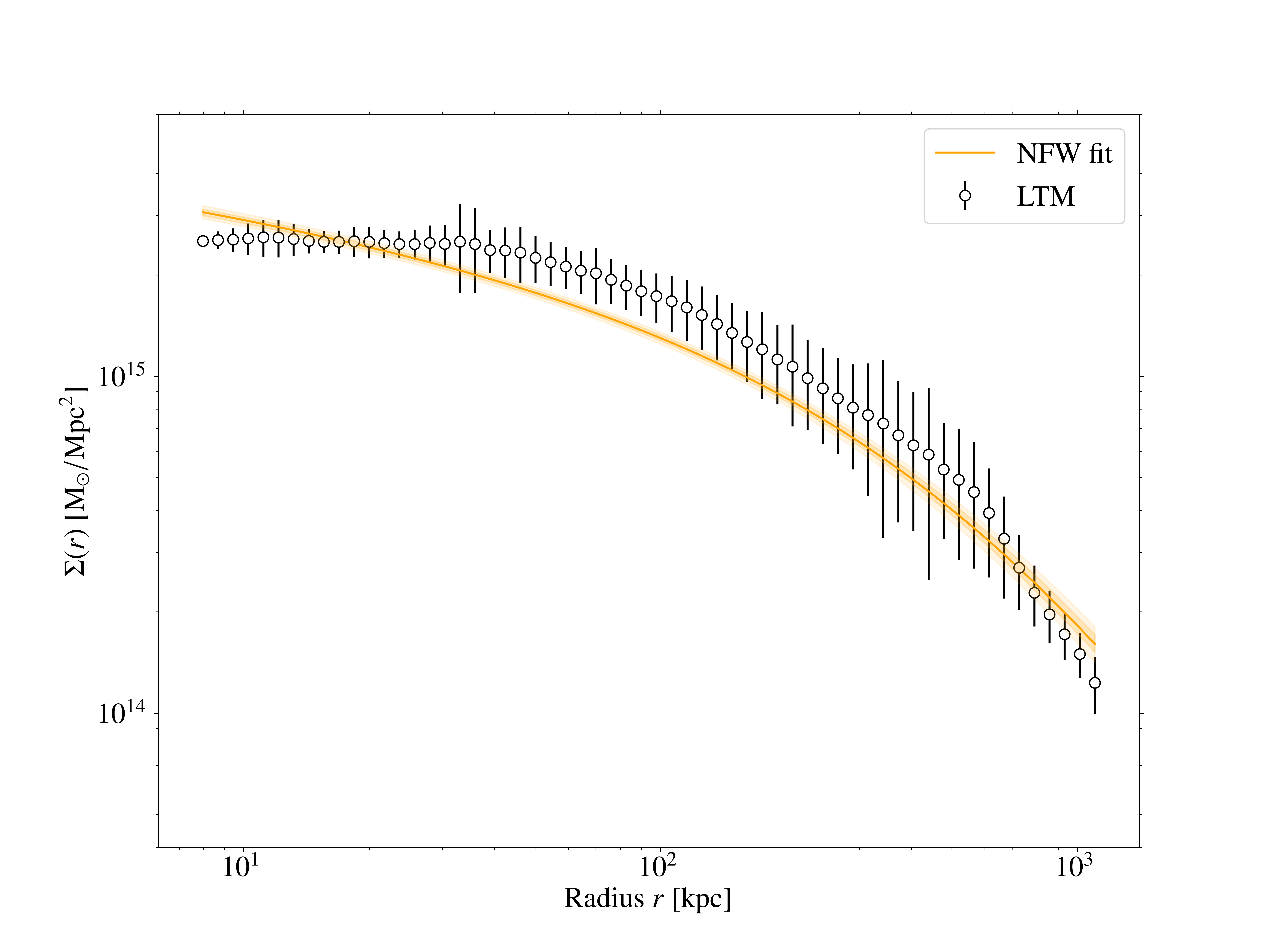}
        \end{minipage}
        \begin{minipage}[b]{0.4\textwidth}
        \includegraphics[trim={1cm 1cm 1cm 1cm},scale=0.3]{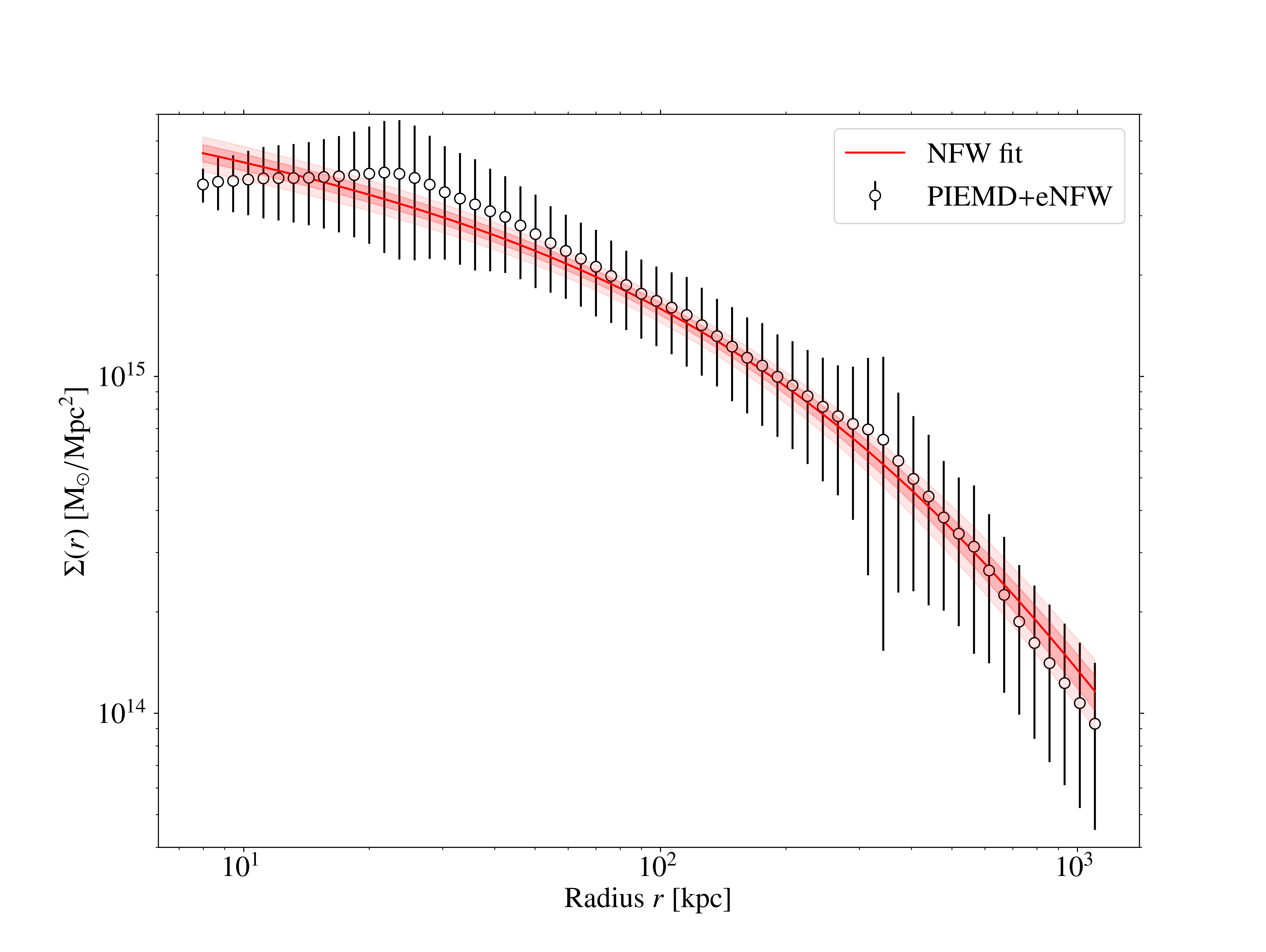}
        \end{minipage}
    \caption{Projected mass density profiles obtained from CLASH convergence maps for the LTM (left) and PIEMD+eNFW (right) models. We also show the best-fit NFW model (orange and red lines) and the 2.5th, 16th, 84th and 97.5th percentiles (shaded area).}
    \label{fig:clashmassprofs}
    \end{figure*}
    
\section{Lensing mass}
\label{othermasses}
\label{sec:lensmass}

Lensing masses, by contrast to HSE ones, probe the total mass without the assumptions on the dynamical state of the cluster. For this reason, it is of great interest to compare the HSE masses to lensing estimates. In this section we present the lensing mass estimate for CL~J1226.9+3332.

\subsection{Lensing data}
We use the CLASH convergence maps (hereafter $\kappa\text{-maps}$) obtained from the weak and strong lensing analysis by \citet{zitrin1}. In the analysis the authors reconstructed the $\kappa\text{-maps}$ for the 25 massive CLASH clusters \citep{postman} using two different lensing models: 1) Light Traces Mass, LTM and 2) Pseudo Isothermal Elliptical Mass Distribution plus an elliptical NFW dark matter halo, PIEMD+eNFW. These models have been detailed in previous studies \citep{zitrin4,zitrin2,zitrin3,zitrin1}. In this work we compute the mass estimate for both $\kappa\text{-maps}$ in order to account for differences in the convergence map modelling.
    
\subsection{Lensing mass density profile}    
For the lensing mass profile reconstruction we followed the approach described in \citet{ferragamo}. The convergence maps describe the projected mass density of the cluster, $\Sigma$, in critical density units, $\kappa = \Sigma / \Sigma_{crit}$ , with
\begin{equation}
    \Sigma_{crit} = \frac{c^2}{4\pi G}\frac{D_{s}}{D_{l}D_{ls}}
\end{equation}
Here $D_{s}$, $D_{l}$ and $D_{ls}$ correspond to the angular diameter distance between the observer and the source, the observer and the lens (the cluster) and between the source and the lens, respectively. The publicly available CLASH $\kappa\text{-maps}$ \citep{zitrin1} have been normalised to $D_{s}/D_{ls}$ = 1. 
    
    To estimate the lensing mass profile of CL~J1226.9+3332, we fit a mass density model to the $\Sigma\text{-map}$.  We assume spherical symmetry and a NFW density profile. We choose to directly fit 
    the analytical projected NFW density profile (Eq.~5 in \citet{ferragamo} derived from \citet{navarro,bartelman}) to the radially averaged projected profiles of the $\Sigma\text{-maps}$. We consider as free parameters $r_{s}$ and $c_{200}$ (see Eq.~\ref{eq:nfw}). The fit is performed via a MCMC analysis using the \texttt{emcee} software and the \texttt{NFW} python package. We center the projected mass profiles at the same position as for the pressure and electron density profiles in Sect.~\ref{sec:szmodel} and \ref{xray}, the X-ray center. Uncertainties were computed from the dispersion in each radial bin and for this analysis we have not considered the possible correlation between the bins.
    
    We show in Fig.~\ref{fig:clashmassprofs} the radial profiles of the projected mass density for CL~1226.9+3332 as obtained from the CLASH LTM (left) and PIEMD+eNFW (right) convergence maps. We present for both profiles the projected best-fit NFW density model and percentiles (shaded area). We observe that for the LTM convergence map the best-fit NFW model underestimates the data except for cluster core and that the uncertainties in the model do not fully account for this. By contrast, the fit for the PIEMD+eNFW succeeds in representing the data. 
    
    From these results we conclude that the main uncertainties in the reconstruction of the lensing density profile comes from the reconstruction of the convergence map. Thus, in the following we will account for those.

    
    \subsection{Lensing mass estimates}
\begin{figure}
    \centering
    \includegraphics[trim={0cm 0cm 0cm 0cm},scale=0.48]{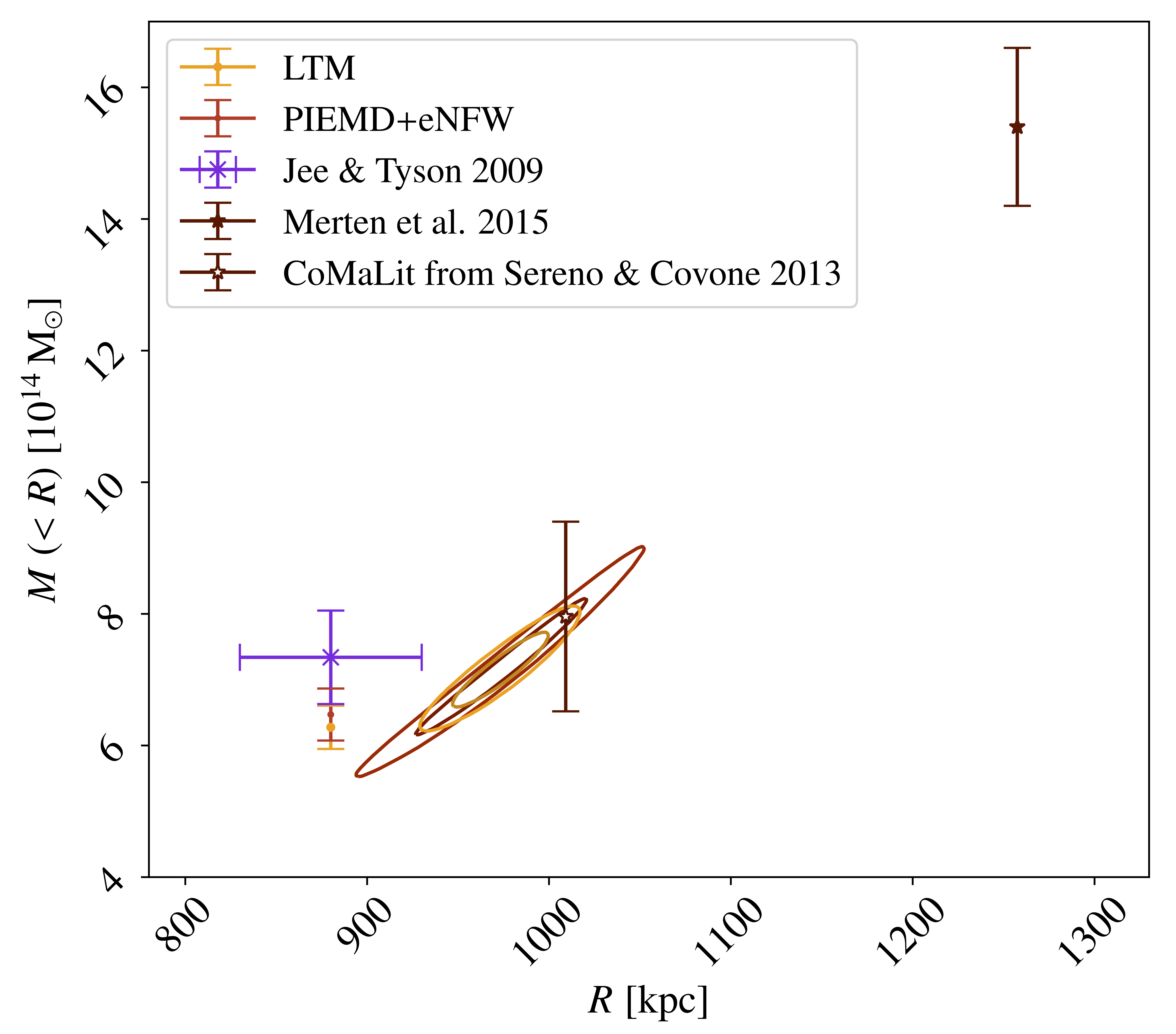}
    \caption{Probabiliy distribution for $M_{500}^{\mathrm{lens}}$ and $R_{500}^{\mathrm{lens}}$ obtained from the fit of the NFW density profile model on the CLASH PIEMD+eNFW (red) and LTM (orange) convergence maps. We show 1 and $2\sigma$ C.L. contours. The purple cross corresponds to the lensing mass estimate in \protect\citet{jeetyson}. We also show for comparison lensing masses estimated at 880 kpc for the PIEMD+eNFW (red) and LTM (orange) as vertical error bars at $1\sigma$ C.L. The brown stars correspond to $M_{500}^{\mathrm{lens}}$ from literature. We differentiate the mass from \protect\citet{merten2015} (full) and \protect\citet{serenocomalit, serenocovone2013} (empty).}
    \label{fig:m500-lens}
    \end{figure}

    From the obtained NFW density profiles we can reconstruct the lensing mass profiles (Eq.~\ref{eqn:intmass}) and subsequently the $M_{500}^{\mathrm{lens}}$ and $R_{500}^{\mathrm{lens}}$ probability distributions. We present in Fig.~\ref{fig:m500-lens} the main results of this analysis for both convergence map models: LTM and PIEMD+eNFW. 
    We obtain consistent results and accounting for their combined posterior distributions we measure $M_{500}^{\mathrm{lens}} = 7.35^{+0.53}_{-0.51} \times 10^{14} \hspace{2pt} \mathrm{M}_{\odot} $ 
    mean value and the 84th and 16th percentiles. 
    
    
    The \citet{jeetyson} weak-lensing analysis did not provide the direct $M_{500}^{\mathrm{lens}}$, but evaluated the lensing mass at the $R_{500}$ from \citet{maughan1}. This result,  shown in purple in Fig.~\ref{fig:m500-lens}, is consistent within uncertainties with our results when evaluating the lensing mass at the same radius for the LTM and PIEMD+eNFW analyses (see orange and red error bars in the figure). \citet{merten2015} performed an independent analysis of the CLASH data, reconstructing their own convergence map. The projected mass density profile presented in Fig.~16 in \citet{merten2015} shows a denser cluster than the profiles from the convergence maps used in this work (Fig.~\ref{fig:clashmassprofs}). For this reason, \citet{merten2015} obtained, also with a NFW density fit, $35\%$ larger masses than in \citet{jeetyson}. The corresponding $M_{500}^{\mathrm{lens}}$ is shown with a full brown star in Fig.~\ref{fig:m500-lens}. The disturbed state of CL~J1226.9+3332 could be the reason, according to \citet{merten2015}, for the different lensing mass estimates. Moreover, the high redshift of the cluster makes more difficult the precise reconstruction of the convergence map. The empty brown star in Fig.~\ref{fig:m500-lens} corresponds to the lensing mass from CoMaLit \citep{serenocomalit} obtained from the analysis in \citet{serenocovone2013}. In this case, the result is compatible with our mass estimates.
    
\section{Comparison of mass estimates}
\label{sec:discussions}

\subsection{CL~J1226.9+3332 mass at R$_{500}$}
\label{sec:masscomparison}
    \begin{figure*}
        \includegraphics[trim={0cm 0cm 0.5cm 0cm},scale=0.66]{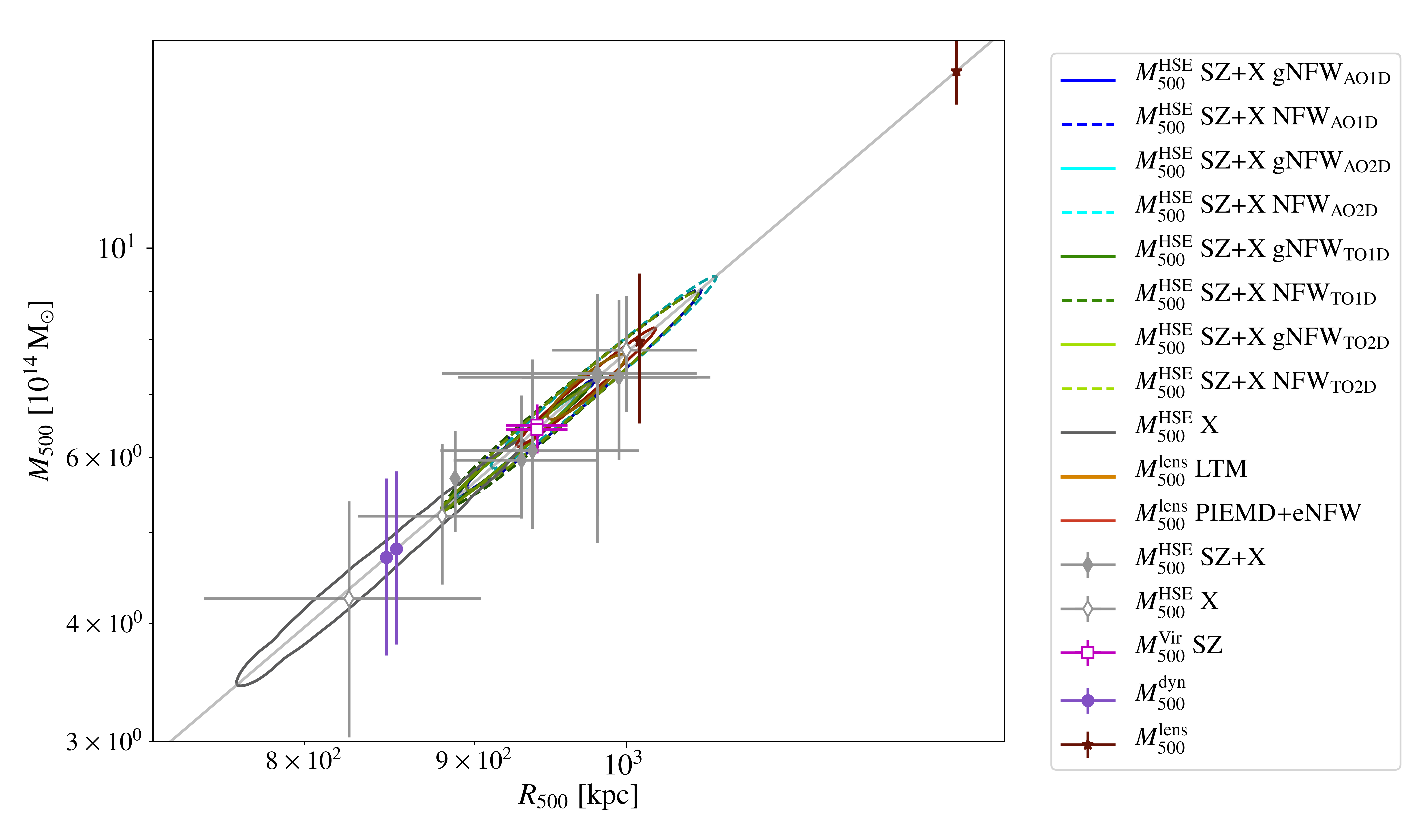}
    \caption{$R_{500}-M_{500}$ plane summarizing the results for CL~J1226.9+3332. In the case of the HSE mass estimates the blue and green $1 \sigma$ contours show the results obtained in this work combining tSZ and X-ray data for the four NIKA2 analyses. The solid and dashed lines are for the gNFW pressure and the NFW density models, respectively. The grey contour corresponds to the HSE mass estimate for the XMM-\textit{Newton} data-only. In the case of literature data the full (empty) grey diamonds represent HSE masses from the combination of tSZ and X-ray data (X-ray-only results). The red and orange contours correspond to the lensing mass estimates obtained from the CLASH LTM and PIEMD+eNFW convergence maps in this work, respectively. Pink squares show the tSZ-only mass assuming Virial relation, purple circles are dynamical mass estimates and the brown stars the lensing estimates. The diagonal bright grey line defines the $R_{500}-M_{500}$ relation. Slight deviations from this line are only due to differences in the cosmological model used in each work.}

    \label{fig:finalmasses}
    \end{figure*}     
    The comparison of different mass estimates is difficult and can lead to wrong physical conclusions. In particular, when comparing integrated masses  the radius at which the mass is computed has a significant impact: $R_{500}$ and $M_{500}$ being constrained at the same time, we are affected by a degeneracy. For this reason, in Fig.~\ref{fig:finalmasses} we show, in the $R_{500}-M_{500}$ plane, the results from the literature with the ones obtained in this work. 

    The green and blue contours show the $R_{500}^{\mathrm{HSE}} - M_{500}^{\mathrm{HSE}}$ results obtained in this work for the  gNFW (solid lines) and NFW (dashed lines) tSZ and X-ray data combined analyses. For comparison the full grey diamonds correspond to the results from the literature presented in Fig.~\ref{fig:masses_literature} also for combined tSZ and X-ray data. We observe that the results in this paper are compatible with previous analyses within $1\sigma$, centered  around $\sim 7 \times 10^{14}\hspace{2pt}\mathrm{M}_{\odot}$. 
    Regarding X-ray-only results, the HSE mass estimates obtained in this work with XMM-\textit{Newton} data (grey contours) suggest mass values centered at $\sim 5 \times 10^{14}\hspace{2pt}\mathrm{M}_{\odot}$. This is in agreement with the lowest estimates from the literature (open grey diamonds) presented in \citet{bulbul} and \citet{maughan1}. On the contrary, the results from \citet{mantz2010} and \citet{mroczkowski2009} show larger masses. However, the $M_{500}^{\mathrm{HSE}}$ in \citet{mantz2010} is not a direct measurement, but an extrapolation from a gas mass measured at $R_{2500}$ converted into total mass, making this result less reliable. 
    Overall, for CL~J1226.9+3332 the HSE masses obtained only from X-ray data tend to lower values than those from the combination of tSZ and X-rays. 
 
    
    The result from \citet{planck2016} (lowest full diamond) is also a special case, as it is not a direct mass measurement, but a mass obtained from the X-ray-derived scaling relation (Eq.~7 in \citet{planck2014a}) applied to the SZ measurement. This may explain why it lies at the border between the X-ray-only data and the tSZ+X combined results. The differences observed between X-ray-only and the combined tSZ+X results could have a physical and observational origin. For such a high redshift cluster X-ray observations become challenging. If the southwestern sub-clump in the cluster is really a hot but not dense structure (as suggested by \citet{jeetyson}), the electron density measurements from X-ray observations might be a struggle.

    We also show in Fig.~\ref{fig:finalmasses} the lensing, dynamical and Virial mass estimates. The lensing mass estimates from this work for the CLASH LTM and PIEMD+eNFW convergence maps are presented as dark orange and red contours, respectively. We observe that they are consistent with the lensing mass from \citet{serenocomalit, serenocovone2013} as well as with HSE mass estimates, but very different from the \citet{merten2015} lensing estimate for the reasons explained in Sect.~\ref{sec:lensmass}. The Virial masses estimated in \citet{mroczkowski2011} and \citet{mroczkowski2012} without X-ray data are shown as pink squares. They rely on the Virial relation and on given pressure and density profile models to relate directly the integrated tSZ flux to the mass (Eq.~15 in \citet{mroczkowski2011}). This kind of analysis seems a good alternative to the HSE mass 
    for clusters without X-ray data. The dynamical mass estimates (purple circles), which we would expect to be larger than the HSE estimate, appear particularly low for CL~J1226.9+3332 \citep{aguado2021}. According to the $M^{SZ}_{500}-M^{dyn}_{500}$ scaling relation obtained from the analysis of 297 \textit{Planck} galaxy clusters in \citet{aguado2021} (Eq.~8 and Table~2) and considering $M^{SZ}_{500}$ the value in \citet{planck2016}, the dynamical mass corresponding to CL~J1226.9+3332 should be in a range between $6-7.5 \times 10^{14} \hspace{2pt} \mathrm{M}_\odot$, thus more in agreement with our lensing mass estimates. Nonetheless, the obtained mass from the velocity dispersion measured on $\sim 50$ galaxy members should be a safe estimate and we do not find a clear reason to explain such low masses. The orientation of the merger could be one possible answer: if the merger is happening in the plane of the sky, the dispersion, and thus the mass, are lower. 
    
    \subsection{Hydrostatic-to-lensing mass bias}
     \label{sec:bias}
    \subsubsection{Hydrostatic mass bias problem}
    We do not expect the HSE to be fulfilled by all galaxy clusters in the Universe. We define the hydrostatic mass bias as
    \begin{equation}
         b = (M^{\mathrm{true}} - M^{\mathrm{HSE}})/M^{\mathrm{true}}
    \end{equation}
    where $M^{\mathrm{true}}$ is the total true mass of the cluster.

    From the observational point of view there are hints of a non null hydrostatic bias. As mentioned in Sect.~\ref{sec:intro}, one example is the tension observed between the cosmological parameters derived from \textit{Planck} cluster number counts and those from the CMB analyses \citep{planck2014b}. A possible explanation for this tension is that cluster masses are underestimated.
    According to \citet{planck2016a} the bias needed to reconcile the cosmological constraints obtained from the CMB power spectra to the cluster counts is $1 - b = M_{500}^{\rm{HSE}}/M_{500}^{\rm{true}} = 0.58 \pm 0.04$. A compatible value was obtained from the updated analysis in \citet{salvati2019}, $1 - b=M_{500}^{\rm{HSE}}/M_{500}^{\rm{true}} = 0.62 \pm 0.05$.
    
    The hydrostatic bias has also been the topic of a large number of studies on numerical simulations \citep[see][and references therein]{ansarifard2020,gianfagna}. 
    However, simulation-based analyses agree on values in the range of 0.75 to 0.9 for $1-b$, not able, therefore, to reconcile the mentioned tension.

    In a hierarchical formation scenario one would expect clusters at higher redshift to be more disturbed and therefore the hydrostatic equilibrium hypothesis to be less valid  \citep{neto2007, angelinelli2019}. Thus, this possible redshift evolution has been studied in the literature: e.g. \citet{salvati2019} find a modest hint of redshift dependence for the bias. However, other works based on cluster observations \citep{McDonald2017} do not find traces of evolution of the morphological state with redshift. To date, the possible bias dependence with redshift is not confirmed in simulations \citep[see][and references therein]{gianfagna}.


   
    \subsubsection{
    Hydrostatic-to-lensing mass bias estimates}

    From the observational side the real HSE bias is unachievable, as one cannot access the true mass of a cluster. But it can be approximated using mass estimates that do not rely on the HSE hypothesis and trace the total mass of the cluster, for instance the lensing mass. In this work we have computed the hydrostatic-to-lensing mass bias using the results obtained in Sect.~\ref{sec:hsemass} and \ref{sec:lensmass} (see \citet{serenoettoricomalit} for an analysis of the CoMaLit samples). For the lensing mass $M_{500}^{\mathrm{lens}}$ we combine the probability distributions of both lensing models in Fig.~\ref{fig:m500-nika2charles}. On the contrary, we consider the different HSE mass estimates independently. Assuming that HSE and lensing masses are uncorrelated estimates, we have combined their probability distributions and computed the ratio, $M_{500}^{\rm{HSE}}/M_{500}^{\rm{lens}} = 1 - b_{\mathrm{HSE/lens}} $. 
    
\begin{figure}
        \centering
        \includegraphics[trim={1cm 0cm 1.5cm 0cm},scale=0.33]{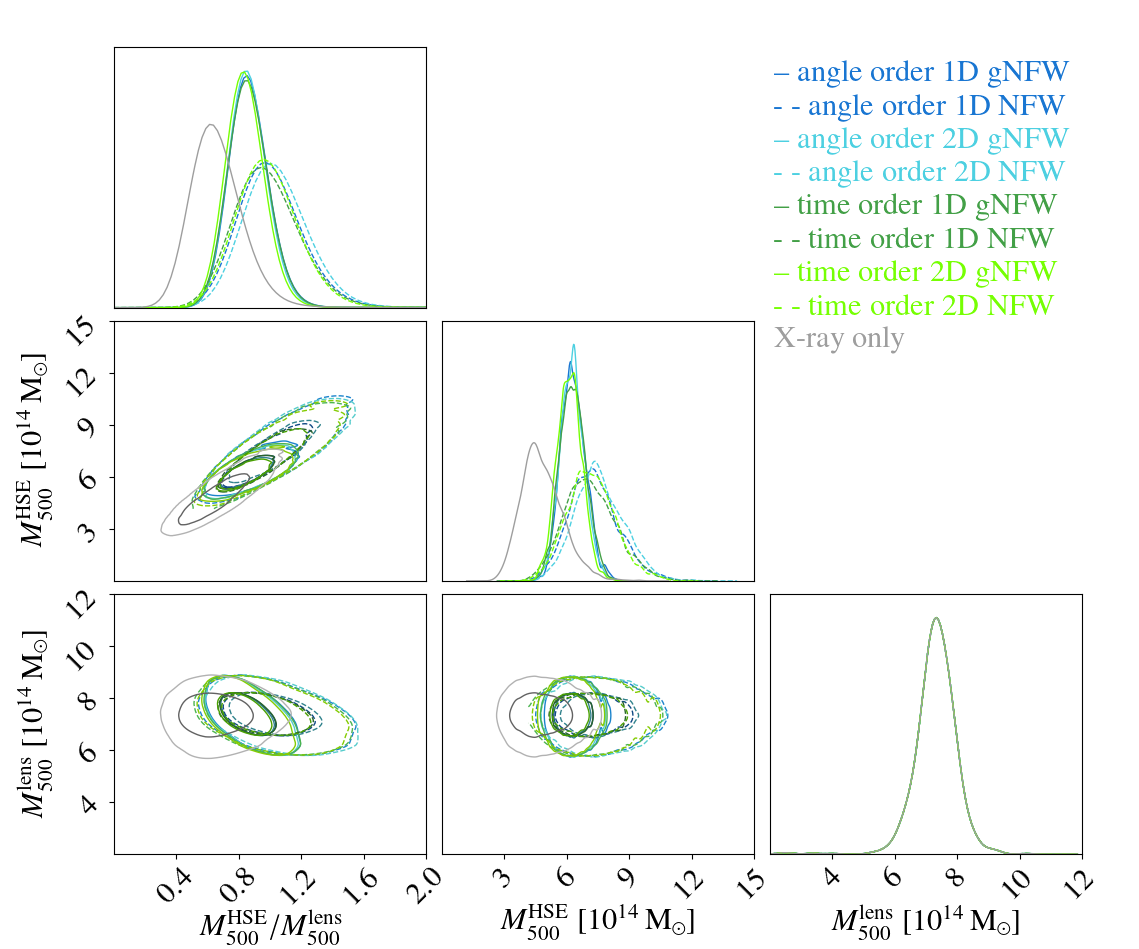}
        \caption{HSE mass estimates with respect to lensing ones at $R_{500}$. Blue and green solid lines correspond to the HSE masses obtained from the gNFW pressure model fit and dashed lines to the NFW method. Grey line correspons to the HSE mass obtained with XMM-\textit{Newton} only data. We present the $1\sigma$ and $2\sigma$ C.L. contours. The lensing mass distribution is the combination of the results for the PIEMD+eNFW and LTM analyses.}
        \label{fig:bias}
    \end{figure}

    We present in Fig.~\ref{fig:bias} the hydrostatic-to-lensing mass ratio at $R_{500}$. The same color code as in previous figures is used to distinguish the HSE estimates that have been obtained with each of the NIKA2 noise and filtering estimators. We draw with solid lines the results obtained modeling the pressure profile with the gNFW model and with dashed lines the results from the NFW fit. All of them are inferred from the NIKA2, R18 and XMM-\textit{Newton} data. The grey contours show the bias using the X-ray-only HSE mass results.  We present in Table~\ref{tab:bias} the marginalized hydrostatic-to-lensing mass ratio and uncertainties for the different cases considered. The results for the gNFW and NFW cases correspond to a combination of the four NIKA2 analyses. As concluded in \citet{ferragamo}, we observe that the value of the bias is sensitive to the considered data sets and modeling choices. 
    
    \begin{table}
    \centering
    \caption{Hydrostatic-to-lensing mass ratio for different HSE mass estimates. We present the mean value and the 84th and 16th percentiles.}
    \label{tab:bias}       
    \begin{tabular}{ll}
      \hline
      \hline
    HSE mass estimates &  $1 - b_{\mathrm{HSE/lens}} $  \\\hline
    (tSZ+X-ray)$_{\mathrm{gNFW}}$ & $0.86^{+0.10}_{-0.11}$\\
    (tSZ+X-ray)$_{\mathrm{NFW}}$ &  $1.01^{+0.17}_{-0.19}$\\
    X-ray &  $0.66^{+0.14}_{-0.14}$\\ \hline
    \end{tabular}
    \vspace*{0.5cm}  
    \end{table}

\section{Summary and conclusions}
\label{conclusion}

The precise estimation of the mass of single clusters appears extremely complicated and affected by multiple systematic effects \citep{pratt2019}, but it is the key for building accurate scaling relations for cosmology. The NIKA2 SZ Large Program aims at providing a SZ-mass scaling relation from the combination of NIKA2 and XMM-\textit{Newton} data. Within this program, we presented a thorough study on the mass of the CL~J1226.9+3332 galaxy cluster. 

We obtained NIKA2 150 and 260~GHz maps, that allowed us to reconstruct the \textit{Radially-binned} pressure profile for the ICM from the tSZ data. To characterize the impact of the data processing, we repeated the whole analysis for two pipeline-filtering transfer functions and noise estimates for the 150~GHz map. We accounted for the presence of point sources that contaminate the negative tSZ signal at 150~GHz. The reconstructed NIKA2 pressure bins are compatible, within
the angular scales accessible to NIKA2, with the profiles obtained from three independent instruments in R18. This validates the pressure reconstruction procedure that will be used for the whole sample analysis in the NIKA2 SZ Large Program.

We compared two approaches to estimate the HSE mass from the combination of tSZ-obtained pressure and X-ray electron density profiles. We considered either a gNFW pressure model (traditionally used for this kind of analyses) or an integrated NFW density model. The second seems a promising approach to ensure radially increasing HSE mass estimates. 
However, other density models should be tested, in order to describe more satisfactorily the shape of the pressure profile. Both methods give completely compatible HSE mass profiles and integrated $M^{\mathrm{HSE}}_{500}$. 
From the comparison of the different mass estimates, we also conclude that for the moment, when estimating the HSE mass in the NIKA2 SZ Large Program, the error budget is dominated by model dependence rather than by the instrumental and data processing systematic effects that we investigated.
In addition, these results are in agreement with the X-ray-only HSE mass estimate obtained in this paper from the XMM-\textit{Newton} electron density and temperature profiles. Nevertheless, the latter prefers lower mass values than the combined tSZ+X-ray results. 

We have also shown that our results are compatible with all the HSE mass estimates found in the literature within uncertainties, which are large. We think that the only way to reduce the current uncertainties is to precisely constrain the slope of the mass profile at $\sim R_{500}$, as we have proved that very similar mass profiles overall can result in significant differences at $M_{500}$.  

From two differently modeled CLASH convergence maps, we have reconstructed the lensing mass profile of CL~J1226.9+3332 and measured the hydrostatic-to-lensing mass bias. We have found that for CL~J1226.9+3332 the bias is $b_{\mathrm{HSE/lens}} \sim 0.1$ for tSZ+X-ray combined HSE masses and $\sim 0.4$ for X-ray-only estimates, with $\sim 0.1$ uncertainties depending on the models. In spite of the large uncertainties, we have the sensitivity to measure the bias of a single cluster, even for the highest-redshift cluster of the NIKA2 SZ Large Program. This is the second cluster (the first one was studied in \citet{ferragamo}) in the NIKA2 SZ Large Program with such a measurement and the analysis of the HSE-to-lensing bias for a larger sample will be the topic of a forthcoming work (as done in \citet{bartalucci2018} with X-ray data). Measuring the hydrostatic-to-lensing mass bias associated with LPSZ clusters will bring a new perspective to the bias from the tSZ side for spatially resolved clusters at intermediate and high redshifts.

\begin{acknowledgements}
We would like to thank the IRAM staff for their support during the campaigns. The NIKA2 dilution cryostat has been designed and built at the Institut N\'eel. In particular, we acknowledge the crucial contribution of the Cryogenics Group, and in particular Gregory Garde, Henri Rodenas, Jean-Paul Leggeri, Philippe Camus. This work has been partially funded by the Foundation Nanoscience Grenoble and the LabEx FOCUS ANR-11-LABX-0013. This work is supported by the French National Research Agency under the contracts ``MKIDS'', ``NIKA'' and ANR-15-CE31-0017 and in the framework of the ``Investissements d’avenir'' program (ANR-15-IDEX-02). This work has benefited from the support of the European Research Council Advanced Grant ORISTARS under the European Union's Seventh Framework Programme (Grant Agreement no. 291294). This work is based on observations carried out under project number 199-16 with the IRAM 30m telescope. IRAM is supported by INSU/CNRS (France), MPG (Germany) and IGN (Spain). E. A. acknowledges funding from the French Programme d’investissements d’avenir through the Enigmass Labex. A. R. acknowledges financial support from the Italian Ministry of University and Research - Project Proposal CIR01\_00010. M. D. P. acknowledges support from Sapienza Universit\`a di Roma thanks to Progetti di Ricerca Medi 2021, RM12117A51D5269B. G. Y. would like to thank the Ministerio de Ciencia e Innovaci\'on (Spain) for financial support under research grant PID2021-122603NB-C21. This project was carried out using the python libraries \texttt{matplotlib} \citep{Hunter2007}, \texttt{numpy} \citep{Harris2020} and \texttt{astropy} \citep{Astropy2013, Astropy2018}.
\end{acknowledgements}

\bibliographystyle{aa} 
\bibliography{mybibliography}
%
%

%

%

\begin{appendix}
   
\section{Masses from the literature}
We present in Table~\ref{tab-masses} all the mass estimates found in literature for \clj\ described in Sect.~\ref{sec:allmasses}. We differentiate the masses reconstructed from ICM observables, from the lensing effect on background sources and from the study of the dynamics of member galaxies. 

Most of the masses have been computed from spherical models and we give the radius at which each mass is evaluated when available. When the mass has been evaluated at a given $R = R_{\Delta}$ we also present the value of the density contrast $\Delta$.

    \begin{table*}
    \centering
    \caption{Mass estimates found in literature for CL~J1226.9+3332.}
    \label{tab-masses}       
    \small
    \begin{tabular}{llllll}
      \hline
      \hline
    Observable & $R$  & $\Delta$ & $M \hspace{1pt}(<R)$ & Reference & Notes  \\
    ~ & [kpc] & ~ &  [$10^{14}$ M$_{\odot}$] &~ &~\\ \hline
    ICM &  ~ & ~ & ~ & ~& ~\\\hline
    ~& $340 \; h_{100}^{-1}$ & - & $3.9\pm 0.5$ & \citet{joy} & ~\\
    ~ & 1000 &  - & $14^{+6}_{-4}$ & \citet{jeetyson} & Projected \\
    ~ & $730\pm 40$ & 1000 & $6.1^{+0.9}_{-0.8} $ &\citet{maughan2}& ~\\
    ~ & $1660\pm 340$ & 200 & $14 \pm 4 $ &\citet{maughan2}& ~\\
    ~ & $880\pm 50$ & 500 & $5.2^{+1.0}_{-0.8} $ & \citet{maughan1}& ~\\
    ~ & $1000\pm 50$ & 500 & $7.8 \pm 1.1 $ & \citet{mantz2010}& ~\\
    ~ & $980^{+100}_{-70}$ & 500 & $7.37^{+2.50}_{-1.57} $ & \citet{mroczkowski2009}& ~\\
    ~ & $410^{+10}_{-10}$ & 2500 & $2.67^{+0.29}_{-0.27} $ & \citet{mroczkowski2009}& ~\\
    ~ & $980^{+90}_{-70}$ & 500 & $7.30^{+2.10}_{-1.51} $ & \citet{mroczkowski2009}& ~\\
    ~ & $420^{+40}_{-30}$ & 2500 & $2.98^{+0.90}_{-0.63} $ & \citet{mroczkowski2009}& ~\\
    ~ & $940^{+20}_{-20}$ & 500 & $6.49^{+0.34}_{-0.34} $ & \citet{mroczkowski2011, mroczkowski2012}& ~\\
    ~ & $390^{+10}_{-10}$ & 2500 & $2.35^{+0.15}_{-0.16} $ & \citet{mroczkowski2011, mroczkowski2012}& ~\\
    ~ & $940^{+20}_{-20}$ & 500 & $6.42^{+0.36}_{-0.36} $ & \citet{mroczkowski2011, mroczkowski2012}& ~\\
    ~ & $400^{+10}_{-10}$ & 2500 & $2.53^{+0.14}_{-0.15} $ & \citet{mroczkowski2011, mroczkowski2012}& ~\\
    ~& $1140^{+100}_{-80}$ & 200 & $7.19^{+1.33}_{-0.92} $ &
    \citet{muchovej}&~\\
    ~& $310^{+30}_{-20}$ & 2500 & $1.68^{+0.37}_{-0.26} $ &
    \citet{muchovej}&~\\
    ~& $812^{+71}_{-81}$ & 500 & $4.25^{+1.22}_{-1.14} $& \citet{bulbul}&  ~\\
    ~& $379^{+37}_{-41}$  & 2500 & $2.16^{+0.69}_{-0.63} $& \citet{bulbul}& ~\\   
    ~ & - & 500 & $5.7^{+0.63}_{-0.69}$ & \citet{planck2016} & Scaling relation \\
    ~ & $930^{+50}_{-43}$&500 &$5.96^{1.02}_{-0.79}$ & \citet{adam2} & ~\\
    ~ & $937^{+72}_{-58}$&500 &$6.10^{1.52}_{-1.06}$ & \citet{adam2} & ~\\
    ~ & $995^{+65}_{-65}$&500 &$7.30^{1.52}_{-1.34}$ & \citet{adam2} & ~\\\hline
    Lensing  &  ~ & ~ & ~ & ~& ~\\\hline
    ~ & $1640 \pm 100$ & 200 &$13.8 \pm 2.0$ & \citet{jeetyson} & ~\\
    ~ & $880 \pm 50$ & - & $7.34 \pm 0.71$ & \citet{jeetyson} & ~\\
    ~ & 155 & - &$1.3 \pm 0.1$ & \citet{jeetyson} & Projected \\
    ~ & 155 & - &$0.85 \pm 0.06$ & \citet{jeetyson} & Projected (SW clump)\\
    ~ & $1680^{+100}_{-90}$ & 200 & $13.7^{+2.4}_{-2.0}$ & \citet{jee2011} & ~\\
    ~ & - & 200 & $22.3 \pm 1.4$ & \citet{merten2015} & ~\\
    ~ & - & 500 & $15.4 \pm 1.2$ & \citet{merten2015} & ~\\
    ~ & - & 2500 & $6.1 \pm 1.0$ & \citet{merten2015} & ~\\
    ~ & - & 200 & $10.0\pm 2.4$ & \citet{serenocovone2013} & ~\\
    ~ & - & 200 & $11.114\pm 2.442$ & \citet{serenocovone2013, serenocomalit}$^{\ref{comalitref}}$ & ~\\
    ~ & - & 500 & $7.96\pm1.44$ & \citet{serenocovone2013,serenocomalit}$^{\ref{comalitref}}$ & ~\\
    ~ & - & 2500 & $3.45\pm 0.37$ & \citet{serenocovone2013,serenocomalit}$^{\ref{comalitref}}$ & ~\\
    ~ & 500 & - & $3.947\pm 0.285$ & \citet{serenocovone2013,serenocomalit}$^{\ref{comalitref}}$ & ~\\
    ~ & 1000 & - & $7.882\pm 1.013$ & \citet{serenocovone2013,serenocomalit}$^{\ref{comalitref}}$ & ~\\
    ~ & 1500 & - & $10.938\pm 1.784$ & \citet{serenocovone2013,serenocomalit}$^{\ref{comalitref}}$ & ~\\

    \hline
    Galaxy dynamics &  ~ & ~ & ~ & ~& ~\\\hline
    ~ & - & 500 & $4.7\pm 1.0$ & \citet{aguado2021} & ~ \\
    ~ & - & 500 & $4.8\pm 1.0$ & \citet{aguado2021} & ~ \\
    \hline
    \end{tabular}
    \vspace*{0.5cm}  
    \end{table*}

\section{Pressure bins and point sources fluxes at 150~GHz from joint fit}
We present in Tables~\ref{tab-3} and \ref{tab-4} the obtained point source fluxes at 150~GHz from the joint fit of the point sources with the pressure profile of the cluster in Sect.~\ref{sec:pressure}. We show the results for the nine sources (PS1 to PS9) and each column corresponds to the value obtained for each of the four analyses. The reconstructed fluxes are consistent within all the considered cases.

\begin{figure*}[b]
        \hfill
        \begin{minipage}[b]{0.24\textwidth}
        \includegraphics[trim={0cm 0cm 0cm 0cm},scale=0.33]{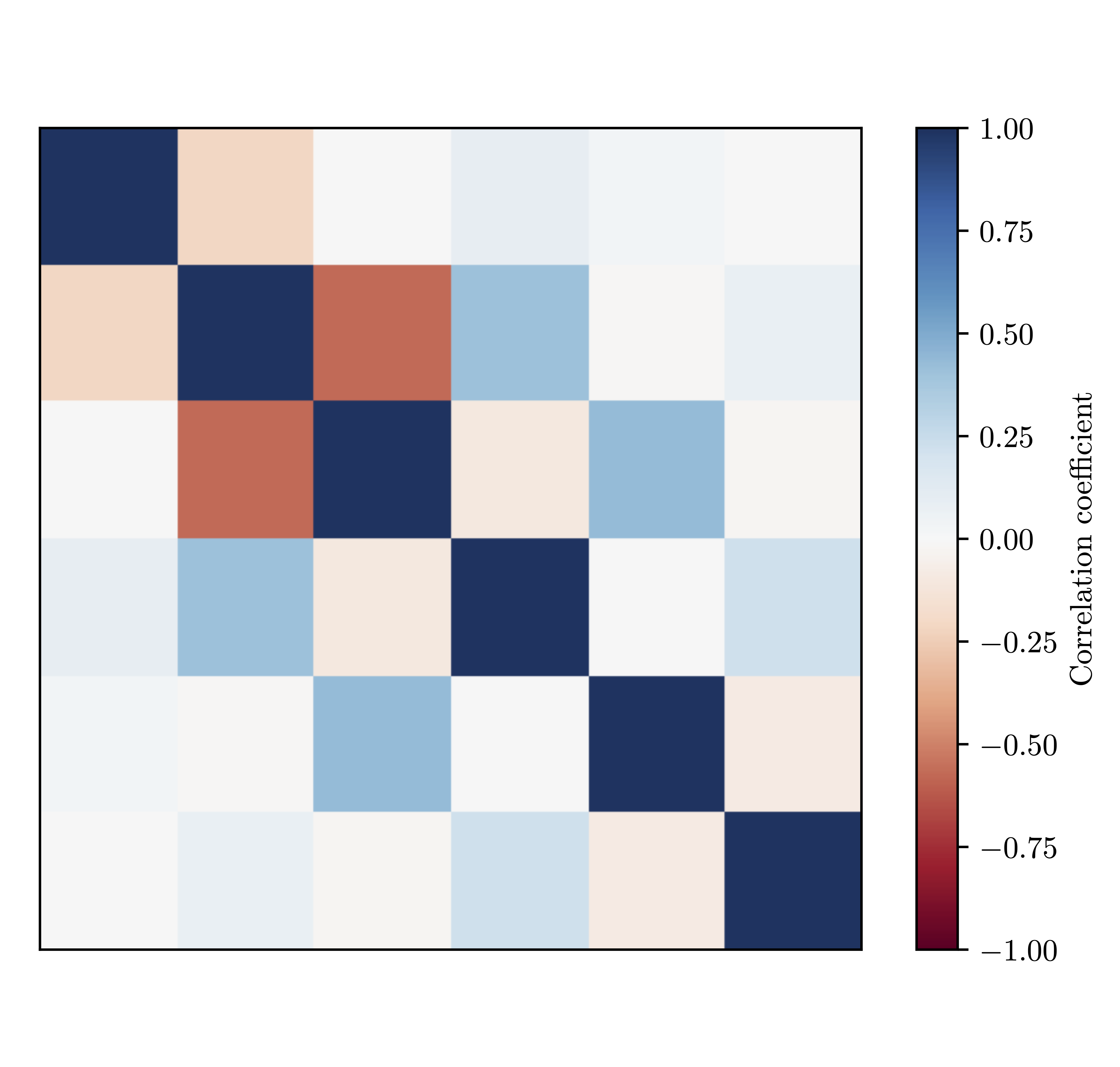}
        \end{minipage}
        \begin{minipage}[b]{0.24\textwidth}
        \includegraphics[trim={0cm 0cm 0cm 0cm},scale=0.33]{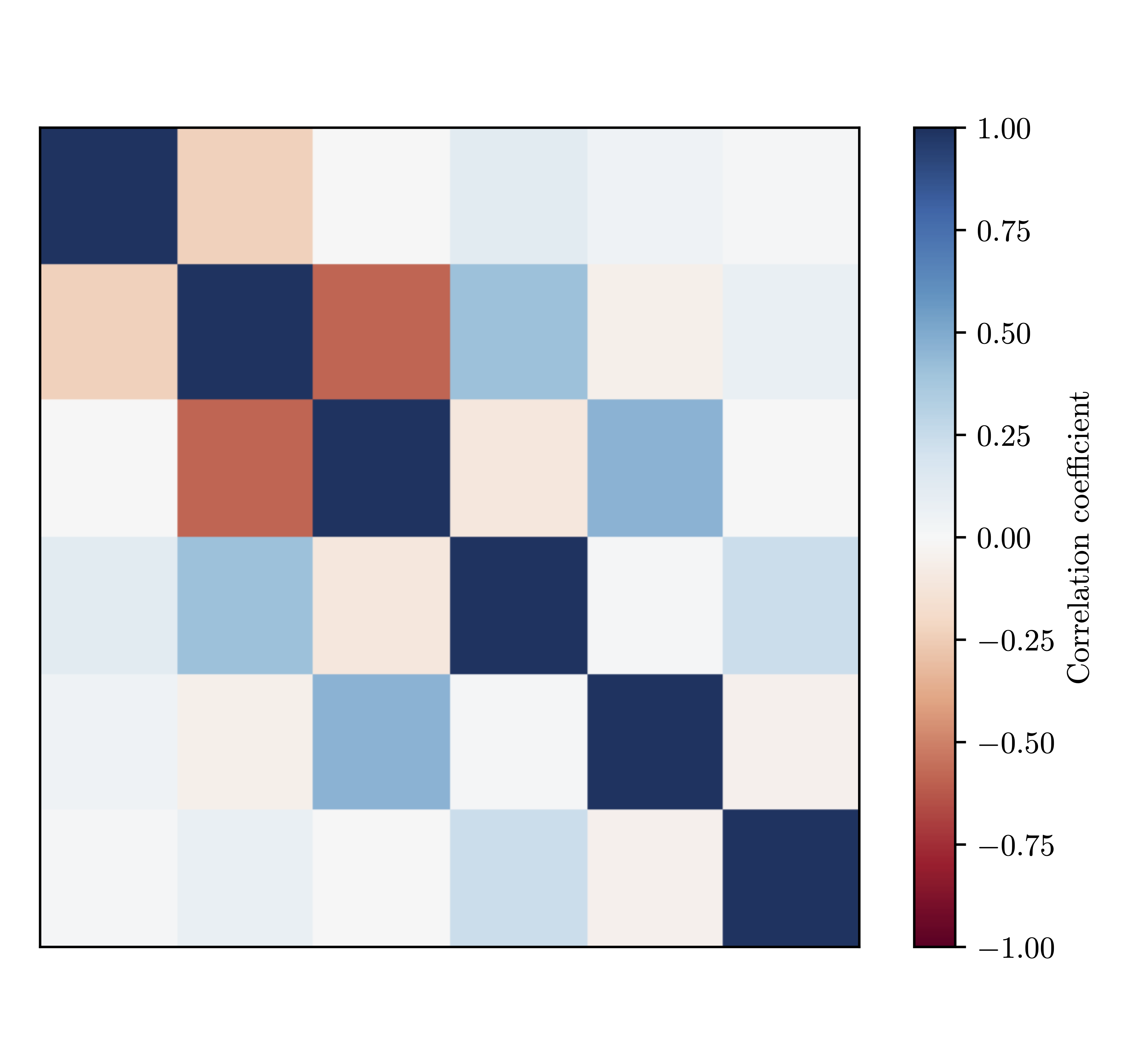}
        \end{minipage}
        \begin{minipage}[b]{0.24\textwidth}
        \includegraphics[trim={0cm 0cm 0cm 0cm},scale=0.33]{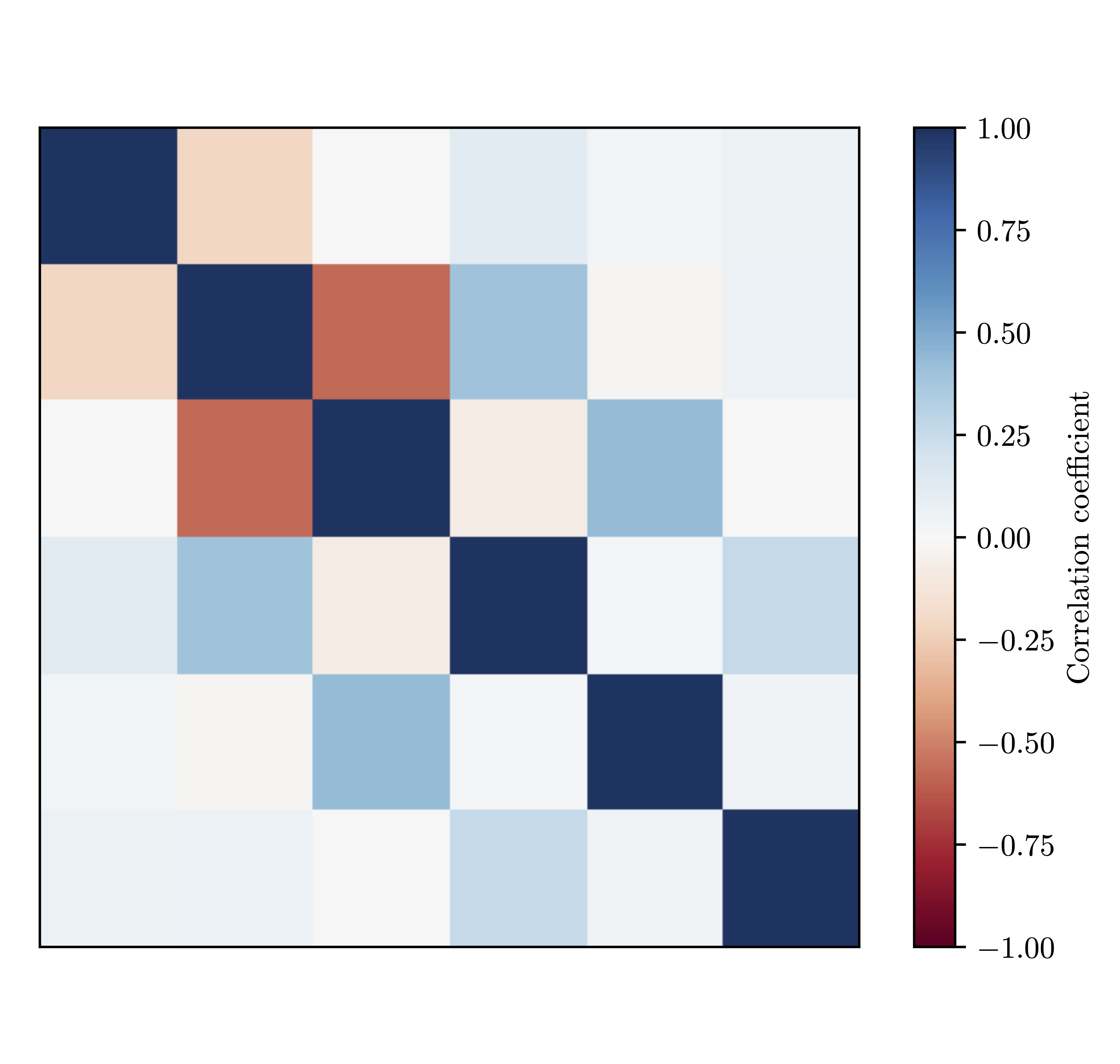}
        \end{minipage}
        \begin{minipage}[b]{0.24\textwidth}
        \includegraphics[trim={0cm 0cm 0cm 0cm},scale=0.33]{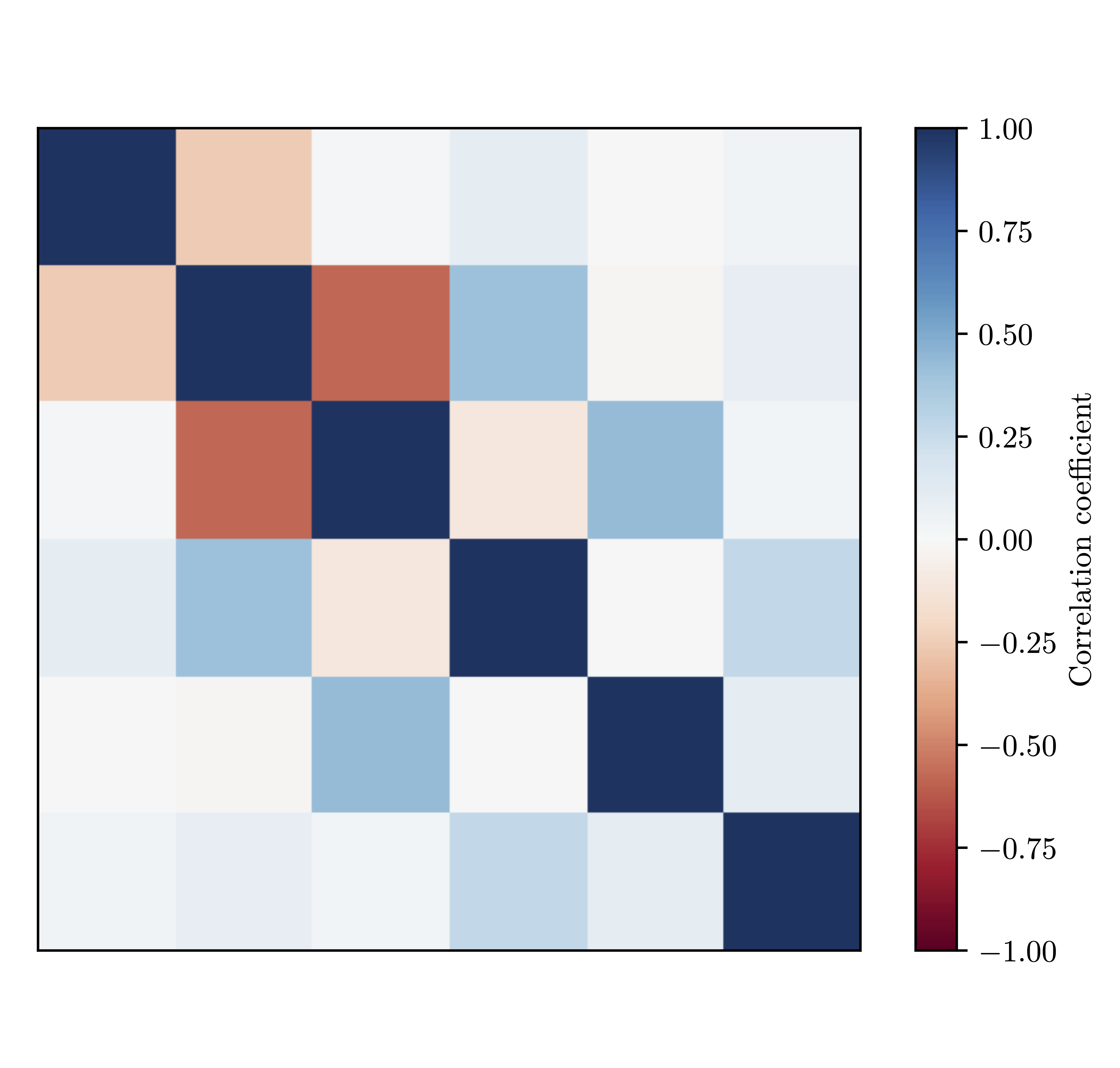}
        \end{minipage}
    \caption{Correlation matrices of \textit{Radially-binned} electron pressure bins. From left to right results obtained with different transfer function and noise estimates: AO1D, TO1D, AO2D and TO2D. }  
    \label{fig:bincorrelations}
    \end{figure*}

The binned pressure reconstructions are also tabulated in Table~\ref{tab:pressurebins} and the bin correlation matrices are shown in Fig.~\ref{fig:bincorrelations}. 

    \begin{table*}
    \centering
    \caption{Submillimetric point sources fluxes. Fluxes at 150 GHz obtained from the joint point sources and pressure profile fits in Sect.~\protect\ref{sec:pressure}.}
    \label{tab-3}       
    \begin{tabular}{llllll}
      \hline
      \hline
    Source & 150 GHz & 150 GHz & 150 GHz & 150 GHz   \\
    ~&  [``angle order'' 1D] &  [``time order'' 1D] &  [``angle order'' 2D] &[``time order'' 2D] \\
    ~ & [mJy] & [mJy] & [mJy] & [mJy] \\\hline
    PS1 &  2.0$^{+0.1}_{-0.2}$ & 1.9$^{+0.2}_{-0.2}$ & 2.0$^{+0.2}_{-0.2}$ & 1.9$^{+0.2}_{-0.2}$\\
    PS2 &  0.9$^{+0.5}_{-0.1}$ & 0.9$^{+0.2}_{-0.1}$ & 0.9$^{+0.1}_{-0.2}$ & 0.9$^{+0.2}_{-0.1}$\\
    PS3 &  1.3$^{+0.2}_{-0.1}$ & 1.3$^{+0.2}_{-0.1}$ & 1.4$^{+0.2}_{-0.2}$ & 1.3$^{+0.2}_{-0.2}$\\
    PS4 &  0.4$^{+0.1}_{-0.1}$ & 0.38$^{+0.10}_{-0.07}$ & 0.38$^{+0.11}_{-0.07}$& 0.39$^{+0.09}_{-0.08}$\\
    PS5 &  0.6$^{+0.1}_{-0.1}$ & 0.5$^{+0.2}_{-0.1}$ & 0.6$^{+0.2}_{-0.1}$ & 0.5$^{+0.2}_{-0.1}$ \\
    PS6 &  0.3$^{+0.2}_{-0.1}$& 0.2$^{+0.2}_{-0.1}$ & 0.3$^{+0.1}_{-0.2}$ & 0.2$^{+0.2}_{-0.1}$\\ 
    PS7 &  0.03$^{+0.08}_{-0.03}$& 0.03$^{+0.08}_{-0.03}$& 0.03$^{+0.08}_{-0.03}$ & 0.04$^{+0.08}_{-0.04}$\\
    PS8 &  0.5$^{+0.1}_{-0.1}$ & 0.45$^{+0.12}_{-0.09}$& 0.45$^{+0.13}_{-0.09}$& 0.44$^{+0.14}_{-0.07}$ \\\hline
    \end{tabular}
    \vspace*{0.5cm}  
    \end{table*}
    
    \begin{table*}
    \centering
    \caption{Radio point source identified in the center of the cluster. Fluxes obtained from the joint fits with the pressure profile of the cluster.}
    \label{tab-4}       
    \begin{tabular}{lllll}
      \hline
      \hline
    Source  & 150 GHz & 150 GHz & 150 GHz & 150 GHz    \\
    ~&  ["angle order" 1D] &  ["time order" 1D] &  ["angle order" 2D] &["time order" 2D]\\
    ~  & [mJy] & [mJy] & [mJy] & [mJy]  \\\hline
    PS9 &  0.06$^{+0.11}_{-0.05}$ & 0.07$^{+0.11}_{-0.06}$& 0.06$^{+0.09}_{-0.05}$& 0.06$^{+0.10}_{-0.04}$\\\hline
    \end{tabular}
    \vspace*{0.5cm}  
    \end{table*}

    \begin{table*}
    \centering
    \caption{\textit{Radially-binned} pressure profile fits. Mean values and the $1\sigma$ uncertainties. Last bin results are trimmed.}
    \label{tab:pressurebins}       
    \begin{tabular}{lllll}
      \hline
      \hline
    $r$   & $P_{e}$ & $P_{e}$ &  $P_{e}$ & $P_{e}$ \\
    ~ &  [``angle order'' 1D] &  [``time order'' 1D] &  [``angle order'' 2D] &[``time order'' 2D] \\
    $\mathrm{[kpc]}$ & [$10^4$ keV cm$^{-3}$] & [$10^4$ keV cm$^{-3}$] & [$10^4$ keV cm$^{-3}$] & [$10^4$ keV cm$^{-3}$] \\\hline
    35.0  & $115 \pm 1088$    & $190 \pm 1161$ & $371 \pm 1054$ &  $1689 \pm 1103$ \\
    69.8 & $1086 \pm 367 $    &  $921 \pm 395$ &  $730 \pm 329$ &  $519 \pm 360$ \\
    209.5  & $648 \pm 113$    &  $641 \pm 115$ & $730 \pm 111$ &  $671 \pm 110$    \\
    341.7 & $262 \pm 32$     &  $244 \pm 33$ &  $261 \pm 32$ &  $244 \pm 33$\\
    900.0 & $30 \pm 4$    &  $30 \pm 4$ & $29 \pm 4$ & $28 \pm 4$  \\
    1500.0 & $ 2 \pm 2$    &  $4 \pm 2$ &  $4 \pm 3$ & $8 \pm 3$  \\\hline
    \end{tabular}
    \vspace*{0.5cm}  
    \end{table*}

\section{Pressure profile fits}
\label{sec:gnfwparamappend}
\begin{figure*}
    \centering
    \includegraphics[scale=0.8]{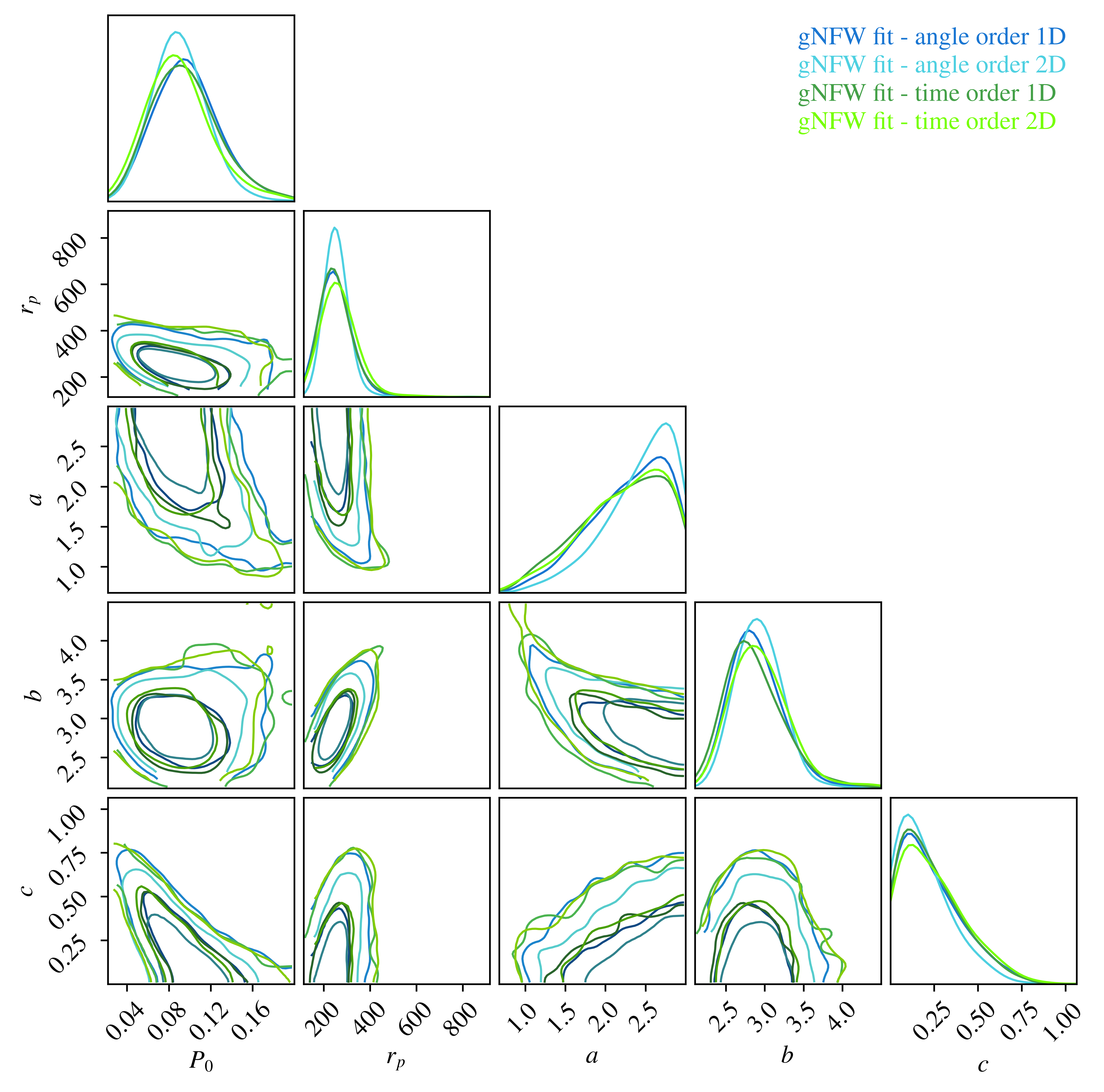}  
   
    \caption{Posterior distributions of the parameters obtained in the fit of the NIKA2 and R18 pressure bins for the gNFW pressure model.}
    \label{fig:gnfwparams}
\end{figure*}

\begin{figure*}
        \centering
        \includegraphics[scale=0.28]{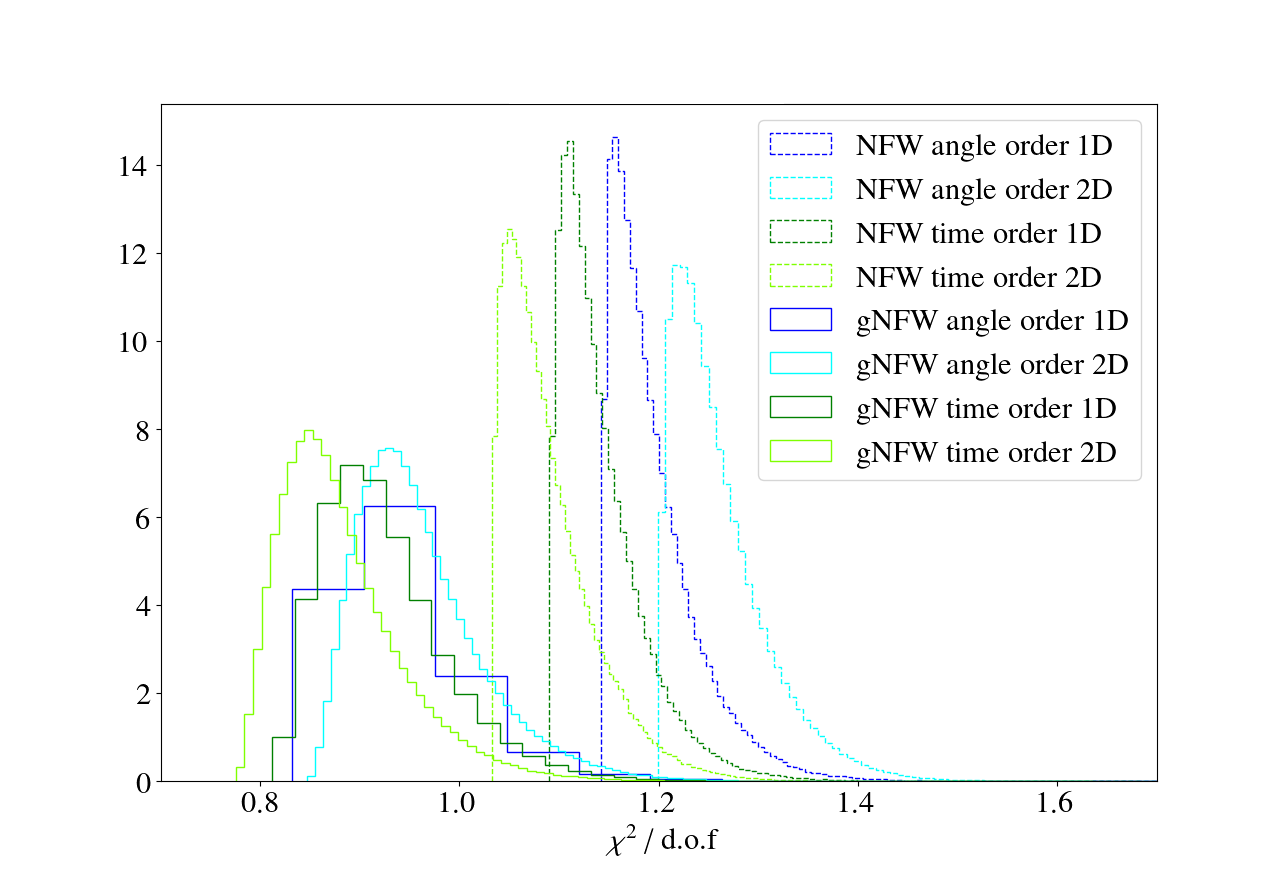}
        \caption{Reduced $\chi^2$ of the gNFW (solid) and NFW (dashed) model fits to the NIKA2 and R18 pressure bins. The different colors show the NIKA2 bins that have been used.}
        \label{chi2}
\end{figure*}

\begin{figure*}
    \centering
    
      \includegraphics[scale=0.80]{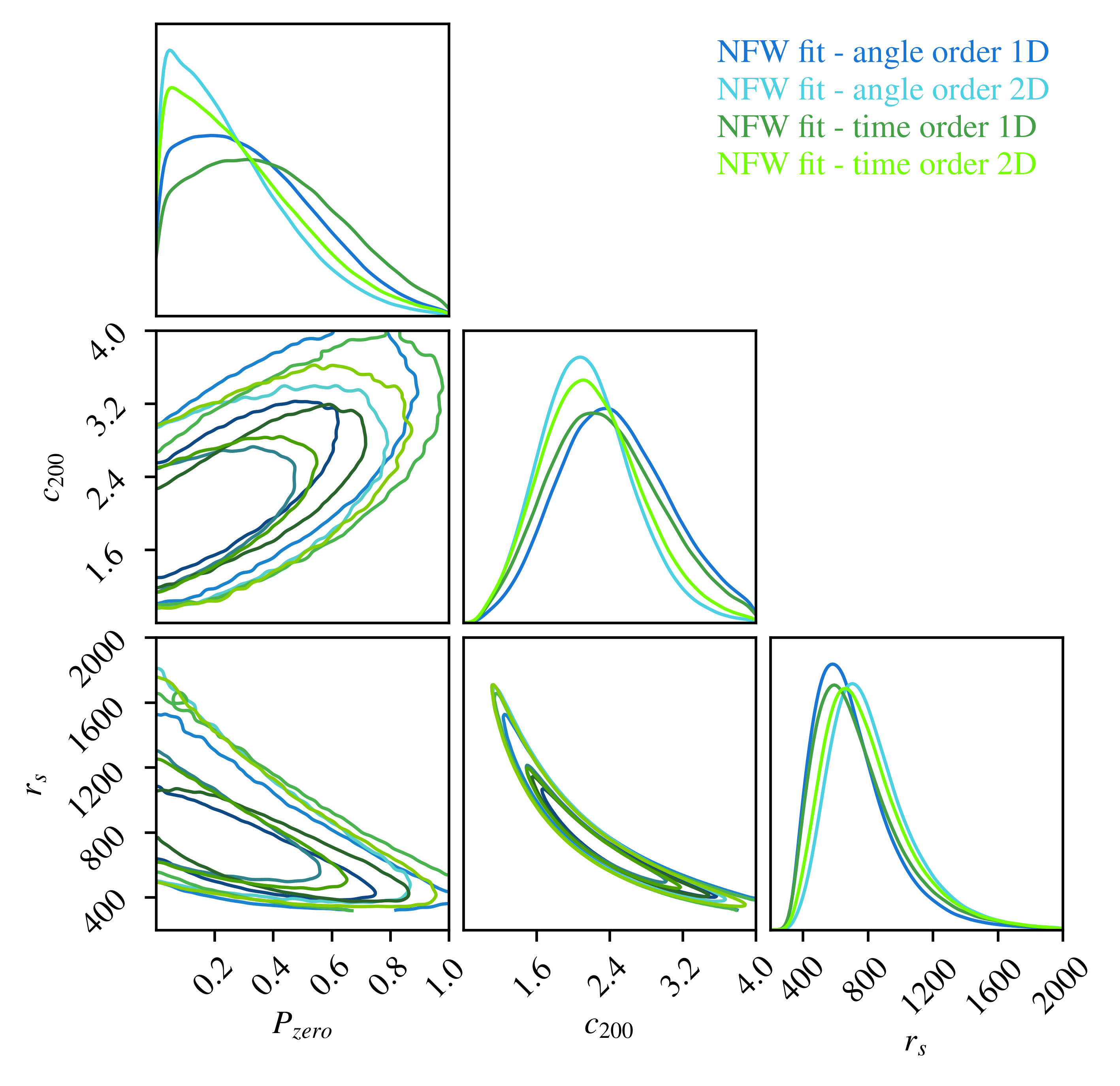}
   
    \caption{Posterior distributions of the parameters obtained in the fit of the NIKA2 and R18 pressure bins, combined with the XMM-\textit{Newton} electron density, for the NFW pressure model. The values of $P_{zero}$ have been multiplied by $10^3$.}  
    \label{fig:nfwparams}
\end{figure*}

We show in figures \ref{fig:gnfwparams} and \ref{fig:nfwparams} the posterior distributions of the parameters obtained from the fit of the pressure bins in Sect.~\ref{Sect:gnfw} and \ref{sec:nfwhsemass}. The reduced $\chi^{2}$ distributions corresponding to these fits are shown in Fig.~\ref{chi2}. 

\begin{figure*}
        \hfill
        \begin{minipage}[b]{0.51\textwidth}
        \includegraphics[trim={1cm 0cm 0cm 0cm},scale=0.25]{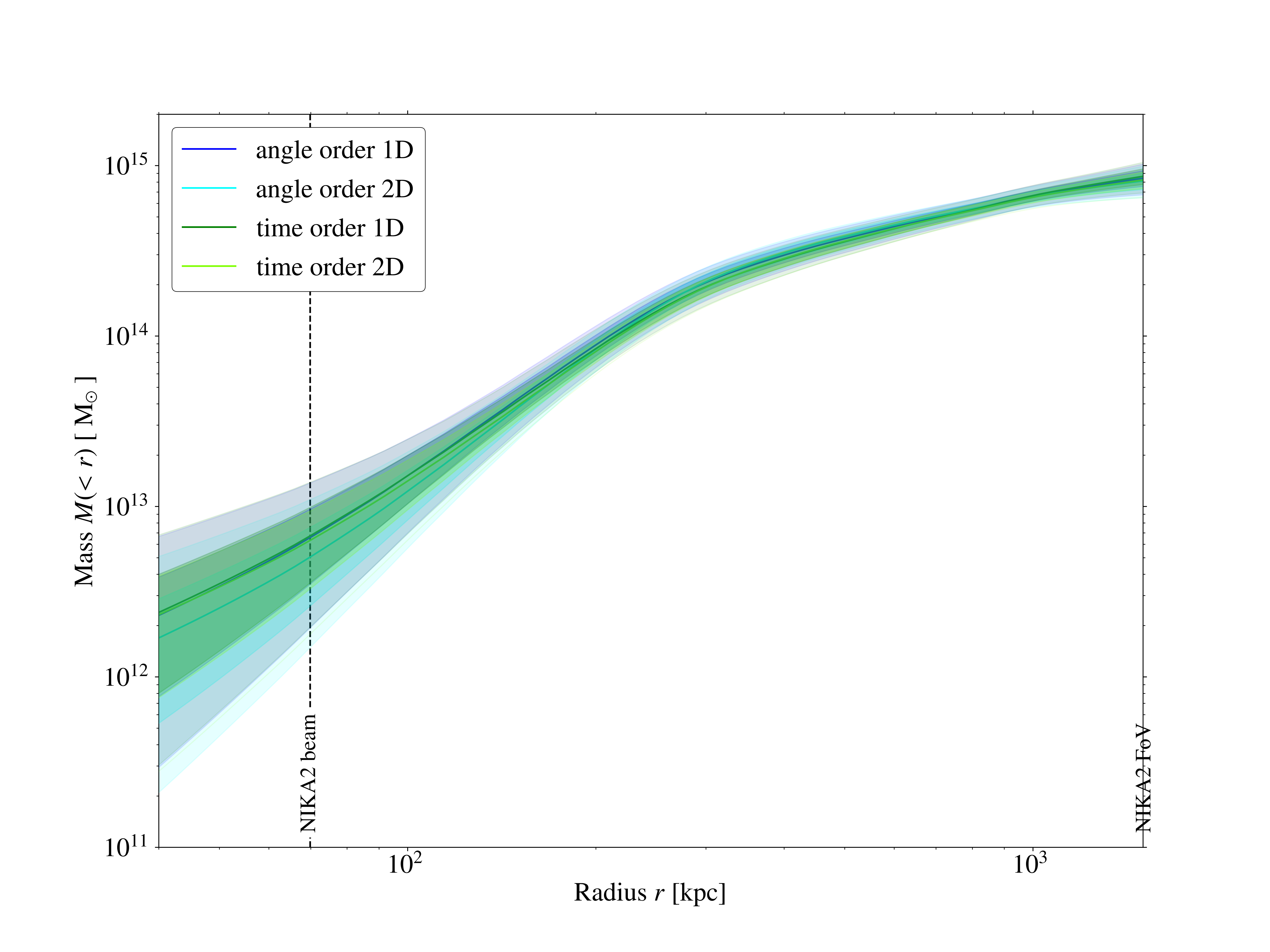}
        \end{minipage}
        \begin{minipage}[b]{0.48\textwidth}
        \includegraphics[trim={2cm 0cm 5cm 0cm},scale=0.25]{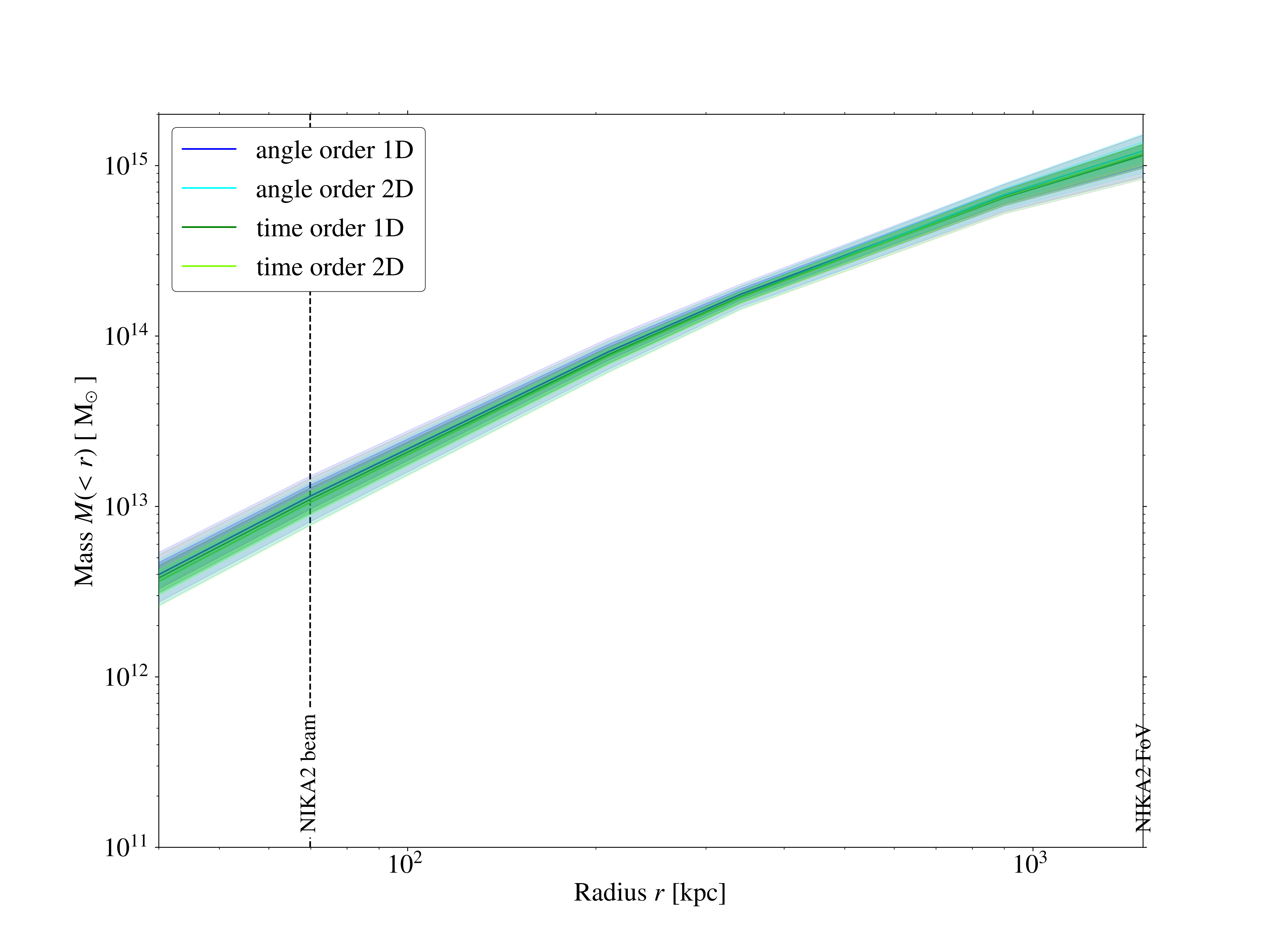}
        \end{minipage}
    \caption{HSE mass profiles obtainted with the gNFW (left) and NFW (right) fit methods. Each color represents the profile obtained using the corresponding NIKA2 pressure bins.}  
    \label{fig:massprofs}
    \end{figure*}

In Fig.~\ref{fig:massprofs} we present the HSE mass profiles resulting from each of the fits. 


\end{appendix}
\end{document}